\documentclass[amsmath,aps,prd,amssymb, tightenlines, superscriptaddress, nofootinbib, notitlepage ]{revtex4-2}

\usepackage[utf8]{inputenc}
\usepackage{float}
\usepackage{graphics,graphicx}
\usepackage{dcolumn}
\usepackage{bm}
\usepackage{amsmath}
\usepackage{tensor}
\usepackage{slashed}
\usepackage{amsmath,amssymb,amsfonts,mathrsfs,bbold}
\usepackage{yfonts}
\usepackage{xcolor}
\usepackage{mathtools}
\usepackage{hyperref} 
\usepackage[framemethod=TikZ]{mdframed}
\usepackage{color}

\usepackage{multirow}
\usepackage{mathbbol}
\usepackage{mathtools}

\usepackage{slashed}
\usepackage{simplewick}
\usepackage{physics}

\usepackage[normalem]{ulem}

\usepackage{stackrel}
\usepackage[most]{tcolorbox}

\usepackage{caption}
\usepackage{subcaption}

\def\eq#1{{Eq.~(\ref{#1})}}

\def\fig#1{{Fig.~\ref{#1}}}
\newcommand{\thalf}{\tfrac{1}{2}}

\newcommand{\as}{\alpha_s}

\newcommand{\cc}{\mbox{c.c.}}

\def\sec#1{{Sec.~\ref{#1}}}

\def\eq#1{{Eq.~(\ref{#1})}}
\def\fig#1{{Fig.~\ref{#1}}}
\newcommand{\ben}{\begin{eqnarray*}}
\newcommand{\een}{\end{eqnarray*}}

\newcommand{\un}[1]{\underline{#1}}

\newcommand{\half}{{1\over 2}}

\newcommand{\tord}{\textrm{T} \:}

\newcommand{\pd}{\partial}

\usepackage{ragged2e} 
\DeclareCaptionJustification{justified}{\justifying}

\newcommand{\oone}{
\begin{picture}(10,8)
\put(5,5){\circle{8}}
\put(2.9,2.5){{\scriptsize 1}}
\end{picture}
}

\newcommand{\otwo}{
\begin{picture}(10,8)
\put(5,5){\circle{8}}
\put(2.9,2.5){{\scriptsize 2}}
\end{picture}
}

\begin{document}

\title{On the Two $R$-Factors in the Small-$x$ Shockwave Formalism}

\author{Yuri V. Kovchegov} 
         \email[Email: ]{kovchegov.1@osu.edu}
         \affiliation{Department of Physics, The Ohio State
           University, Columbus, OH 43210, USA}

\author{M. Gabriel Santiago}
        \email[Email: ]{melvin.santiago@temple.edu}
        \affiliation{Center for Nuclear Femtography, SURA, 12000 Jefferson Avenue Newport News, VA 23606}
       \affiliation{Department of Physics, Old Dominion University, Norfolk, VA 23529}
        \affiliation{Jefferson Lab, Newport News, VA 23606}
        \affiliation{Department of Physics, Temple University, Philadelphia, PA 19122}

\author{Huachen Sun} 
         \email[Email: ]{sun.2885@osu.edu}
	\affiliation{Department of Physics, The Ohio State
           University, Columbus, OH 43210, USA}
           
\begin{abstract}
   There are two $R$-factors frequently used in the phenomenology of exclusive processes at small values of the Bjorken $x$ variable. One $R$-factor takes into account the effects of non-zero longitudinal momentum transfer, which is assumed to be zero in the dipole scattering amplitude. Another $R$-factor accounts for the real part of the elastic scattering amplitude which is often neglected, with the standard dipole scattering amplitude giving only the imaginary part of the elastic amplitude. 
   
   In this work we present two new theoretical developments aimed at eliminating the need for the two $R$-factors. We argue that the $R$-factors can be replaced by (i) modifying the argument of the dipole scattering amplitude and by (ii) augmenting the initial conditions for its non-linear small-$x$ evolution. Specifically, we show that to account for the effects of non-zero skewness $\xi$, one has to replace the rapidity argument $Y = \ln (1/x)$ of the eikonal dipole amplitude $N$ and the odderon dipole amplitude $\cal O$ by $Y = \ln \min \left\{ 1/|x|, 1/|\xi|\right\}$. The prescription applies to the elastic scattering cross sections, as well as for calculations of the Generalized Parton Distributions and Generalized Transverse Momentum Dependent parton distributions at small $x$ and at small but non-zero skewness $\xi$. We also show that the real part of the scattering amplitude, proportional to Im~$N$, which is intimately connected to the signature factor of the amplitude, can be accounted for by a more careful evaluation of the initial condition for the evolution and by writing the non-linear evolution equation in an integral form. One can similarly construct Im~$\cal O$ for the odd-signature odderon amplitude. We hope that future implementation of our prescriptions presented here will eliminate the need for both phenomenological $R$-factors.  
\end{abstract}

\maketitle

\tableofcontents


\section{Introduction}
\label{sec:int}

Mapping out the spatial three-dimensional structure of hadrons in terms of their quark and gluon degrees of freedom has been a longstanding problem in Quantum Chromodynamics (QCD). The three-dimensional hadronic structure is quantified in terms of the generalized parton distributions (GPDs) \cite{Muller:1994ses, Ji:1996ek, Radyushkin:1996nd, Ji:1996nm, Ji:1998pc}, which, unlike the parton distribution functions (PDFs) dependent on the momentum scale $Q^2$ and the longitudinal momentum fraction $x$, also depend on the Mandelstam variable $t$ and skewness $\xi$, quantifying the transverse and longitudinal momentum transfer between the probe and the target hadron. While in recent decades a lot of progress has been made on understanding GPDs at large and moderate $x$ \cite{Berger:2001xd,Polyakov:2002wz,Ivanov:2002jj,Diehl:2003ny,Guidal:2004nd,Belitsky:2005qn,Goloskokov:2005sd,Mueller:2005ed,Enberg:2006he,Kumericki:2007sa,Kumericki:2009uq,Goldstein:2010gu,ElBeiyad:2010pji,Gonzalez-Hernandez:2012xap,Kumericki:2016ehc,Dupre:2016mai,Berthou:2015oaw,Boussarie:2016qop,Pedrak:2017cpp,Duplancic:2018bum,Moutarde:2019tqa,Lin:2020rxa,Pedrak:2020mfm,Hashamipour:2021kes,Dutrieux:2021wll,Grocholski:2021man,Grocholski:2022rqj,Qiu:2022bpq,Guo:2022upw,Duplancic:2022ffo,Qiu:2023mrm,Deja:2023ahc,Duplancic:2023kwe,Guo:2023ahv,Qiu:2024mny,Goharipour:2024atx,Siddikov:2024blb,Almaeen:2024guo,Guo:2024wxy,Guo:2025muf}, their behavior at small $x$ has been explored to a much lesser degree. (We note the progress made in the recent works \cite{Hatta:2022bxn, Bhattacharya:2025fnz} along with our own \cite{Kovchegov:2025yyl}, which, in an abbreviated form, contains some of the material to be presented below.) The goal of the present work is to understand the small-$x$ asymptotics of spin-independent quark and gluon GPDs with small but non-zero skewness $\xi$ and to improve our understanding of high-energy elastic processes, which can be used to study GPDs.

In high-energy exclusive processes, the hadronic (or nuclear) target acts as a dense color charge medium, being dominated by quasi-classical gluon fields \cite{Gribov:1984tu,Iancu:2003xm,Weigert:2005us,JalilianMarian:2005jf,Gelis:2010nm,Albacete:2014fwa,Kovchegov:2012mbw,Morreale:2021pnn,Wallon:2023asa}. The leading contribution to the scattering amplitude can be described in the small-$x$ / shock wave formalism \cite{Gribov:1984tu,Iancu:2003xm,Weigert:2005us,JalilianMarian:2005jf,Gelis:2010nm,Albacete:2014fwa,Kovchegov:2012mbw,Morreale:2021pnn,Wallon:2023asa}, where it is given by Wilson line correlators which encode the eikonal scattering of a quark--anti-quark color dipole (and of other color structures, such as quadrupoles \cite{JalilianMarian:2004da}) off the hadron's gluon fields. These Wilson line correlators include information on the three dimensional distribution of gluons in the target, and indeed are related to the GPDs which describe the same scattering amplitudes in the formalism of collinear factorization. The GPDs are defined as off-forward matrix elements of bi-linear quark and gluon operators, and they encode various properties of the bulk hadron state, including the tomographic distribution of quarks and gluons \cite{Burkardt:2000za,Burkardt:2002hr,Ji:2003ak,Belitsky:2003nz}, the distribution of the hadron's mass and angular momentum among its constituents \cite{Ji:1996ek,Ji:1994av}, as well as various mechanical properties like stresses and pressures within the hadron \cite{Polyakov:2002wz,Polyakov:2002yz,Burkert:2018bqq,Kumericki:2019ddg,Ji:2025qax}. There exists a body of literature containing systematic treatments in the shock wave formalism of various PDFs and Transverse Momentum Dependent Parton Distribution Functions (TMD PDFs or TMDs) at small $x$ in \cite{Mueller:1999wm, Marquet:2009ca, Dominguez:2011wm, Kovchegov:2015zha, Kotko:2015ura, Marquet:2016cgx, Hatta:2016aoc, vanHameren:2016ftb, Kovchegov:2015pbl, Kovchegov:2018znm, Kovchegov:2017lsr, Kovchegov:2021iyc, 
Chirilli:2021lif, Cougoulic:2022gbk, Kovchegov:2022kyy, Kovchegov:2024aus, Hauksson:2024bvv,
Balitsky:2024ozy, Borden:2024bxa, Kovchegov:2025gcg, Balitsky:2026nux}, which has been recently extended to include GPDs as well as Generalized Transverse Momentum Dependent Parton Distributions (GTMDs) in \cite{Bhattacharya:2025fnz, Kovchegov:2025yyl}. Note that there is recent work on extending the shock wave formalism beyond the usual small-$x$ kinematics for exclusive processes \cite{Boussarie:2023xun}.

The calculation of the scattering amplitudes for exclusive processes within the shock wave formalism  typically ignores two major features of such processes, namely the non-zero longitudinal momentum transfer (termed skewness in the collinear factorization framework) between the probe and the target hadron and the real part of the scattering amplitude \cite{Kovchegov:1999ji,Kovner:2001vi,Hentschinski:2005er,Kovner:2006ge,Hatta:2006hs,Kowalski:2006hc,Rezaeian:2012ji,Hatta:2017cte,Mantysaari:2021ryb,Mantysaari:2022kdm,Hatta:2022bxn}. These contributions can be restored with significant phenomenological corrections \cite{Kowalski:2006hc,Toll:2012mb,Mantysaari:2016jaz}, but a calculation purely within the shock wave formalism has been missing. In this work, we will establish a prescription for including each of these contributions within the small-$x$ / shock wave formalism. Furthermore, we will lay out a framework for calculating the GTMDs and GPDs within the shock wave formalism starting from their operator definitions in factorization theorems. Although we will restrict our calculations to the unpolarized GTMDs/GPDs, our framework allows for the calculation of quantities which are energy suppressed and thus absent in the usual shock wave calculations, including spin-dependent parton distributions. The authors have previously studied the simplification of GTMD/GPD operators at small-$x$ and the inclusion of skewness in the leading logarithmic approximation (LLA) of their small-$x$ evolution in \cite{Kovchegov:2025yyl}. Here we will give a detailed derivation of the results obtained in \cite{Kovchegov:2025yyl} and introduce a prescription for systematically including the correction due to the real part of the scattering amplitude into the small-$x$ shock wave formalism.

The paper is structured as follows. In \sec{sec:gpds_exp} we take the operator definitions of the unpolarized gluon and quark GTMDs and GPDs and convert them to shock wave formalism expressions written in terms of Wilson line correlators. These Wilson line correlators have an unspecified rapidity with respect to the target hadron, which we address in the following \sec{sec:skew}. There, we consider the effect of keeping a non-zero longitudinal momentum transfer in the diagrams which yield the small-$x$ evolution of the dipole scattering amplitude. (To be more specific, we are talking about the ``plus" momentum transfer for the mostly $x^+$-moving target in the light-cone coordinates notation.) We consider general diagrams including non-linear terms where a gluon emission within one dipole splits it into two, generating two independent gluon cascades, and find that the effect of including longitudinal momentum transfer is to set the rapidity $Y$ of the evolved dipole amplitude $N_{10} (Y)$ (along with the odderon dipole amplitude ${\cal O}_{10} (Y)$)  to $Y = \ln \min \left\{ 1/|x| , 1/|\xi| \right\}$ for $x$ the average parton longitudinal momentum fraction in the GTMD/GPD matrix element and $\xi$ the skewness variable denoting the shift in the parton longitudinal momentum fraction. 

After establishing the expressions for the GTMDs/GPDs at small-$x$ and the prescription for including skewness in their small-$x$ evolution, we then derive a prescription for including the real part of the scattering amplitude through the small-$x$ evolution in \sec{sec:R_fac}. We find that the imaginary correction to the dipole scattering amplitude (which yields the real part of the exclusive-process scattering amplitude) is obtained by modifying the Glauber-Gribov-Mueller/McLerran-Venugopalan (GGM/MV) \cite{Mueller:1989st, McLerran:1993ka, McLerran:1993ni, McLerran:1994vd} initial condition / inhomogeneous term in the integral form of the small-$x$ evolution equation \cite{Balitsky:1995ub, Balitsky:1998ya, Kovchegov:1999yj, Kovchegov:1999ua} to\footnote{Here everything in the square brackets in the exponent is a modification advocated below, absent in  the original GGM/MV expression except for the $\theta$-function, which is often implicitly assumed.} 
\begin{align}\label{IC}
    N_{10}^{(0)} (s) = 1 - \exp \left\{ - \frac{1}{4} \, x_{10}^2 \, Q_{s0}^2 \, \ln \left( \frac{1}{x_{10} \, \Lambda}  \right)  \, \left[ \theta \left(s - \Lambda^2 \right) + \frac{i}{\pi} \,  \ln \left( \frac{|s - \Lambda^2|}{|s + \Lambda^2 |} \right) 
 \right] \right\} .
\end{align}
Here $\un{x}_1$ and $\un{x}_0$ are the transverse positions of the quark and anti-quark in the dipole with $\un{x}_{10} = \un{x}_1 - \un{x}_0$ and $x_{10} = |\un x_{10}|$, 
$Q_{s0}$ is the initial saturation scale of the hadronic or nuclear target, and $\Lambda$ is an infrared (IR) cutoff scale. Note that \eq{IC} is written for $x_{10} < 1/\Lambda$ and with $s >0$ the center of mass energy squared for the scattering of the dipole off the target hadron or nucleus.

We summarize our findings in \sec{sec:conc}. 
In this manuscript we use light cone coordinates defined as $u^{\mu} = (u^+, u^-, \un{u})$ for an arbitrary 4-vector $u^\mu$ with $u^{\pm} = (u^0 \pm u^3) / \sqrt{2}$ and the scalar product $u \vdot v = u^+ v^- + u^- v^+ - \un{u} \vdot \un{v}$. Transverse vectors are denoted by $\un{u} = (u^1, u^2)$, while the transverse integration measures are given by $\dd[2]{u}_{\perp}$. We also write the magnitude of a transverse vector as $u_T = |\un u|$. For all of the calculations in this manuscript, we take the hadron to move mainly in the $x^+$ direction.


\section{Unpolarized GPDs and GTMDs at small $x$}
\label{sec:gpds}


\subsection{Expressions for unpolarized GPDs and GTMDs at small $x$}
\label{sec:gpds_exp}

We begin with the operator definitions of the gluon and quark GPDs as defined from collinear factorization \cite{Muller:1994ses,Ji:1996nm, Radyushkin:1996nd, Ji:1998pc}. At leading-twist, the gluon GPDs are given by an off-forward matrix element of a non-local product of two gluon field strength operators $F_a^{\mu \nu} = \partial^{\mu} A^{a \, \nu} - \partial^{\mu} A^{a \, \nu} + g_s \, f^{abc} \, A^{b \, \mu} \, A^{c \, \nu}$, such as
\begin{align}
    \int\limits_{-\infty}^\infty \frac{\dd{z}^-}{2\pi} e^{ix \bar{P}^+ z^-} \!\! \bra{P', S'} F_b^{+i} \left(-\tfrac{z}{2} \right) U^{ba}_{\un 0} \left[-\tfrac{z^-}{2}, \tfrac{z^-}{2} \right] F_a^{+j} \left(\tfrac{z}{2} \right) \ket{P, S} .
\end{align}
One has an analogous definition in the quark case using quark field operators
\begin{align}
    \int\limits_{-\infty}^\infty \frac{\dd{z}^-}{2\pi} e^{ix \bar{P}^+ z^-} \bra{P', S'} \, \bar{\psi} \left(-\tfrac{z}{2} \right) V_{\un 0} \left[-\tfrac{z^-}{2}, \tfrac{z^-}{2} \right] \, \Gamma \, \psi \left(\tfrac{z}{2} \right) \ket{P, S} .
\end{align}
In these definitions, $z^\mu = (0^+, z^-, {\un 0})$ is the separation vector between the field operators, $i,j = 1,2$, $\bar{P}^+ \equiv (P'^+ + P^+)/2$ is the average momentum of the hadron, $\Gamma$ is a Dirac matrix, and $U_{\un 0}^{ba} [-z^-/2, z^-/2]$ is a light-cone Wilson line in the adjoint representation, defined as
\begin{align}
    U_{\un x} [x_f^- , x_i^-] \equiv \mathcal{P} \exp \left[i g_s \int\limits_{x_i^-}^{x_f^-} \dd{x}^{-} A^{a \, +} (0^+, x^-, {\un x} ) \, T^a \right] .
\end{align}
Here $g_s$ is the strong coupling constant, while the SU($N_c$) generator matrices in the adjoint representation are $(T^a)_{bc} = -i f_{abc}$  and $\mathcal{P}$ is the path-ordering operator. ($N_c$ is the number of quark colors.) For the quark correlator, we also employ $V_{\un 0} [-z^-/2,z^-/2]$, a light-cone Wilson line in the fundamental representation,
\begin{align}
    V_{\un x} [x_f^- , x_i^-] \equiv \mathcal{P} \exp \left[i g_s \int\limits_{x_i^-}^{x_f^-} \dd{x}^{-} A^{a \, +} (0^+, x^-, {\un x} ) \,  t^a \right] ,
\end{align}
with generator matrices $t^a$ in the fundamental representation. We denote by $x$ the average fraction of the longitudinal momentum of the hadron which is carried by the quark or gluon.


\subsubsection{Unpolarized Gluon GPD and GTMD at small $x$}

Following the pioneering work \cite{Hatta:2022bxn} (and the more recent \cite{Bhattacharya:2025fnz}), and employing the methods used for PDFs \cite{Mueller:1999wm, Kovchegov:2015zha, Kotko:2015ura, Marquet:2016cgx, Hatta:2016aoc, vanHameren:2016ftb, Kovchegov:2015pbl, Kovchegov:2018znm, Kovchegov:2017lsr, Kovchegov:2021iyc, 
Chirilli:2021lif, Cougoulic:2022gbk, Kovchegov:2022kyy, Kovchegov:2024aus, Borden:2024bxa, Kovchegov:2025gcg}, the first step in studying the small-$x$ asymptotics of GPDs is to generalize the non-local operator to that of GTMDs. This introduces a transverse separation between the two field operators which allows one to convert the operators into a dipole scattering amplitude. Here we focus on the GPDs for unpolarized partons in unpolarized hadrons, with the relevant unpolarized dipole gluon GTMD $F_{1,1}^g$ defined by \cite{Bhattacharya:2018lgm}
\begin{align}\label{glueGTMD}
      &\frac{1}{2} \sum_S \int \frac{\dd{z^-}\dd[2]{z_{\perp}}}{(2\pi)^3} \, e^{ix \bar{P}^+ z^- - i \un{k} \vdot \un{z}} \, \delta^{ij}     \, \bra{P', S} 2 \tr \Big[ F^{+i} \left(-\tfrac{z}{2} \right) {V}^{[+]} \left[ - \tfrac{z}{2} , \tfrac{z}{2} \right] F^{+j} \left(\tfrac{z}{2} \right) {V}^{[-]} \left[ \tfrac{z}{2} , - \tfrac{z}{2} \right] \Big] \ket{P, S}  \\
      &= \frac{1}{2} \sum_S \frac{\bar{P}^+}{2 M} \, \bar{u} (P',S) \, F_{1,1}^g (x, \xi, \un{\Delta}, \un{k}) \, u (P,S)  ,  \notag
\end{align}
where now $z^\mu = (0^+, z^-, {\un z})$, the trace and field strength operators are in the fundamental representation, $u(P,S)$ and $\bar{u} (P', S)$ are on shell spinors for the hadron of mass $M$, $\Delta = P' - P$ is the momentum transferred to the hadron, $x$ is the average longitudinal momentum fraction of the gluons, and $\xi = - \Delta^+/2 \bar{P}^+$. There is an important caveat to our notation: Lorentz invariance requires that the GTMDs only depend on scalar products of the available transverse vectors, namely $k_T^2, \Delta_T^2$ and $\un k \cdot \un \Delta$. We write the arguments of the GTMDs as the full transverse momenta $\un{\Delta}$ and $\un{k}$ only for compactness. Similarly, we suppress the renormalization scale dependence in the arguments of GTMDs. 
In \eq{glueGTMD}, the gauge links 
\begin{align}
    V^{[\pm]} [x_f, x_i] = V_{\un{x}_f} [x^-_f, \pm \infty^-] V_{\pm \infty^-} [\un{x}_f,\un{x}_i] V_{\un{x}_i} [\pm \infty^-, x_i^-]
\end{align}
are fundamental-representation staples in the positive and negative light-cone minus directions with $V_{\pm \infty^-} [\un{x}_f,\un{x}_i]$ the transverse gauge links at $x^-$-infinities, defined by
\begin{align}\label{tr_link}
    V_{\pm \infty^-} [\un{x}_f,\un{x}_i] \equiv \mathcal{P} \exp \left[- i g_s \int\limits_{\un{x}_i}^{\un{x}_f} \dd{x}^{i} A^{a \, i} (0^+, \pm \infty^-, {\un x} ) \,  t^a \right].
\end{align}
This means that we are considering the dipole--type gluon GTMDs, while the Weizsacker-Williams--type gluon GTMDs could also be analyzed and would result in expressions in terms of four Wilson line correlators. We leave such investigations to future work.

From \eq{glueGTMD}, we read off the unpolarized gluon GTMD 
\begin{align}\label{glueGTMD111}
& F_{1,1}^g (x, \xi, \un{\Delta}, \un{k}) =  \sqrt{1-\xi^2} \, \sum_S \int \frac{\dd{z}^- \dd[2]{z_{\perp}}}{(2 \pi)^3 \, 2 \, \bar{P}^+} \, e^{ix \bar{P}^+ z^- - i \un{k} \vdot \un{z}}  \\
    & \times  \, \bra{P', S} 2 \tr \Big[ F^{+i} \left(-\tfrac{z}{2} \right) {V}^{[+]} \left[ - \tfrac{z}{2} , \tfrac{z}{2} \right] F^{+i} \left(\tfrac{z}{2} \right) {V}^{[-]} \left[ \tfrac{z}{2} , - \tfrac{z}{2} \right] \Big] \ket{P, S} .  \notag 
\end{align}

Next, we simplify the expression \eqref{glueGTMD111} for the gluon GTMD at small $x$ and small $\xi$ by employing the Light Cone Operator Treatment (LCOT) formalism developed in \cite{Kovchegov:2015pbl,Kovchegov:2017lsr, Kovchegov:2018znm, Kovchegov:2018zeq, Kovchegov:2021iyc}, assuming that one does not choose the $A^+ =0$ light-cone gauge of the hadron so that we can put the transverse links at infinities \eqref{tr_link} to one. We want to switch to the shock wave picture, so we need to define the notation for the off-forward ``averaging" of operators in the background fields of the target (cf. \cite{Kovchegov:2019rrz}). For forward matrix elements, the standard unpolarized color glass condensate (CGC)/saturation average of a two-point operator $\hat{\cal O} \left( {x}_0, {x}_1 \right)$, depending on positions  $x_p^\mu = (0^+, x_p^-, {\un x}_p)$ with $p=0,1$, is defined as \cite{McLerran:1993ka, McLerran:1993ni, McLerran:1994vd, Kovchegov:1996ty, Jalilian-Marian:1997dw}
\begin{align}\label{for_ave}
    \frac{1}{2} \sum_S \bra{P,S} \hat{\mathcal{O}} (x_0, x_1) \ket{P,S} = 2 P^+ \int \dd[2]{b}_{\perp} \dd{b}^- \langle \hat{\mathcal{O}} (x_0, x_1) \rangle 
\end{align}
(and denoted by $\langle \ldots \rangle$), where the factor of $2P^+$ results from the normalization of the single-particle states 
\begin{align}
    \bra{P}\ket{P} = 2 P^+ (2\pi)^3 \delta (0^+) \delta^{2} (\un{0}) = 2 P^+ V^-
\end{align}
with the (infinite) volume $V^- = \int \dd[2]{x}_{\perp} \dd{x}^-$. The impact parameter is defined by 
\begin{align}
    b^\mu = \alpha \, x_1^\mu + (1-\alpha) \, x_0^\mu
\end{align}
with $\alpha$ an arbitrary real number while $x_{10}^\mu = x_1^\mu - x_0^\mu$, such that 
\begin{align}
    x_1^\mu = b^\mu + (1-\alpha) \, x_{10}^\mu, \ \ \ x_0^\mu = b^\mu - \alpha \, x_{10}^\mu. 
\end{align}
Since
\begin{align}
    \hat{\mathcal{O}} (x_0, x_1)  = \hat{\mathcal{O}} (b - \alpha \, x_{10}, b + (1-\alpha) \, x_{10}) = e^{i {\hat{P} \cdot b} } \,  \hat{\mathcal{O}} (- \alpha \, x_{10},  (1-\alpha) \, x_{10}) \, e^{- i {\hat{P} \cdot b} }
\end{align}
with $\hat P$ the momentum operator, the left-hand side of \eq{for_ave} is, indeed, independent of $b$.

We are now ready to generalize \eq{for_ave} to the off-forward case. This yields 
\begin{align}\label{sat_ave}
 \frac{1}{2} \sum_S \bra{P' = P + \Delta, S} \hat{\cal O} \left( - \alpha \, {x}_{10}, (1-\alpha) \, {x}_{10} \right) \ket{P, S}  = 2 \bar{P}^+ \, \int db^- \, d^2 b_\perp \, e^{- i \Delta^+ \, b^- + i {\un \Delta} \cdot {\un b}} \, \left\langle \hat{\cal O} \left( {x}_0, {x}_1 \right) \right\rangle  
\end{align}
(see \cite{Wu:2017rry, Cougoulic:2020tbc, Kovchegov:2013cva, Kovchegov:2015zha} along with Appendix~A of \cite{Kovchegov:2019rrz} for the relation between the CGC/saturation averaging and operator matrix elements using Wigner distributions, which results in the expression above).
The sign difference in the exponent compared to \cite{Wu:2017rry, Cougoulic:2020tbc} is due to our impact parameter $b$ denoting the position of the dipole with respect to the target nucleon, while in those references the impact parameter denotes the position of the nucleon or a large-$x$ parton in a nucleus.

We note that the definitions \eqref{glueGTMD} and \eqref{glueGTMD111} for the gluon GTMD correspond to the choice of $\alpha = 1/2$ in \eq{sat_ave}: however, Eqs.~\eqref{glueGTMD} and \eqref{glueGTMD111} can be rewritten for any other value of $\alpha$ by replacing $z/2 \to (1-\alpha) \, z$ and $-z/2 \to - \alpha \, z$ in them, which would result in different GTMD (see \cite{Hatta:2017cte} for a related discussion). Employing \eq{sat_ave} with $\alpha = 1/2$, and neglecting the transverse gauge links at $x^- \to \pm \infty$ \cite{Collins:1992kk, Collins:2002kn, Brodsky:2002rv, Kovchegov:1996ty, Jalilian-Marian:1997xn, Kovchegov:1997pc, Kovchegov:1998bi, Brodsky:2002ue, Belitsky:2002sm, Chirilli:2015fza}, we rewrite \eq{glueGTMD111} as
\begin{align}\label{glueGTMD1111}
& F_{1,1}^g (x, \xi, \un{\Delta}, \un{k}) =  \sqrt{1-\xi^2} \, \int \frac{\dd{x}_1^- \dd{x}_0^- \dd[2]{x_{1\perp} \dd[2]{x_{0\perp}}} }{2 \pi^3 }  \, e^{ix \bar{P}^+ x_{10}^- + i \bar{P}^+ \xi (x_1^- + x_0^-) - i \un{k} \cdot \un{x}_{10} + i \un \Delta \cdot \un b} \\
    & \times  \, \left\langle \tr \Big[ V_{\un x_0} [-\infty , x_0^-] F^{+i} \left( x_0 \right)  V_{\un x_0} [x_0^-, \infty] \, V_{\un x_1} [\infty, x_1^-] \, F^{+i} \left( x_1 \right) \, V_{\un x_1} [ x_1^-, - \infty]  \Big] \right\rangle .  \notag 
\end{align}
Expanding \eq{glueGTMD1111} in $x$ and $\xi$ to the lowest non-trivial order (cf.~\cite{Hatta:2016aoc, Kovchegov:2017lsr, Cougoulic:2022gbk} for the helicity-dependent PDFs case) by using
\begin{align}\label{x-exp}
    & \int\limits_{-\infty}^\infty \dd{x_1^-} e^{i (x +\xi)  \bar{P}^+ \, x_1^-} V_{\un x_1} [\infty, x_1^-] \,
    F^{+i} (0^+, x_1^- , {\un x}_1) \, V_{\un x_1} [x_1^-, - \infty] \notag \\
    &   = - \int\limits_{-\infty}^\infty \dd{x_1^-} e^{i (x +\xi)  \bar{P}^+ \, x_1^-}  V_{\un x_1} [\infty, x_1^-] 
    \left[ \pd^i A^+ + i (x+\xi) \bar{P}^+ \, A^i \right]  V_{\un x_1} [x_1^-, - \infty] = \frac{i}{g_s} \, \pd^i V_{\un x_1} + {\cal O} (x) + {\cal O} (\xi) 
\end{align}
along with its Hermitian conjugate with $\xi \to - \xi$, we obtain (cf. \cite{Hatta:2022bxn})
\begin{align}\label{glueGTMD11}
    &F_{1,1}^g (|x| \ll 1, |\xi| \ll 1, \un{\Delta}, \un{k}) = - \frac{N_c}{8 \pi^4 \, \as}  \, \int \dd[2]{x}_{1 \, \perp} \dd[2]{x}_{0 \, \perp} \,  e^{- i \un{k} \vdot \un{x}_{10} + i \un{\Delta} \vdot {\un b}} \, \left( \un \nabla_{{x}_1} \vdot  \un \nabla_{{x}_0} \right)   \left[ N_{10} (Y) - i \, {\cal O}_{10} (Y) \right] , 
\end{align}
where and we have employed the dipole $S$-matrix 
\begin{align}\label{S}
D_{10} (Y) \equiv \frac{1}{N_c} \left\langle \tr\Big[ V_{\un{x}_1} V_{\un{x}_0}^\dagger \Big] \right\rangle = 1 - N_{10} (Y) + i \, {\cal O}_{10} (Y) 
\end{align}
and defined the unpolarized C-even dipole scattering amplitude \cite{Balitsky:1995ub, Kovchegov:1999yj}
\begin{align}\label{Ndef}
    N_{10} (Y) \equiv 1 - \frac{1}{2 N_c} \, \left\langle \tr \left[ V_{\un{x}_1} V_{\un{x}_0}^{\dagger} \right] + \tr \left[ V_{\un{x}_0} V_{\un{x}_1}^{\dagger} \right]  \right\rangle (Y)
\end{align}
along with the C-odd odderon amplitude \cite{Kovchegov:2003dm, Hatta:2005as} (see also \cite{Kovchegov:2012ga})
\begin{align}\label{Odef}
    {\cal O}_{10} (Y) \equiv \frac{1}{2 i N_c} \, \left\langle \tr \left[ V_{\un{x}_1} V_{\un{x}_0}^{\dagger} \right] - \tr \left[ V_{\un{x}_0} V_{\un{x}_1}^{\dagger} \right]  \right\rangle (Y)
\end{align}
with the infinite fundamental light-cone Wilson lines denoted by $V_{\un{x}} = V_{\un{x}} [\infty^-, -\infty^-]$ and $Y$ the rapidity of the dipole with respect to the target. Here $\as = g_s^2/(4\pi)$. We stress again that \eq{glueGTMD11} is derived here for
\begin{align}
\un b = \frac{\un{x}_1 + \un{x}_0}{2}    
\end{align}
in it, but applies for any definition of the impact parameter $\un b = \alpha \un x_1 + (1-\alpha) \un x_0$, as long as the definition \eqref{glueGTMD} is modified accordingly. The latter modification would, indeed, change the gluon GTMD $F_{1,1}^g$.

The unpolarized gluon GPD at large $Q^2$ is given, up to power corrections and $\mathcal{O} (\alpha_s)$ scheme choice corrections, by \cite{Boussarie:2023izj,delRio:2024vvq}
\begin{align}\label{gGPD}
    & H^g (|x| \ll 1, |\xi| \ll 1, t, Q^2) = \int\limits^{Q^2}  \dd[2]{k}_{\perp}  F_{1,1}^g (|x| \ll 1, |\xi| \ll 1, \un{\Delta}, \un{k}) \notag \\
    & = - \frac{N_c}{2 \pi^2 \, \as} \,  \int \dd[2]{b}_{\perp} \,  e^{i \un{\Delta} \vdot {\un b}} \, \left[ \left( \nabla_{\un{x}_1} \vdot  \nabla_{\un{x}_0} \right)   N_{10} (Y) \right]_{x_{10}^2 = 1/Q^2} . 
\end{align}
As the odderon contribution to the gluon GTMD \eqref{glueGTMD11} is odd under $\un k \to - \un k$, it does not contribute to the gluon GPD in \eq{gGPD}.


\subsubsection{Unpolarized Quark GPD and GTMD at small $x$}

The unpolarized quark GTMD is defined by \cite{Meissner:2009ww}
\begin{align}\label{qGTMD}
    &\frac{1}{2} \sum_S \int \frac{\dd{z^-}\dd[2]{z_{\perp}}}{2 (2\pi)^3} e^{ix \bar{P}^+ z^- - i \un{k} \vdot \un{z}}  \, \bra{P', S} \bar{\psi} \left(-\tfrac{z}{2} \right) V^{[+]} \left[ -\tfrac{z}{2},\tfrac{z}{2} \right] \gamma^+ \psi \left( \tfrac{z}{2} \right) \ket{P, S}  \\ &  = \frac{1}{4 M} \sum_S \bar{u} (P',S) \, F_{1,1}^q (x, \xi, \un{\Delta}, \un{k}) \, u(P,S) . \notag
\end{align}
Here, we chose the future-pointing Wilson line staple, as used in semi-inclusive deep inelastic scattering (SIDIS). 
To simplify the quark GTMD at $|x|, |\xi| \ll 1$, we start with the definition \eqref{qGTMD}, employ \eq{sat_ave}
and insert a complete set of states to obtain (while approximating $\sqrt{1-\xi^2} \approx 1$ at $|\xi| \ll 1$)
\begin{align}\label{qGTMD_smallx}
    F_{1,1}^q (|x| \ll 1, |\xi| \ll 1, \un{\Delta}, \un{k}) & \, = \frac{\bar{P}^+}{(2 \pi)^3}  \int \dd[2]{x}_{1 \, \perp} \dd{x}_1^- \dd[2]{x}_{0 \, \perp} \dd{x}_0^- \times e^{i (x + \xi) \bar{P}^+ x_1^- - i (x - \xi) \bar{P}^+ x_0^-} e^{ - i \un{k} \vdot \un{x}_{10}  + i \un{\Delta} \vdot (\un{x}_1 + \un{x}_0)/2 } \, \left( \gamma^+ \right)_{\alpha\beta} \notag \\
    &\times \sum_X\Big\langle  \bar{\psi}_{\alpha} (x_0) V_{\un{x}_0} [x_0^-, \infty] \ket{X} \bra{X} | V_{\un{x}_1} [\infty, x_1^-] \psi_{\beta} (x_1) \Big\rangle  
\end{align}
with the implied summation over the quark spinor indices $\alpha, \beta$.
We can calculate this matrix element diagrammatically, following the calculation detailed in \cite{Kovchegov:2021iyc}: we take contractions of the two quark field operators, making use of the anti-quark propagator in the background field
\begin{align}\label{aquark_prop}
    \contraction
    {}
    {\bar\psi^i_\alpha}
    {(x_0) \:}
    {\psi^j_\beta}
    \bar\psi^i_\alpha (x_0) \: \psi^j_\beta (x_1)
    &= \int \dd[2]{w}_{\perp} \frac{\dd[4]{k}_1\dd[4]{k}_2}{(2\pi)^8} e^{ik_1^+ x_1^-} e^{i \un{k}_1 \vdot (\un{w} - \un{x}_1)} e^{-ik_2^+ x_0^-} e^{i \un{k}_2 \vdot (\un{x}_0 - \un{w})} \\
    &\times \Bigg{\{} \Bigg[ \frac{i\slashed{k}_1}{k_1^2 + i \epsilon} \Bigg] \Bigg[ \Big(\hat{V}_{\un{w}}^\dagger \Big)^{ji} (2\pi) \delta (k_1^- - k_2^-)  \Bigg] \Big[ \slashed{k}_2 (2\pi) \delta (k_2^2) \Big] \Bigg{\}}_{\beta \alpha} . \notag
\end{align}
The operator $\hat{V}_{\un{w}}$ is both a Dirac matrix and a color matrix, which we will simplify in the eikonal limit. The quarks are assumed to be massless, for simplicity. Anticipating that in our calculation we will put the $k_1$ anti-quark line on mass shell, we replace the $\slashed{k}_1$ and $\slashed{k}_2$ factors with polarization sums $\sum_\sigma v_\sigma (k_{1(2)}) \bar{v}_\sigma (k_{1(2)}) $ in terms of $\pm$-reversed BL spinors \cite{Kovchegov:2018znm,Kovchegov:2018zeq}
\begin{align}\label{anti-BLspinors}
u_\sigma (p) = \frac{1}{\sqrt{p^-}} \, [p^- + m \, \gamma^0 +  \gamma^0 \, {\un \gamma} \cdot {\un p} ] \,  \rho (\sigma), \ \ \ v_\sigma (p) = \frac{1}{\sqrt{p^-}} \, [p^- - m \, \gamma^0 +  \gamma^0 \, {\un \gamma} \cdot {\un p} ] \,  \rho (-\sigma),
\end{align}
with $p^\mu = \left( \frac{{\un p}^2+ m^2}{p^-}, p^-, {\un p} \right)$ and
\begin{align}
  \rho (+1) \, = \, \frac{1}{\sqrt{2}} \, \left(
  \begin{array}{c}
      1 \\ 0 \\ -1 \\ 0
  \end{array}
\right), \ \ \ \rho (-1) \, = \, \frac{1}{\sqrt{2}} \, \left(
  \begin{array}{c}
        0 \\ 1 \\ 0 \\ 1
  \end{array}
\right) .
\end{align} 
The action of the operator $\hat{V}_{\un{w}}$ at eikonal order is defined by matrix elements with these spinors as
\begin{align}
    \bar{v}_{\sigma} (p) \Big(\hat{V}_{\un{w}} \Big) v_{\sigma'} (p') = 2 \sqrt{p^- p'^-} \, \delta_{\sigma \sigma'} \, V_{\un{w}} ,
\end{align}
so the eikonal propagator becomes
\begin{align}
    \contraction
    {}
    {\bar\psi^i_\alpha}
    {(x_0) \:}
    {\psi^j_\beta}
    \bar\psi^i_\alpha (x_0) \: \psi^j_\beta (x_1)
    &= \int \dd[2]{w}_{\perp} \frac{\dd[4]{k}_1\dd[4]{k}_2}{(2\pi)^8} e^{ik_1^+ x_1^-} e^{i \un{k}_1 \vdot (\un{w} - \un{x}_1)} e^{-ik_2^+ x_0^-} e^{i \un{k}_2 \vdot (\un{x}_0 - \un{w})} \\
    &\hspace{2cm} \times \Bigg[ \frac{i \, (2\pi)^2 \, 2 k_1^- \, \delta (k_1^- - k_2^-) \, \delta (k_2^2) }{k_1^2 + i \epsilon} \Bigg] \Bigg{\{} \sum_\sigma  v_\sigma (k_1) \Big(V_{\un{w}}^\dagger  \Big)^{ji} \bar{v}_{\sigma} (k_2)\Bigg{\}}_{\beta \alpha} . \notag
\end{align}


\begin{figure}[ht]
    \captionsetup[subfigure]{justification=centering}
    \centering
    \begin{subfigure}[b]
    {0.45\linewidth}
        \includegraphics[width=\linewidth]{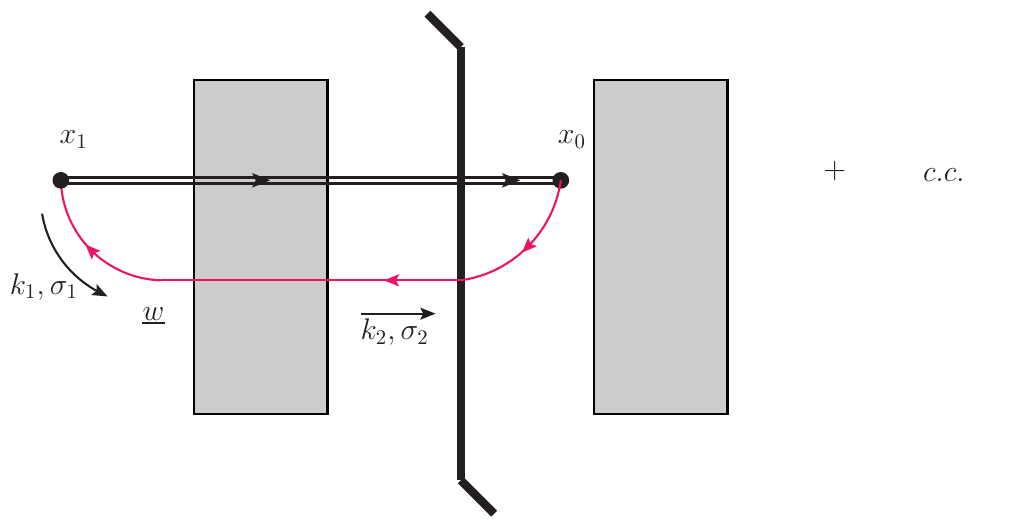}
        \caption*{B}
        \label{FIG:classb}
    \end{subfigure}
    ~
    \begin{subfigure}[b]
    {0.4\linewidth}
        \includegraphics[width=\linewidth]{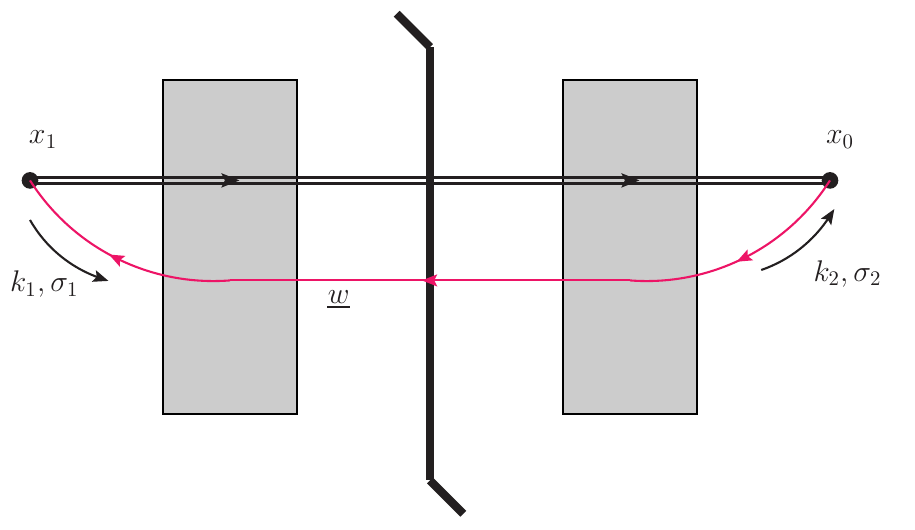}
        \caption*{C}
        \label{FIG:classc}
    \end{subfigure}
    ~
    \begin{subfigure}[b]{0.3\linewidth}
        \includegraphics[width=\linewidth]{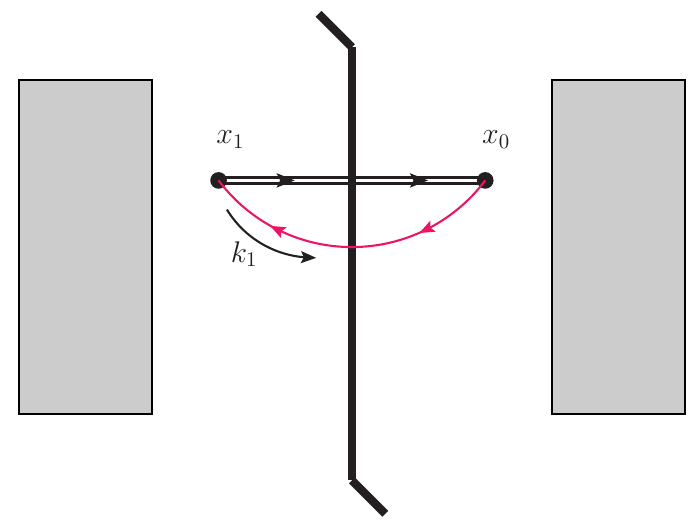}
        \caption*{D}
        \label{FIG:classd}
    \end{subfigure}
    \caption{Eikonal diagrams contributing to the unpolarized quark GTMD at small-$x$ and small-$\xi$. Double lines represent eikonal, light-cone Wilson lines arising from the gauge links in the GTMD staple, and single lines with arrows represent the anti-quark propagator from contracting the quark field operators, while the thick vertical line represents the final state cut. Shaded rectangles denote the shock wave.  
    }
    \label{FIG:classall}
\end{figure}


For the leading diagrams contributing to the GTMD, we allow the light cone minus positions of the quark field operators $x_0^-$ and $x_1^-$ to be either before or after the shock wave (in \cite{Kovchegov:2021iyc} the quark fields were allowed to be taken inside the shock wave as sub-eikonal corrections were being included), and we truncate the sum over intermediate states at the leading Fock state, that being an anti-quark $\ket{\bar{q}} \bra{\bar{q}}$. This gives us three diagrams as depicted in \fig{FIG:classall}, labeled in the same way as in \cite{Kovchegov:2021iyc}. In the diagram B we have the case where the the anti-quark propagator and staple gauge link cross the shock wave on only one side of the final state cut. In the diagram C we have the propagator and gauge link crossing the shock wave on both sides of the final state cut. Finally, in the diagram D neither the propagator nor the gauge links cross the shock wave (a pure non-interaction term). 

Using the propagator in \eq{aquark_prop}, we calculate the diagram B (and its cojugate) by taking $x_0^- > 0$, $x_1^- < 0$ (or $x_0^- < 0$, $x_1^- > 0$ for the conjugate diagram), setting the gauge link along $x_1$ to unity as it does not cross the shock wave, and inserting the propagator in \eq{qGTMD_smallx} to obtain
\begin{align}
    B &= \frac{\bar{P}^+}{(2 \pi)^3}  \int \dd[2]{x}_{1 \, \perp} \dd{x}_1^- \dd[2]{x}_{0 \, \perp} \dd{x}_0^- \dd[2]{w}_{\perp} \frac{\dd[4]{k}_1\dd[4]{k}_2}{(2\pi)^8}  e^{i (x + \xi) \bar{P}^+ x_1^- - i (x - \xi) \bar{P}^+ x_0^-} e^{ - i \un{k} \vdot \un{x}_{10}  + i \un{\Delta} \vdot (\un{x}_1 + \un{x}_0)/2 }  \\
    &\times e^{ik_1^+ x_1^-} e^{i \un{k}_1 \vdot (\un{w} - \un{x}_1)} e^{-ik_2^+ x_0^-} e^{i \un{k}_2 \vdot (\un{x}_0 - \un{w})} \, \Bigg[ \frac{i \, (2\pi)^2 \, 2 k_1^- \, \delta (k_1^- - k_2^-) \, \delta (k_2^2) }{k_1^2 + i \epsilon} \Bigg] \Bigg\langle  \Big( \sum_\sigma \bar{v}_{\sigma} (k_2) \, \gamma^+ \, v_\sigma (k_1) \Big)  \tr \Big[ V_{\un{x}_1} V_{\un{w}}^\dagger \Big] \Bigg\rangle  +  \cc .   \notag
\end{align}
Employing the spinor matrix element
\begin{align}
    \bar{v}_\sigma (p) \gamma^+ v_{\sigma'} (p') = \frac{1}{2\sqrt{p^- p'^-}} \Big[ \delta_{\sigma \sigma'} \un{p} \vdot \un{p}' - i \sigma \delta_{\sigma \sigma'} \un{p} \cross \un{p}' \Big] ,
\end{align}
integrating over $x_0^-, \, x_1^-, \, \un{x}_0, \, \un k_2, $ and $k_2^-$, defining the longitudinal momentum fraction $z \equiv k_1^- / q^- = 2 k_1^- \bar{P}^+ / s$ of some projectile's large momentum $p_2^-$ carried by the anti-quark in \fig{FIG:classall} with $s \approx 2 \bar{P}^+ q^-$ the projectile-target center of mass energy squared, and finally relabeling $\un{w}$ as $\un{x}_0$ we obtain
\begin{align}
    &B=\frac{2 N_c \, s}{(2\pi)^4} \int \dd[2]{x}_{1 \, \perp} \dd[2]{x}_{0 \, \perp} \frac{\dd[2]{k}_{1 \perp}}{(2\pi)^2} \, e^{-i(\un{k}_1 + \un{k})\vdot \un{x}_{10} + i \un{\Delta} \vdot (\un{x}_1 + \un{x}_0)/2} \\
   &\times \int\limits_{\Lambda^2/s}^1 \dd{z}  \, \frac{1}{N_c} \left\langle \tr \Big[ V_{\un{x}_1} V_{\un{x}_0}^\dagger \Big] \right\rangle \frac{ \un{k}_1 \vdot \left( \un{k} + \thalf \un \Delta \right) }{\left[ (x+\xi) \, z\, s + \un{k}_1^2 - i \epsilon \right] \left[ (x-\xi) \, z \, s + \left( \un{k} + \thalf \un \Delta \right)^2 - i \epsilon \right]} + \cc , \notag
\end{align}
with $\Lambda$ an IR cutoff scale describing the target. The lower integration limit on $z$ stems from the shock wave approximation, which is valid for the $x^-$-coherence length of the anti-quark being much larger than the width of the shock wave, that is, $2 k_1^- /{\un k}_1^2 > 1/ \bar{P}^+$, such that $k_1^- > {\un k}_1^2 /(2 \bar{P}^+) \ge \Lambda^2 /(2 \bar{P}^+)$. The upper cutoff on the $z$-integral, $z<1$ results from our interpretation here that $z$ is the fraction of some projectile's momentum $q^-$. Note that, unlike the sub-eikonal distributions considered, for instance, in \cite{Kovchegov:2015pbl, Kovchegov:2018znm, Kovchegov:2017lsr, Kovchegov:2021iyc, Chirilli:2021lif, Cougoulic:2022gbk, Kovchegov:2022kyy}, where the $z$-integral is logarithmic and both upper and lower regulators are necessary, here the $k_1^-$ integral is finite and does not require regularization. This allowed the authors of \cite{Bhattacharya:2025fnz} to integrate over $k^-_1$ from $0$ to $+\infty$ while assuming that the Wilson line correlator is only mildly dependent on $k_1^-$, obtaining a zero result for the diagram C.  

For the diagram C we have both gauge links crossing the shock wave, so both Wilson lines are kept as interacting. Meanwhile, the gluon exchanges along the anti-quark propagator will cancel between the two sides of the final state cut, so we have a free propagator given by setting $(\hat{V}_{\un{w}}^{\dagger})^{ji} = \delta^{ji} \gamma^-$. Plugging the non-interacting propagator into \eq{qGTMD_smallx} and performing several of the integrals yields
\begin{align}
    C &= \frac{2 N_c \, s}{(2\pi)^4} \int \dd[2]{x}_{1 \, \perp} \dd[2]{x}_{0 \, \perp} \frac{\dd[2]{k}_{1 \perp}}{(2\pi)^2} \, e^{-i(\un{k}_1 + \un{k})\vdot \un{x}_{10} + i \un{\Delta} \vdot (\un{x}_1 + \un{x}_0)/2} \\
   &\times \int\limits_{\Lambda^2/s}^1 \dd{z}  \, \frac{1}{N_c} \left\langle \tr \Big[ V_{\un{x}_1} V_{\un{x}_0}^\dagger \Big] \right\rangle \frac{ \un{k}_1^2 }{\left[ (x+\xi) \, z\, s + \un{k}_1^2 - i \epsilon \right] \left[ (x-\xi) \, z \, s + \un{k}_1^2 + i \epsilon \right]}  . \notag
\end{align}

Finally, we have the diagram D. Here the anti-quark propagator and both gauge links are non-interacting. We can read this contribution off from diagram C by taking the Wilson lines to be identity operators, which gives
\begin{align}
    D &= \frac{2 N_c \, s}{(2\pi)^4} \int \dd[2]{x}_{1 \, \perp} \dd[2]{x}_{0 \, \perp} \frac{\dd[2]{k}_{1 \perp}}{(2\pi)^2} \, e^{-i(\un{k}_1 + \un{k})\vdot \un{x}_{10} + i \un{\Delta} \vdot (\un{x}_1 + \un{x}_0)/2} \\
   &\times \int\limits_{\Lambda^2/s}^1 \dd{z}  \,  \frac{ \un{k}_1^2 }{\left[ (x+\xi) \, z\, s + \un{k}_1^2 - i \epsilon \right] \left[ (x-\xi) \, z \, s + \un{k}_1^2 + i \epsilon \right]}  . \notag
\end{align}

Next, we rewrite the quark dipole $S$-matrix using \eq{S}.
We note that the non-interacting terms (arising from 1 in \eq{S}) in the diagrams B, C, and D add up to yield a non-zero disconnected contribution for
an anti-quark passing through the hadron without interactions, which is possible in the $x<-|\xi|$ region. Such disconnected terms are subtracted in the TMD operator definition \cite{Collins:2011zzd}, such that the non-interacting terms from the three diagrams fully cancel. 
Performing a similar subtraction in the GTMD case at hand, we see that the final expression for the quark GTMD at small-$x$ and small-$\xi$ is (cf.~\cite{Bhattacharya:2025fnz}),
\begin{align}\label{q_GTMD_1}
   &F_{1,1}^q (|x| \ll 1, |\xi| \ll 1, \un{\Delta}, \un{k})= - \frac{2 N_c \, s}{(2\pi)^4} \int \dd[2]{x}_{1 \, \perp} \dd[2]{x}_{0 \, \perp} \,  \int\frac{\dd[2]{k}_{1 \perp}}{(2\pi)^2} \, e^{-i(\un{k}_1 + \un{k})\vdot \un{x}_{10} + i \un{\Delta} \vdot (\un{x}_1 + \un{x}_0)/2} \, \int\limits_{\Lambda^2/s}^1 \dd{z}  \, \left[ N_{10} (Y) - i \, {\cal O}_{10} (Y) \right] \notag \\
   &\times  \Bigg\{ \frac{ \un{k}_1 \vdot \left( \un{k} + \thalf \un \Delta \right) }{\left[ (x+\xi) \, z\, s + \un{k}_1^2 - i \epsilon \right] \left[ (x-\xi) \, z \, s + \left( \un{k} + \thalf \un \Delta \right)^2 - i \epsilon \right]}  + \frac{ \un{k}_1 \vdot \left( \un{k} - \thalf \un \Delta \right) }{\left[ (x+\xi) \, z\, s + \left( \un{k} - \thalf \un \Delta \right)^2 + i \epsilon \right] \left[ (x-\xi) \, z \, s + \un{k}_1^2 + i \epsilon \right]} \notag \\
   & + \frac{\un{k}_1^2}{\left[ (x+\xi) \, z\, s + \un{k}_1^2 - i \epsilon \right] \left[ (x-\xi) \, z \, s + {\un k}_1^2 + i \epsilon \right]}  \Bigg\} .  
\end{align}
Since $N_{10} (Y) = N_{01} (Y)$ and ${\cal O}_{10} (Y) = - {\cal O}_{01} (Y)$, one can readily see that our expression \eqref{q_GTMD_1} for quark GTMD satisfies $\left[ F_{1,1}^q (x, \xi, \un{\Delta}, \un{k}) \right]^* = F_{1,1}^q (x, - \xi, - \un{\Delta}, \un{k})$, in agreement with Eq.~(3.18) in \cite{Meissner:2009ww} following from Hermiticity, as long as $N_{10} (Y)$ and ${\cal O}_{10} (Y)$ are even functions of $\xi$. (We will see below that this is the case in the LLA.)

The unpolarized quark GPD at $|x|, |\xi| \ll 1$, corresponding to the GTMD in \eq{q_GTMD_1}, is given by
\begin{align}\label{qGPD}
    H^q (|x|\ll 1, |\xi| \ll 1, t, Q^2) = \int\limits^{Q^2}  \dd[2]{k}_{\perp} F_{1,1}^q (|x| \ll 1, |\xi| \ll 1, \un{\Delta}, \un{k}) . 
\end{align}
While this is not our main goal here, the expression \eqref{qGPD} for $H^q$ can be simplified after we substitute $F_{1,1}^q$ into it from \eq{q_GTMD_1}. This is done in Appendix~\ref{sec:Hq}, assuming weak dependence of $N_{10}$ and ${\cal O}_{10}$ on $z$, which is justified in the LLA. The calculation in Appendix~\ref{sec:Hq} gives
\begin{align}\label{qGPD_simp3}
   H^q(|x|\ll 1, |\xi| \ll 1, t, Q^2) & \, = \frac{2 N_c \, }{(2\pi)^4} \int \dd[2]{x}_{1 \, \perp} \dd[2]{x}_{0 \, \perp} \frac{e^{i \un{\Delta} \vdot  \un{x}_1}}{x_{10}^4} \, \frac{1}{\xi^3} \\
   & \times \, \left\{ N_{10} (Y) \left[ 4 \, x \, \xi +  ( x^2 - \xi^2 ) \left[ \ln \left( \frac{x - \xi - i \epsilon}{x + \xi - i \epsilon} \right) + \ln \left( \frac{x - \xi + i \epsilon}{x + \xi + i \epsilon} \right)  \right] \right] \right. \notag \\
   & \left. - i \, \mathcal{O}_{10} (Y) \, ( x^2 - \xi^2 ) \left[ \ln \left( \frac{x - \xi - i \epsilon}{x + \xi - i \epsilon} \right) - \ln \left( \frac{x - \xi + i \epsilon}{x + \xi + i \epsilon} \right)  \right]  \right\} . \notag
\end{align}
We can take $\xi>0$ without loss of generality due to the symmetry of GPDs under $\xi \rightarrow -\xi$. Then we have four regions to consider: the two Dokshitzer-Gribov-Lipatov-Altarelli-Parisi (DGLAP) \cite{Dokshitzer:1977sg,Gribov:1972ri,Altarelli:1977zs} regions  $x>\xi$ with $x>0$ or $x<-\xi$ with $x<0$, and the two Efremov-Radyushkin-Brodsky-Lepage (ERBL) \cite{Efremov:1978rn,Lepage:1979zb} regions $x<\xi$ with $x>0$ or $x > - \, \xi$ with $x<0$. Simplifying $H^q$ in these regions yields 
\begin{align}\label{qGPD_simp4}
    H^q &(|x|\ll 1, |\xi| \ll 1, t, Q^2) = \frac{4 N_c \, }{(2\pi)^4} \int \dd[2]{x}_{1 \, \perp} \dd[2]{x}_{0 \, \perp} \frac{e^{i \un{\Delta} \vdot  \un{x}_1}}{x_{10}^4} \, \frac{1}{\xi^3} \\
   & \times \, \left\{ N_{10} (Y) \left[ 2 \, x \, \xi + (x^2-\xi^2) \,  \ln \left| \frac{x - \xi}{x+\xi} \right| \right] + \mathcal{O}_{10} (Y) \, \pi \, (\xi^2 - x^2) \, \theta (|\xi| - |x|) \right\}. \notag
\end{align}
We see that, within the simplifying assumption that $N_{10}$ and ${\cal O}_{10}$ are relatively slowly-varying functions of $z$, the odderon contribution to the unpolarized quark GPD $H^q$ appears to be non-zero, but only within the ERBL region. \\

Let us summarize our calculations so far. We have obtained expressions \eqref{glueGTMD11} and \eqref{q_GTMD_1} (or \eqref{qGPD_simp4}) for, respectively, the gluon and quark unpolarized GTMDs at small $x$ and $\xi$ in terms of the impact parameter-dependent dipole scattering amplitudes; these GTMDs give their corresponding GPDs after integrating over the parton transverse momentum (see Eqs.~\eqref{gGPD} and \eqref{qGPD}). However, the rapidity variable of the dipole amplitudes in these formulas is not yet fixed. Usually, for PDFs and TMDs at small $x$, one takes $Y = \ln (1/x)$ (within LLA). The question we would like to address next in this work is what $Y$ should one use in Eqs.~\eqref{glueGTMD11}, \eqref{gGPD}, \eqref{q_GTMD_1} and \eqref{qGPD}? 
In the next Section we will examine the role of $\xi$ in the evolution of the dipole amplitude and establish a prescription for including effects of non-zero skewness by the appropriate choice of rapidity $Y$ in the argument of the dipole amplitudes obtained by solving the full non-linear evolution equations. We will find that the $x^-$-lifetime ordering required for resumming the leading logarithms will give us an effective rapidity variable $Y$ for the evolved dipole amplitudes $N_{10} (Y)$ and ${\cal O}_{10} (Y)$ which will depend on the skewness.


\subsection{Dipole amplitudes at non-zero skewness}

\label{sec:skew}

The dipole amplitudes $N_{10}$ and ${\cal O}_{10}$ are constructed out of the light-cone Wilson lines and, at first glance, do not seem to account for the longitudinal momentum transfer, which would lead to non-zero skewness. Their impact parameter dependence accounts for the transverse momentum transfer $\un \Delta$ which is Fourier conjugate to the impact parameter of the dipole, as employed in the expressions for the gluon and quark GTMDs derived above and given by Eqs.~\eqref{glueGTMD11} and \eqref{q_GTMD_1}. Our goal here is to allow for non-zero but small skewness, while still staying at the leading power of energy (eikonal) level. It is, therefore, natural to assume that to keep the leading power of energy, we should still employ the light-cone Wilson line definitions \eqref{Ndef} and \eqref{Odef} for the dipole amplitudes, even at non-zero skewness. Below, starting with this eikonal (Wilson-line) assumption, we will derive a prescription for including the longitudinal (plus) momentum transfer into the dipole amplitude $N_{10}$ evolved using the non-linear  Balitsky-Kovchegov (BK) \cite{Balitsky:1995ub, Balitsky:1998ya, Kovchegov:1999yj, Kovchegov:1999ua} and Jalilian-Marian-Iancu-McLerran-Weigert-Leodinov-Kovner (JIMWLK) \cite{Jalilian-Marian:1997jx,Jalilian-Marian:1997gr,Jalilian-Marian:1997dw,Iancu:2001ad,Iancu:2000hn} equations. Our prescription will also apply to the odderon amplitude ${\cal O}_{10}$, whose evolution was derived in \cite{Kovchegov:2003dm,Hatta:2005as} using the $s$-channel/shock wave formalism with the result that was equivalent to the odderon evolution from \cite{Bartels:1999yt} derived in the $t$-channel formalism \cite{Bartels:1980pe, Kwiecinski:1980wb, Jaroszewicz:1980mq} (see also \cite{Kovchegov:2012rz, Janik:1998xj, Caron-Huot:2013fea,Bartels:2013yga, Brower:2008cy,Avsar:2009hc,Brower:2014wha} for the properties of the odderon). Note that in the $t$-channel Balitsky-Fadin-Kuraev-Lipatov (BFKL) \cite{Kuraev:1977fs,Balitsky:1978ic} formalism and for the pomeron exchange the case on non-zero longitudinal momentum transfer was considered earlier in \cite{Bartels:1981jh}, resulting in the same prescription for including non-zero longitudinal momentum transfer as we will advocate for below. Therefore, our results here generalize the prescription from \cite{Bartels:1981jh} to the case of non-linear small-$x$ evolution and to the case of linear and non-linear odderon evolution.


\subsubsection{A simple argument}
\label{sec:simple}

\begin{figure}[ht]
\centering
\includegraphics[width= 0.4\linewidth]{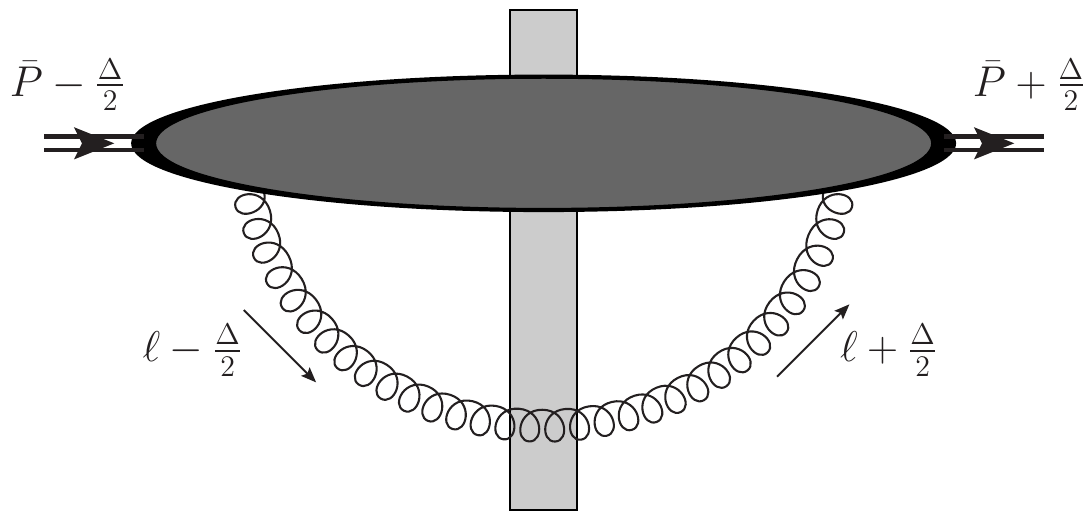}
\caption{A diagram illustrating one step of longitudinally non-forward small-$x$ evolution in the target. Evolution correction is generated by the gluon line going through the shock wave (vertical rectangle).}
\label{FIG:quick}
\end{figure}

To begin, let us ilustrate our argument by employing a simple example. Consider one step of small-$x$ evolution, which is non-forward in the longitudinal momentum. It is pictured in \fig{FIG:quick} using the shock wave formalism: however, our argument can equally be illustrated by using one rung of the ladder in the $t$-channel formalism. Somewhat unusually for the shock wave approach, we run the evolution from the nucleon target, with the shock wave representing the field of the projectile. For the plus-moving projectile, the  logarithm of energy in one step of unpolarized small-$x$ evolution usually arises from the integral over the plus momentum: in the longitudinally non-forward case at hand, the integral is  
\begin{align}\label{ell_int}
\int\limits^{{\bar P}^+ }_{x {\bar P}^+} \frac{\dd \ell^+}{\left( \ell^+ - \frac{\Delta^+}{2} \right) \, \left( \ell^+ + \frac{\Delta^+}{2} \right)} \, \Gamma (\ell^+, \Delta^+),
\end{align}
where $\Gamma (\ell^+, \Delta^+)$ represents the ``gluon--shock wave vertex", the factor coming from the gluon interacting with the field of the shock wave dependent on $\ell^+$ and $\Delta^+$, while the factors in the denominator result from the two gluon lines shown in \fig{FIG:quick}. In the usual forward (in the longitudinal) momentum case one has $\Gamma (\ell^+, \Delta^+ =0) = 2 \ell^+$. In the non-forward case, we can imagine $\Gamma (\ell^+, \Delta^+) = 2 \, \ell^+ + C \, \Delta^+$ with some constant $C$, though, in a complete calculation $\Gamma (\ell^+, \Delta^+)$ consists of several linear in $\ell^+$ and $\Delta^+$ numerators and denominators, with the number of numerators exceeding that of the denominators by one (while still satisfying $\Gamma (\ell^+, \Delta^+ =0) = 2 \ell^+$). Since the gluons' plus momenta cannot be larger than those of the incoming and outgoing proton, the upper limit on the $\ell^+$ integral is ${\bar P}^+$. Similarly, the lower limit of the $\ell^+$ integral is $x \, {\bar P}^+$, as we are calculating a distribution with the average momentum fraction $x$; for simplicity, we assumed that $x>0$. 

Analyzing the expression \eqref{ell_int}, we see that the $\ell^+$ integral is logarithmic only if $\ell^+ \gg |\Delta^+|/2$, or, equivalently, $\ell^+ / {\bar P}^+ \gg |\xi|$. In this regime, $\Gamma (\ell^+, \Delta^+) \approx 2 \, \ell^+$ and the integral \eqref{ell_int} becomes 
\begin{align}\label{ell_int2}
2 \, \int\limits^{{\bar P}^+ }_{\max \{ x, |\xi| \} \, {\bar P}^+} \frac{\dd \ell^+}{\ell^+ } .
\end{align}
This is almost a standard integral in small-$x$ evolution for PDFs and TMDs: the only difference is that now the lower limit has changed from $x$ to $\max \{ x, |\xi| \}$. We conclude that in going from small-$x$ evolution for unpolarized PDFs and TMDs to that for unpolarized GPDs and GTMDs, we need to replace $x \to \max \{ x, |\xi| \}$ to account for non-zero skewness (cf. \cite{Bartels:1981jh}). Our simple example appears to suggest that the argument of the dipole amplitudes $N_{10} (Y)$ and ${\cal O}_{10} (Y)$ should be modified from $Y = \ln (1/x)$ to $Y = \ln (1/\max \{ x, |\xi| \})$ when calculating GPDs and GTMDs: we will show that this is indeed the case in a more detailed analysis below. Another conclusion one can draw from this simple example is that small-$x$ evolution cannot truly be non-forward in the longitudinal direction: in the region where the momenta of the two gluons, $\ell^+ - \frac{\Delta^+}{2}$ and $\ell^+ + \frac{\Delta^+}{2}$, are truly different, which means that $\ell^+ \lesssim \frac{\Delta^+}{2}$, the integral \eqref{ell_int} becomes non-logarithmic and the evolution stops. Thus, skewness $\xi$ enters the dipole amplitudes only as a cutoff on small-$x$ evolution, but plays no role in the evolution otherwise. Finally, let us point out that the above argument can be generalized to include virtual corrections as well.


\subsubsection{A general argument}

\begin{figure}[ht]
\centering
\includegraphics[width= 0.5\linewidth]{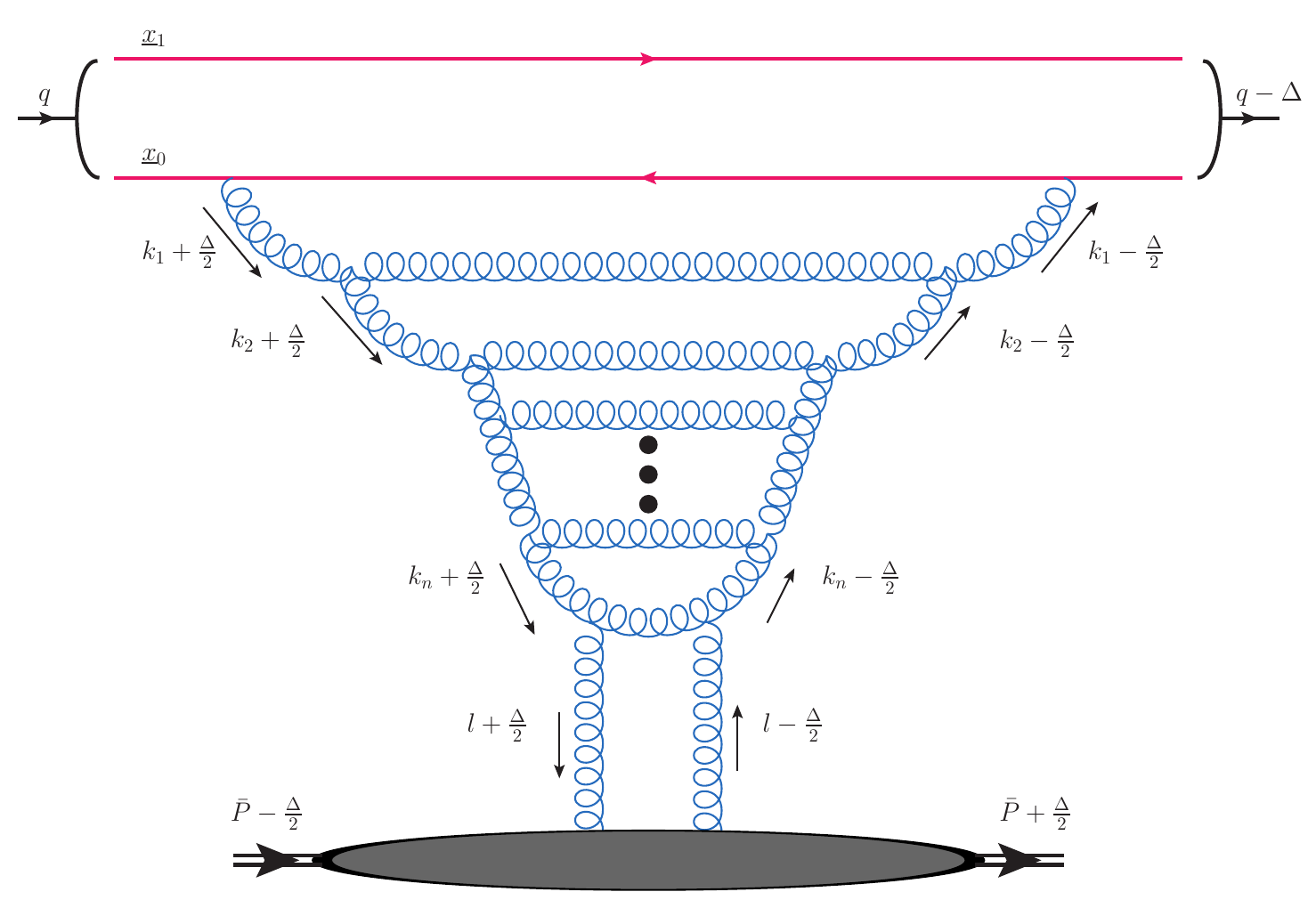}
\caption{Example of a ladder diagram which enters the leading logarithmic small-$x$ evolution of a color dipole in non-forward kinematics for an elastic scattering process.}
\label{FIG:skew_lad}
\end{figure}

Next, let us construct a more general argument, valid for the non-linear evolution. We will first consider the $s$-channel/shock wave formalism's version of the ladder diagrams contributing to a non-forward elastic process, as shown in \fig{FIG:skew_lad}, with the evolution running from the projectile now. We take the incoming projectile to have momentum $q^\mu = (- Q^2/(2 q^-), q^-, {\un 0})$, where $q^-$ is large. Note that while the projectile is not present in the operator definition of the GTMDs and GPDs, above we have related the unpolarized GTMDs and GPDs to the dipole amplitudes. To calculate a dipole amplitude it is convenient to imagine that the quark and anti-quark (giving the light-cone Wilson lines) in it were created in a decay of some projectile, like a virtual photon in deep inelastic scattering (DIS). The large $q^-$ momentum of the virtual photon provides a natural upper limit on the $k^-$ integrals of the gluons generated by small-$x$ evolution, $k^- < q^-$.   
Thus, we take the incoming projectile to produce a color dipole with the quark at position $\un x_1$ and the anti-quark at position $\un x_0$. The dipole, in turn, will produce a cascade of gluons through s-channel emissions, yielding a ladder of gluons with the momenta of the gluons corresponding to the $i$th emission (the ``rails" of the ladder attached to the $i$th ``rung") equal to 
\begin{align}
    k_i^\mu = \left( \frac{\left({\un k}_{i} \pm \tfrac{\un \Delta}{2} \right)^2}{2 (k_i^- \pm \tfrac{\Delta^-}{2})} , k_i^- \pm \frac{\Delta^-}{2} , {\un k}_{i} \pm \frac{\un \Delta}{2} \right).
\end{align} 
We consider these ladder diagrams as formulated within light-cone perturbation theory (LCPT) \cite{Lepage:1980fj, Brodsky:1997de}, where we have a light-cone time $x^-$ ordered picture of the scattering (unlike the light cone time $x^+$ used in the original works \cite{Lepage:1980fj, Brodsky:1997de}). Below, we will also employ the average fractions of the projectile's ``minus" light cone momentum  $z_i = k_i^- / q^-$ and the fractional minus-momentum transfer $\zeta = \Delta^- / 2 q^-$. 

We are interested in resumming the leading-logarithmic, high energy contribution from the gluon cascade, so we must impose the following conditions (cf. \cite{Kovchegov:2016zex, Cougoulic:2019aja}). First of all, the light-cone ``minus" momenta have to be ordered, 
\begin{align}\label{k_minus_ordering}
    q^- \gg k_1^- \pm \frac{\Delta^-}{2} \gg k_2^- \pm \frac{\Delta^-}{2} \gg \ldots \gg k_n^- \pm \frac{\Delta^-}{2} . 
\end{align}
Note also that $k_i^- \pm \tfrac{\Delta^-}{2} >0$ by the LCPT rules: the particles are moving forward in the light-cone ``minus" direction and must thus have positive light-cone ``minus" momentum. On either side of the shock wave, the contribution of the $i$th gluon emission must dominate the light-cone energy denominator for the intermediate state immediately following that emission. This leads to the lifetime ordering condition
\begin{align}\label{lifetime1}
    \frac{Q^2}{2q^-}, |\Delta^+| \ll \frac{(\un{k}_1 \pm \un{\Delta}/2)^2}{2(k_1^- \pm \Delta^-/2)}  \ll \frac{(\un{k}_2 \pm \un{\Delta}/2)^2}{2(k_2^- \pm \Delta^-/2)}  \ll \ldots 
    \ll \frac{(\un{k}_{n-1} \pm \un{\Delta}/2)^2}{2(k_{n-1}^- \pm \Delta^-/2)}  \ll \frac{(\un{k}_n \pm \un{\Delta}/2)^2}{2(k_n^- \pm \Delta^-/2)}  . 
\end{align} 
Finally, as is usual for unpolarized small-$x$ evolution, we take all the transverse momenta to be of the same order (and comparable to $Q$),
\begin{align}\label{t_mom_ord}
    Q \sim \left| \un{k}_1 \pm \frac{\un \Delta}{2}  \right| \sim \left| \un{k}_2 \pm \frac{\un \Delta}{2}  \right| \sim ...  \sim \left| \un{k}_n \pm \frac{\un \Delta}{2}  \right| .
\end{align}

Consider the case where some $k_i^-$ momenta are comparable to or smaller in magnitude than $|\Delta^-|$. Then, either $k_i^- + \Delta^- / 2$ or $k_i^- - \Delta^- / 2$ are negative, and we would have violated the LCPT rule about the minus momentum flowing in the forward (right-ward) direction being positive. Hence, we can rule out this contribution and conclude that $k_i^- \gg \Delta^-$ for all $i$.

Because $\Delta^- / q^- \sim 1/s$ is an energy suppressed quantity, with $s \approx 2 q^- \bar P^+$, we can safely neglect it relative to $z_i$ near the top of the cascade and satisfy our orderings. However, as the light-cone momenta $k_i^-$ decrease through the cascade and approach the minimum value, eventually one has the situation where  $k_i^-  \pm \Delta^- / 2$ become too small and there are no further leading logarithmic emissions allowed by LCPT. We see, again, that one needs to have $k_i^- \gg \Delta^-$ for all $i$. At the same time, the absolute minimum for the light-cone ``minus" momentum fractions of the gluons $z_i$ is set by the shock wave as $\Lambda^2/s$ (with $\Lambda$ the IR cutoff). One can see this by requiring that the $x^-$-lifetime of a gluon is longer than the proton shock wave, $2 k^-/k_\perp^2 \gtrsim 1/\bar P^+$, such that $z = k^-/q^- \gtrsim k_\perp^2/s \ge \Lambda^2/s$. From the above discussion, we see that in the case where $\Lambda^2/s < |\zeta|$, one must terminate the cascade when the gluon light-cone ``minus" momentum fractions reach $z_n \approx |\zeta|$. On the other hand, if $\Lambda^2/s > |\zeta|$ then one can take the momenta down to the usual shock wave cutoff $z>\Lambda^2/s$ and have the usual small-$x$ evolution logarithms. Thus, we find that the proper ordering condition for the light-cone momentum fractions is the cascade is
\begin{align}\label{beta-ordering}
    z_1 \gg z_2 \gg ... \gg z_{n-1} \gg z_n \gg \max \left\{ \frac{\Lambda^2}{s}, |\zeta| \right\} .
\end{align}
The diagram containing $n$ gluon emissions in the cascade (and thus giving $n$ steps of evolution) will give us a leading logarithmic contribution given by
\begin{align}
    \alpha_s^{n} \, \int\limits_{z_{\min}}^{1} \frac{\dd{z_1}}{z_1} \int\limits_{z_{\min}}^{z_2} \frac{\dd{z_2}}{z_2}\dots\int\limits_{z_{\min}}^{z_{n-1}} \frac{\dd{z_{n}}}{z_{n}}
    = \frac{1}{n!} \, \alpha_s^{n} \, \ln^{n} \Bigg(\frac{1}{z_{\min}} \Bigg) 
\end{align}
with $z_{\min} = \max \left\{ \Lambda^2/s, |\zeta| \right\}$. The situation here is very similar to the example considered in Sec.~\ref{sec:simple}.

In this work, we take the total momentum transfer $|t| \approx {\Delta}_\perp^2$ to be small compared to the hard scales in the process, following the typical kinematics for collinear factorization of exclusive processes where $t$ is of the same order as the IR scales of the target: $|t| \sim \Lambda^2 \ll Q^2$. In addition, one can argue by invoking the uncertainty principle that $\Delta_\perp \sim 1/R$ with $R$ the radius of the target nucleon: this makes $|t| \approx \Delta_\perp^2$ similarly small. Thus, the shock wave cutoff $\Lambda^2/s$ coming from the forward limit will generically be of the same order as $\zeta$, since $|\zeta| \sim \Delta_\perp^2/s \approx |t|/s \sim \Lambda^2/s$. Thus, we can neglect the difference between $|\zeta|$ and $\Lambda^2/s$ in $\ln (1/z_\textrm{min})$, putting $z_\textrm{min} = \max \left\{ \Lambda^2/s, |\zeta| \right\} \approx \Lambda^2/s$ with the logarithmic accuracy, and thus ignoring the effect of non-zero $\zeta$ on small-$x$ evolution.

\begin{figure}[ht]
\centering
\includegraphics[width= 0.5\linewidth]{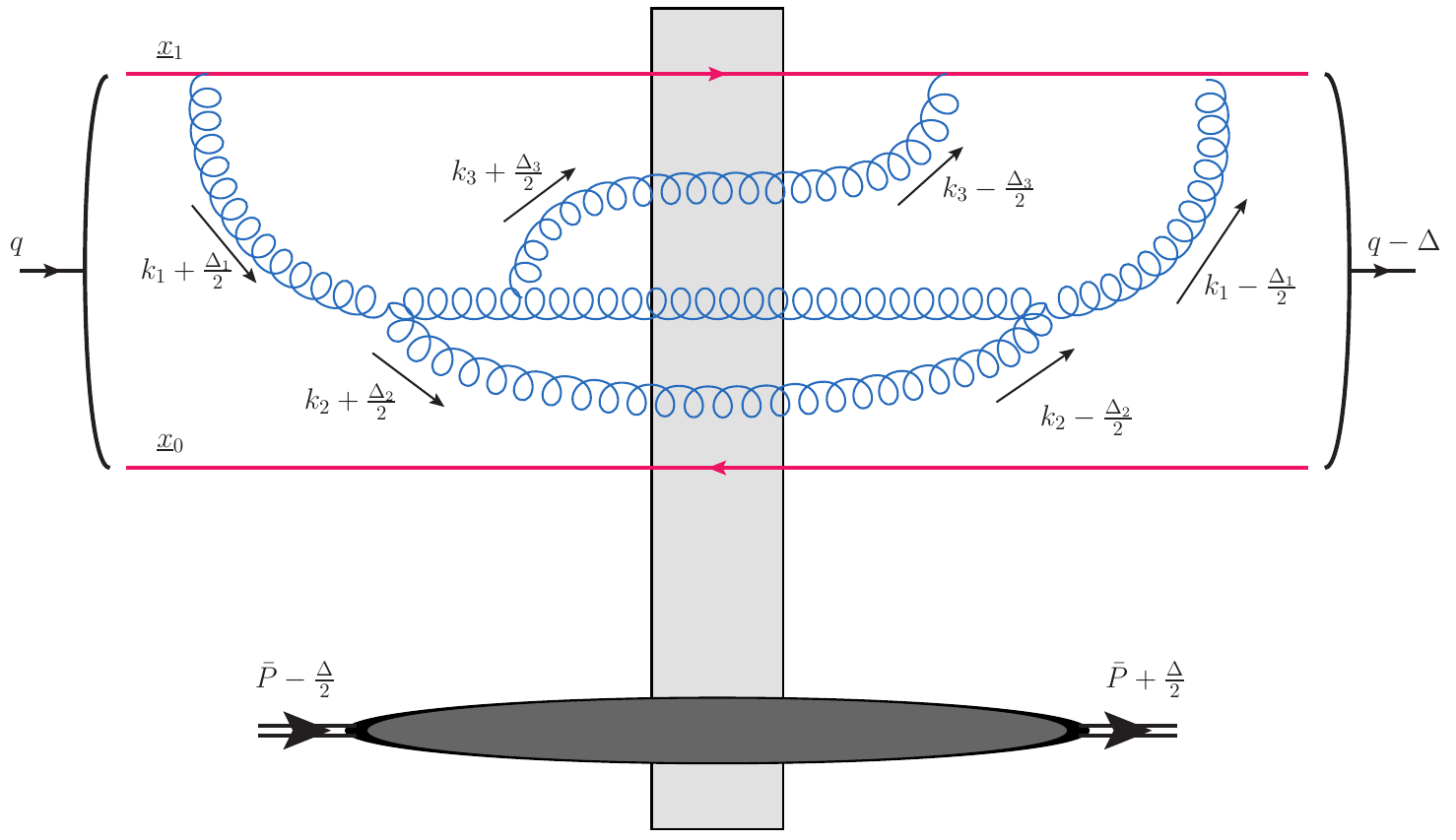}
\caption{A non-ladder gluon cascade in the non-forward elastic scattering case. The rectangle represents the shock wave, comprising subsequent small-$x$ evolution and interaction of the resulting cascade with the target.}
\label{FIG:skew_cascade}
\end{figure}

We have analyzed the effect of non-zero minus longitudinal momentum transfer in the case of ladder diagrams, but even linear BFKL evolution requires going beyond these ladder diagrams within the modern dipole framework \cite{Mueller:1994rr,Mueller:1994jq,Mueller:1995gb}. It appears that, in order to study non-zero plus momentum transfer,  one has to generalize the above discussion to generic gluon or dipole cascade diagrams. Furthermore, we wish to consider non-linear evolution \cite{Balitsky:1995ub, Balitsky:1998ya, Kovchegov:1999yj, Kovchegov:1999ua,Jalilian-Marian:1997jx,Jalilian-Marian:1997gr,Jalilian-Marian:1997dw,Iancu:2001ad,Iancu:2000hn} where successive emissions of gluons generate more and more dipoles which can each have their own cascade and final interaction with the target. This is illustrated in \fig{FIG:skew_cascade}: we see that the original quark anti-quark dipole $10$ emits a gluon labeled by the momentum $k_1$. Then, subsequent emissions of gluons labeled by $k_2$ and $k_3$ can generate a cascade within either the dipole formed by the quark and the gluon with momentum $k_1$ or the dipole formed by the anti-quark and the gluon with momentum $k_1$. All of the gluons in this diagram cross the shock wave denoted by the rectangle, and thus can form dipoles which will generate various further dipole cascades which will then interact with the target. Because each emission/absorption can be non-forward, it must be labeled by a separate momentum transfer $\Delta_i$: 
each $\Delta_i$ is related to an independent loop momentum variable and can thus vary significantly, as it is integrated over. Each gluon line has a momentum labeled as
\begin{align}
    k_i^\mu = \left( \frac{\left({\un k}_{i} \pm \tfrac{\un \Delta_i}{2} \right)^2}{2 \left(k_i^- \pm \tfrac{\Delta^-_i}{2} \right)} , k_i^- \pm \frac{\Delta^-_i}{2} , {\un k}_{i} \pm \frac{\un \Delta_i}{2} \right)
\end{align}
with the plus signs to the left of the shock wave, and the minus signs to its right.
We again require that $k_i^- \pm \Delta_i^-/2 >0$ in order to have valid LCPT diagrams. However, given the independent momentum transfers $\Delta_i$, we might have complications with our resummation when the longitudinal momentum transfer on a given line gets very large or very small. In the case that we have a very large $\Delta_i^-$, the diagrams violate the $k_i^- \pm \Delta_i^-/2 >0$ condition and can be discarded. On the other hand, in the case where we have a vanishingly small $\Delta_i^-$, such that $\Delta_i^- \ll \Delta^-$ with $\Delta^-$ the net minus momentum transfer, further emissions from that gluon line may include contributions where $k_{j>i}^-$ is very small, $\Delta_i^- \ll  k_{j>i}^- \ll \Delta^-$. Such emissions may still be outside the shock wave if $2 (k_i^- \pm \Delta_i^-/2) / \left({\un k}_{i} \pm \tfrac{\un \Delta_i}{2} \right)^2 \gg 1/\bar P^+$, but may be inside the effective shock wave if $2 (k_i^- \pm \Delta_i^-/2) / \left({\un k}_{i} \pm \tfrac{\un \Delta_i}{2} \right)^2 \ll \Delta^- /\Lambda^2 = (\zeta /P^+) \, (s/\Lambda^2)$ with $\zeta = \Delta^- / 2 q^-$ again. This is possible if $\zeta > \Lambda^2 /s$. While such emissions lead to longitudinal logarithms, these logarithms would yield $\ln (\zeta \, s /\Lambda^2) \sim \ln (|t|/\Lambda^2)$, which will not be logarithms of energy and thus will fall outside of the LLA we are working with: again, we discard such contributions. We can thus generalize the ordering conditions from the ladder case \eqref{lifetime1} and \eqref{t_mom_ord} to the non-linear evolution case as
\begin{align}\label{lifetime2}
    \frac{Q^2}{2q^-}, |\Delta^+| \ll \frac{(\un{k}_1 \pm \un{\Delta}_1/2)^2}{2(k_1^- \pm \Delta_1^-/2)}  \ll \frac{(\un{k}_2 \pm \un{\Delta}_2/2)^2}{2(k_2^- \pm \Delta_2^-/2)}  \ll \ldots 
    \ll \frac{(\un{k}_{n-1} \pm \un{\Delta}_{n-1}/2)^2}{2(k_{n-1}^- \pm \Delta_{n-1}^-/2)}  \ll \frac{(\un{k}_n \pm \un{\Delta}_n/2)^2}{2(k_n^- \pm \Delta_n^-/2)} \ll \bar P^+
\end{align} 
and
\begin{align}\label{t_mom_ord2}
    Q \sim \left| \un{k}_1 \pm \frac{\un \Delta_1}{2}  \right| \sim \left| \un{k}_2 \pm \frac{\un \Delta_2}{2}  \right| \sim ...  \sim \left| \un{k}_n \pm \frac{\un \Delta_n}{2}  \right| .
\end{align}
Note that the numbering of the momenta reflects the ordering in $x^-$ light cone time over which the gluons are emitted, such that the gluon with the shortest lifetime (largest $\frac{(\un{k}_n \pm \un{\Delta}_n/2)^2}{2(k_n^- \pm \Delta_n^-/2)} $) is the last emission among all of the cascades at each step in the evolution.

We have analyzed evolution which is logarithmic in the ``minus" light cone momenta of the emitted gluons. However, the variables $x$ and $\xi$ in the GTMDs/GPDs are ``plus" momentum fraction variables. Therefore, to incorporate the effects of non-zero $\xi$ we need to track the plus momenta, as we did in Sec.~\ref{sec:simple}. Note that there is no plus momenta in LCPT ordered along the $x^-$-axis which we employ here (e.g., in \fig{FIG:skew_cascade}). Instead of the plus momenta, this version of LCPT has energy denominators, which are related to the gluon $x^-$-light-cone lifetimes, and are employed in the lifetime ordering condition \eqref{lifetime2}. We will now show how the lifetime ordering in the gluon cascade yields a dependence of the dipole amplitude on the skewness variable $\xi$ in the LLA  small-$x$ evolution. Returning to the energy denominators, the leading logarithms come from the contribution where the last emitted gluon has the largest light-cone energy, dominating the energy denominator. This implies that the light-cone energy (the plus momentum) of each newly emitted gluon should be larger than that for all the pre-existing gluons (and quarks), and also larger than the plus momentum of the particles in the initial state. On the left side of the shock wave this latter requirement leads to
\begin{align}
    \frac{(\un{k}_i + \un{\Delta}_i/2)^2}{2(k_i^- + \Delta_i^-/2)} \gg |q^+| .
\end{align}
This is already satisfied by the lifetime ordering \eqref{lifetime2} we established above, which we see by noting that $q^+ = - Q^2/(2 q^-)$. 
On the right side of the shock wave we similarly require that the gluon light-cone energies are larger than the energy of the outgoing projectile particle (which, when we use the GPD in a scattering process, will be a photon in Deeply Virtual Compton Scattering (DVCS) or a vector meson in Deeply Virtual Meson Production (DVMP)). This gives 
\begin{align}\label{ineq2}
    \frac{(\un{k}_i - \un{\Delta}_i/2)^2}{2(k_i^- - \Delta_i^-/2)} \gg q^+, |\Delta^+ | .
\end{align}
This inequality is included in \eq{lifetime2} as well. In order to satisfy the inequality \eqref{ineq2} with $|\Delta^+|$ on the right we must have
\begin{align}
     \frac{(\un{k}_i - \un{\Delta}_i/2)^2}{2(k_i^- - \Delta_i^-/2)} \gg 2 |\xi| \bar{P}^+ .
\end{align}
Far from the bottom of the cascade we have $k_i^-=z_i q^- \gg \Delta_i^-/2$, so we can neglect $\Delta_i^-$ to find
\begin{align}\label{lto}
    \frac{(\un{k}_i - \un{\Delta}_i/2)^2}{z_is} \gg 2 |\xi| .
\end{align}
The left hand side of \eq{lto} is proportional to the inverse of the $x^-$-lifetime of the $i$th gluon, with the lifetime decreasing as we go down the cascade. The strongest constraint is then at the top of the cascade for the first emitted gluon, $i=1$, where the gluon lives the longest in $x^-$: putting $i=1$ in \eq{lto} yields
\begin{align}
    \frac{(\un{k}_1 - \un{\Delta}_1 /2)^2}{z_1 s} \gg 2 |\xi| ,
\end{align}
where the initial dipole size $x_{10} \sim 1/Q$ sets the typical momentum for the first emitted gluon as $|\un{k}_1 - \un{\Delta}_1 /2| \sim 1/{x_{10}} \sim Q$. We can then rewrite this inequality as
\begin{align}
z_1 \ll \frac{Q^2}{ 2 |\xi| s} .
\end{align}
Finally, remembering that $z_1 \le 1$, because it is a momentum fraction, and, moreover, $z_1 \ll 1$ in the LLA, we write this as
\begin{align}\label{z1_ineq}
    z_1 \ll \min \left\{ 1, \, \frac{x}{2 |\xi|} \right\} ,
\end{align}
where we have defined $x = Q^2/s$ from the longitudinal momentum fraction in the forward limit. 

To justify this value of the average parton momentum fraction $x$, let us go back to the definitions of GTMDs in Eqs.~\eqref{glueGTMD} and \eqref{qGTMD}. There, from the Fourier exponent, we see that the $x^-$-lifetime of the first gluon or (anti-)quark is 
\begin{align}\label{lifetime}
    \Delta x^- \lesssim \frac{1}{x \, \bar P^+}.
\end{align}
Noticing that, on the other hand, the $x^-$-lifetime can be written as 
\begin{align}\label{x-}
    \Delta x^- \approx \frac{2 k^-}{k_\perp^2} \approx \frac{2 k^-}{Q^2}
\end{align}
with the LLA accuracy employed here (with $k$ the momentum of the first gluon for the gluon distribution or the momentum of the anti-quark for the quark distribution). Substituting \eq{x-} into \eq{lifetime}, we arrive at
\begin{align}\label{k1-}
    k^- \lesssim \frac{Q^2}{2 x \bar P^+} 
\end{align}
as the upper cutoff on the $k^-$ momentum components resulting from the definitions of GTMDs (and GPDs). On the other hand, the $q^-$ momentum was introduced at the beginning of this Subsection as the upper cutoff on the $k^-$ momenta. Identifying the right-hand side of \eq{k1-} with $q^-$, and remembering that $s = 2 \bar P^+ q^-$, we arrive at $x = Q^2/s$, as desired.

Taking inequality \eqref{z1_ineq} into account, we find that the logarithmic integral over $z_1$ corresponding to the first step of evolution becomes
\begin{align}\label{zint}
    \int\limits_{\frac{1}{s x_{10}^2}}^{\min \left\{ 1, \frac{x}{2 |\xi|} \right\}} \frac{d z_1}{z_1} \approx \int\limits_{x}^{\min \left\{ 1, \frac{x}{2 |\xi|} \right\}} \frac{d z_1}{z_1} \approx \ln \left( \min \left\{ \frac{1}{x} , \frac{1}{|\xi|} \right\} \right) .
\end{align}
Note that we have taken the lower limit of the $z_1$ integral to be $1/s x_{10}^2$ for simplicity. Recall that we established that this lower bound is equal to  $\Lambda^2/s \approx \zeta$ for kinematics with $|t| \approx \Delta_\perp^2 < Q^2$. Our choice of $1/(s x_{10}^2) \approx Q^2/s \approx x$ for the lower limit of the $z_1$ integration is different from $\Lambda^2/s$: however, the difference by an additive term $\sim \as \, \ln (Q^2 /\Lambda^2)$ is outside of our LLA precision. We similarly neglected a factor of 2 under the logarithm in \eq{zint} within our precision.

We have established the upper bound on the $z_1$ integral corresponding to the first step of evolution. The momentum fractions ordering \eqref{beta-ordering} enforces this upper bound on all further successive emissions. We have therefore confirmed the prescription from the example in Sec.~\ref{sec:simple} that in order to include non-zero skewness, one has to replace
\begin{align}
Y = \ln \frac{1}{x} \to Y = \ln \left( \min \left\{ \frac{1}{x} , \frac{1}{|\xi|} \right\} \right) 
\end{align}
in the arguments of the unpolarized dipole amplitudes. We can then write the gluon GPD at small-$x$ and small-$\xi$ (assuming $|x| \gg |\xi|$ or $|x| \ll |\xi|$) as
\begin{align}\label{gGPD_ev}
    & H^g (|x| \ll 1, |\xi| \ll 1, t, Q^2) = - \frac{N_c}{2 \pi^2 \, \as} \int \dd[2]{b}_{\perp} \,  e^{i \un{\Delta} \vdot {\un b}} \, \left[ \left( \nabla_{\un{x}_1} \vdot  \nabla_{\un{x}_0} \right)   N_{10} \left(Y = \ln \left( \min \left\{\frac{1}{|x|}, \frac{1}{|\xi|} \right\} \right) \right)  \right]_{x_{10}^2 = 1/Q^2} . 
\end{align}
Similarly, one can specify the rapidity in the dipole amplitudes for the quark GPD in \eq{qGPD} and for the gluon and quark GTMDs in Eqs.~\eqref{glueGTMD11} and \eqref{q_GTMD_1}, respectively, to be
\begin{align}\label{Y}
    Y = \ln \left( \min \left\{\frac{1}{|x|}, \frac{1}{|\xi|} \right\} \right) .
\end{align}
Note that we have replaced $x \to |x|$, as $x$ ranges from $-1$ to $1$ in GPDs and GTMDs. Allowing for positive or negative $x$ and using symmetry to work with $|\xi|$ instead of $\xi$, we see that the dipole amplitude rapidity is determined by the intrinsic averaged parton momentum fraction $x$ in the DGLAP region $|x| > |\xi|$, while it is determined by the longitudinal momentum transfer in the ERBL region $|x| < |\xi|$.

Let us restate the main result of this Section: the effect of skewness on the leading logarithmic evolution of the unpolarized GPDs and GTMDs at low $x$ and low $\xi$ is to set the effective rapidity of the unpolarized dipole amplitudes $N_{10} (Y)$ and ${\cal O}_{10} (Y)$ which enter these distributions in Eqs.~\eqref{glueGTMD11}, \eqref{gGPD}, \eqref{q_GTMD_1} and \eqref{qGPD} to equal the logarithm of the minimum of $1/|x|$ and $1/|\xi|$, as shown in \eq{Y}.


\section{The real part of the scattering amplitude $R$-factor} \label{sec:R_fac}


\subsection{Signature factor and the dipole scattering amplitude} 

The developments in the small-$x$/saturation physics over the last decades \cite{Gribov:1984tu, Iancu:2003xm, Weigert:2005us, JalilianMarian:2005jf, Gelis:2010nm, Albacete:2014fwa, Kovchegov:2012mbw, Morreale:2021pnn} using the $s$-channel/shock wave approach \cite{Mueller:1994rr,Mueller:1994jq,Mueller:1995gb,Balitsky:1995ub,Balitsky:1998ya,Kovchegov:1999yj,Kovchegov:1999ua,Jalilian-Marian:1997dw,Jalilian-Marian:1997gr,Weigert:2000gi,Iancu:2001ad,Iancu:2000hn,Ferreiro:2001qy} have largely neglected the signature factor in the Regge theory terminology, with the exception of reference \cite{Kovner:2005qj}. In Regge theory one distinguishes the contributions dependent on either of these two configurations (see, e.g., \cite{Kirschner:1983di})
\begin{align}\label{sig1}
M^\pm (s) = \frac{1}{2} \, \left[ M(s) \pm M(-s) \right],
\end{align}
which are symmetric and antisymmetric under the $s \leftrightarrow u \approx -s$ interchange. Here $M(s)$ is the scattering amplitude averaged over all initial- and final-state helicities (since we are interested in unpolarized scattering). We also assume some projection onto a definite color state. The plus sign in \eq{sig1} corresponds to the positive signature, while the minus sign corresponds to the negative signature. 

At the eikonal order, the scattering amplitude is proportional to the center-of-mass energy squared $s$, that is, $M(s) \sim s$ at the Born level, without small-$x$ evolution. It is customary to work with the rescaled amplitude \cite{Kovchegov:2012mbw}
\begin{align}\label{Adef}
A(s) = \frac{M(s)}{2 |s|}.
\end{align}
We see that 
\begin{align}
M^\pm (s) = |s| \, \left[ A(s) \pm A(-s) \right].
\end{align}

\begin{figure}[ht]
\centering
\includegraphics[width=  \textwidth]{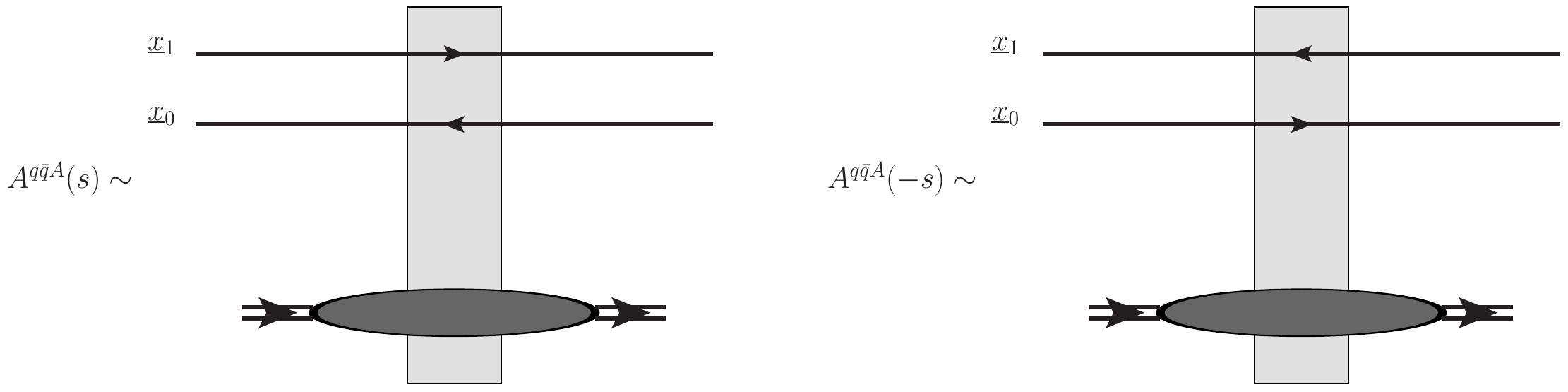}
\caption{Dipole-target scattering diagrams related by the $s \leftrightarrow u \approx -s$ interchange.}
\label{FIG:dipoles}
\end{figure}

Consider scattering of a dipole on some target at the eikonal level. For the rescaled amplitude $A(s)$ we write
\begin{align}\label{Adef}
i \, A^{q {\bar q} A} (s) = \frac{1}{N_c} \, \left\langle \tord  \tr \left[ V_{\un{x}_1} V_{\un{x}_0}^{\dagger} \right] \right\rangle (s) - 1,
\end{align}
where the letter $A$ in the superscript denotes the target, which may be a nucleon or a nucleus. This amplitude corresponds to the incoming quark at $\un x_1$ and an incoming anti-quark at $\un x_0$: see the left panel of \fig{FIG:dipoles} for a diagrammatic representation. The $s \leftrightarrow u$ interchange corresponds to swapping the initial and final state particles in the projectile: hence, $A(-s)$ corresponds to the incoming quark at $\un x_0$ and an incoming anti-quark at $\un x_1$:
\begin{align}
i \, A^{q {\bar q} A} (-s) = \frac{1}{N_c} \, \left\langle \tord  \tr \left[ V_{\un{x}_0} V_{\un{x}_1}^{\dagger} \right] \right\rangle (s) - 1.
\end{align}
This is illustrated in the right panel of \fig{FIG:dipoles}. We conclude that 
\begin{align}
i \, M^\pm_{q {\bar q} A} (s) = \frac{|s|}{N_c} \,  \, \left[  \left\langle \tord  \tr \left[ V_{\un{x}_1} V_{\un{x}_0}^{\dagger} \right] - N_c \right\rangle (s)  \pm \left\langle \tord  \tr \left[ V_{\un{x}_0} V_{\un{x}_1}^{\dagger} \right] - N_c \right\rangle (s) \right].
\end{align}


\subsection{Signature factor in the non-linear small-$x$ evolution}

Consider the dipole $S$-matrix from \eq{S}, now with the time-ordering sign T, 
\begin{align}\label{Ddef}
D_{10} (Y) = \frac{1}{N_c} \, \left\langle \tord  \tr \left[ V_{\un{x}_1} V_{\un{x}_0}^{\dagger} \right] \right\rangle (Y) . 
\end{align}
At large $N_c$, it obeys the standard BK evolution equation \cite{Balitsky:1995ub,Balitsky:1998ya,Kovchegov:1999yj,Kovchegov:1999ua} (in the LLA)
\begin{align}\label{D10}
\pd_Y D_{10} (Y) = \frac{\as \, N_c}{2 \pi^2} \, \int d^2 x_2 \, \frac{x_{10}^2}{x_{21}^2 \, x_{20}^2} \, \left[ D_{12} (Y) \, D_{20} (Y)  - D_{10} (Y)  \right]. 
\end{align}
Interchanging $1$ and $0$ we write
\begin{align}\label{D01}
\pd_Y D_{01} (Y) = \frac{\as \, N_c}{2 \pi^2} \, \int d^2 x_2 \, \frac{x_{10}^2}{x_{21}^2 \, x_{20}^2} \, \left[ D_{02} (Y) \, D_{21} (Y)  - D_{01} (Y)  \right]. 
\end{align}
Defining positive- and negative-signature objects (see, e.g., \cite{Kovchegov:2012ga})
\begin{align}
S_{10} (Y) \equiv \thalf \left[ D_{10} (Y)   + D_{01} (Y)  \right], \ \ \ {\cal O}_{10} (Y) \equiv \tfrac{1}{2 \, i} \left[ D_{10} (Y)   - D_{01} (Y)  \right],
\end{align}
we see that the half-sum of Eqs.~\eqref{D10} and \eqref{D01} yields
\begin{align}\label{Seq0}
\pd_Y S_{10} (Y) = \frac{\as \, N_c}{2 \pi^2} \, \int d^2 x_2 \, \frac{x_{10}^2}{x_{21}^2 \, x_{20}^2} \, \left[ \frac{D_{12} (Y) \, D_{20} (Y) + D_{02} (Y) \, D_{21} (Y)}{2}  - S_{10} (Y)  \right]. 
\end{align}
Since
\begin{align}
D_{12} \, D_{20} + D_{02}  \, D_{21} = \thalf \, \left( D_{12} + D_{21} \right) \, \left( D_{20} + D_{02} \right)  + \thalf \, \left( D_{12} - D_{21} \right) \, \left( D_{20} - D_{02} \right) = 2 \, S_{12} \, S_{20} - 2 \, {\cal O}_{12} \, {\cal O}_{20},
\end{align}
we have (cf. \cite{Kovchegov:2003dm, Hatta:2005as})
\begin{align}\label{Seq}
\pd_Y S_{10} (Y) = \frac{\as \, N_c}{2 \pi^2} \, \int d^2 x_2 \, \frac{x_{10}^2}{x_{21}^2 \, x_{20}^2} \, \left[ S_{12} (Y) \, S_{20} (Y) - S_{10} (Y) - {\cal O}_{12} (Y) \, {\cal O}_{20} (Y)  \right]. 
\end{align}
If we drop the odderon amplitude, we will be back to the BK equation for the positive-signature symmetric $S$-matrix $S_{10} (Y)$. Finally, defining
\begin{align}
N_{10} (Y) =  1 - S_{10} (Y) , 
\end{align}
we rewrite \eq{Seq} as
\begin{align}\label{Neq}
\pd_Y N_{10} (Y) = \frac{\as \, N_c}{2 \pi^2} \, \int d^2 x_2 \, \frac{x_{10}^2}{x_{21}^2 \, x_{20}^2} \, \left[ N_{12} (Y) + N_{20} (Y) - N_{10} (Y) - N_{12} (Y) \, N_{20} (Y)  + {\cal O}_{12} (Y) \, {\cal O}_{20} (Y)  \right]. 
\end{align}

The difference of Eqs.~\eqref{D10} and \eqref{D01}, divided by $2 i$, gives
\begin{align}\label{Oeq0}
\pd_Y {\cal O}_{10} (Y) = \frac{\as \, N_c}{2 \pi^2} \, \int d^2 x_2 \, \frac{x_{10}^2}{x_{21}^2 \, x_{20}^2} \, \left[ \frac{D_{12} (Y) \, D_{20} (Y) - D_{02} (Y) \, D_{21} (Y)}{2 \, i}  - {\cal O}_{10} (Y)  \right],
\end{align}
or, equivalently, 
\begin{align}\label{Oeq}
\pd_Y {\cal O}_{10} (Y) = \frac{\as \, N_c}{2 \pi^2} \, \int d^2 x_2 \, \frac{x_{10}^2}{x_{21}^2 \, x_{20}^2} \, \left[ S_{12} (Y) \, {\cal O}_{20} (Y) + {\cal O}_{12} (Y) \, S_{20} (Y)  - {\cal O}_{10} (Y)  \right]
\end{align}
This is the odderon evolution equation, in the presence of saturation effects \cite{Kovchegov:2003dm, Hatta:2005as}.

From Eqs.~\eqref{Neq} and \eqref{Oeq} we conclude that the evolution of amplitudes of different signature mixes, but in a way that preserves the signature, in the sense that $(-1) \, (-1) = +1$ and $+1 \, (-1) = -1$. Note that $N_{10} (Y)$ has a positive signature, while ${\cal O}_{10} (Y)$ has a negative signature. For the future use, let us rewrite these equations in the integral form as (cf. \cite{Kovchegov:2015pbl,  Kovchegov:2016zex,  Kovchegov:2017lsr, Kovchegov:2018znm, Chirilli:2021lif,  Cougoulic:2022gbk, Borden:2023ugd, Adamiak:2023okq, Borden:2024bxa, Borden:2025ehe})
\begin{align}\label{Neq_int}
& N_{10} (z s) = N_{10}^{(0)} (z s) +  \frac{\as \, N_c}{2 \pi^2} \, \int\limits_{0}^z \frac{d z'}{z'} \,  \int d^2 x_2 \, \frac{x_{10}^2}{x_{21}^2 \, x_{20}^2} \, \left[ N_{12} (z' s) + N_{20} (z' s) - N_{10} (z' s) - N_{12} (z' s) \, N_{20} (z' s) \right.  \\
& \left. \hspace*{13cm} + {\cal O}_{12} (z' s) \, {\cal O}_{20} (z' s)  \right]  \notag
\end{align}
and 
\begin{align}\label{Oeq_int}
{\cal O}_{10} (zs) = {\cal O}^{(0)}_{10} (zs) + \frac{\as \, N_c}{2 \pi^2} \, \int\limits_{0}^z \frac{d z'}{z'}  \, \int d^2 x_2 \, \frac{x_{10}^2}{x_{21}^2 \, x_{20}^2} \, \left[ S_{12} (z' s) \, {\cal O}_{20} (z' s) + {\cal O}_{12} (z' s) \, S_{20} (z' s)  - {\cal O}_{10} (z' s)  \right]. 
\end{align} 
We have switched from rapidity $Y = \ln (zs x_{10}^2) \approx \ln (zs/\Lambda^2)$ as the energy-dependent argument to the center-of-mass energy squared $z s$.
Here, $N_{10}^{(0)}$ and ${\cal O}^{(0)}_{10}$ are the initial conditions for these two dipole amplitudes: if they are energy-independent, then a simple differentiation of Eqs.~\eqref{Neq_int} and \eqref{Oeq_int} with respect to $\ln z \sim Y$ reduces them to Eqs.~\eqref{Neq} and \eqref{Oeq}, respectively, above. However, such a simplification is not possible if $N_{10}^{(0)}$ and ${\cal O}^{(0)}_{10}$ have a non-trivial energy dependence.  In this way, Eqs.~\eqref{Neq_int} and \eqref{Oeq_int} are more general than Eqs.~\eqref{Neq} and \eqref{Oeq}: they allow for energy-dependent inhomogeneous terms/initial conditions (cf. \cite{Itakura:2003jp,  Kovchegov:2015pbl, Kovchegov:2016zex, Kovchegov:2017lsr, Kovchegov:2018znm, Chirilli:2021lif, Kovchegov:2021iyc, Cougoulic:2022gbk, Kovchegov:2022kyy, Borden:2024bxa} for sub-eikonal small-$x$ evolution). 

Note that the lower limit of the $z'$ integrals in Eqs.~\eqref{Neq_int} and \eqref{Oeq_int} is 0 and not $\Lambda^2/s$ coming from the region of validity of the shockwave picture \cite{Kovchegov:2015pbl,  Kovchegov:2016zex,  Kovchegov:2017lsr, Kovchegov:2018znm, Chirilli:2021lif,  Cougoulic:2022gbk, Borden:2023ugd, Adamiak:2023okq, Borden:2024bxa, Borden:2025ehe}) (with $\Lambda$ the IR cutoff). We are, thus, extending the integration a little bit outside of the allowed region. Below we will argue how one can modify the initial conditions $N_{10}^{(0)}$ and ${\cal O}^{(0)}_{10}$ to make the $z'$ integrals convergent near $z'=0$ by effectively cutting them off at $z' \approx \Lambda^2/s$ while generating the proper signature factors.


\subsection{Initial conditions in the non-linear small-$x$ evolution for positive and negative signature factor amplitudes}

\begin{figure}[ht]
\centering
\includegraphics[width= 0.7 \textwidth]{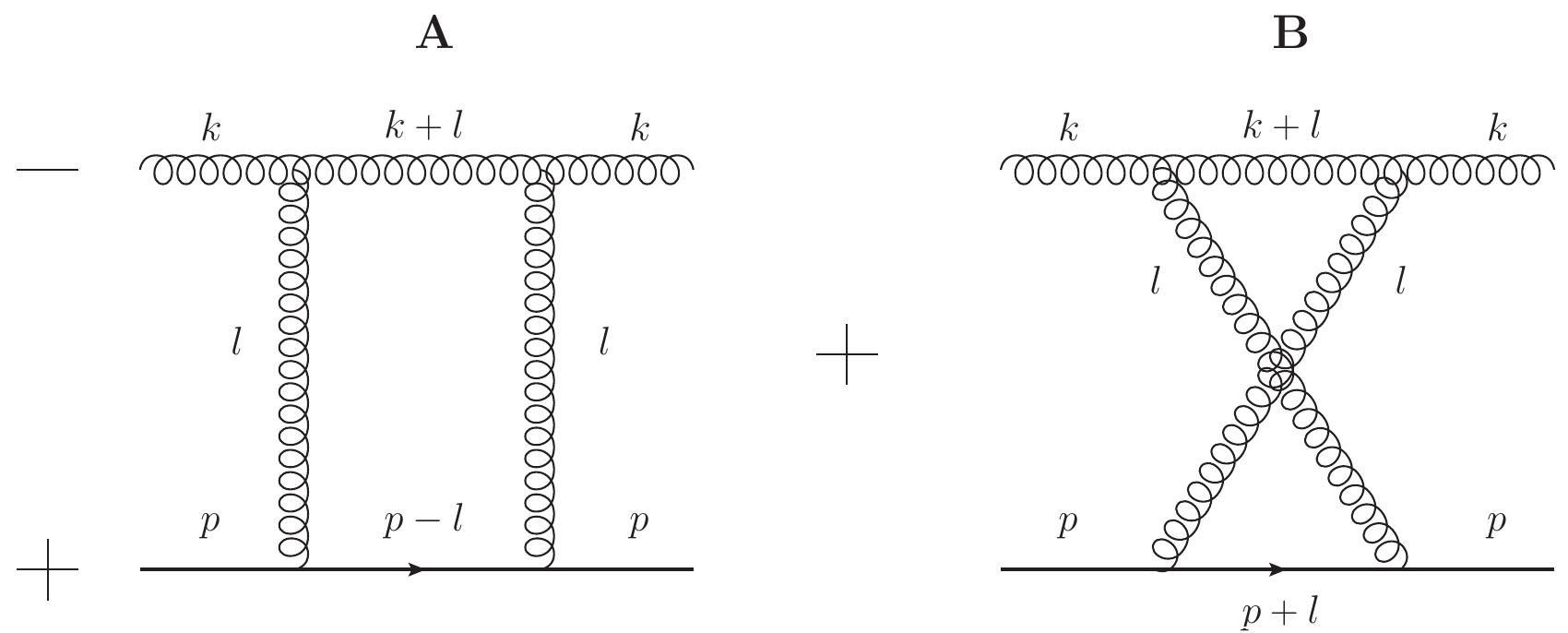}
\caption{High-energy scattering of a gluon generated by small-$x$ $s$-channel evolution on a quark in the target.}
\label{FIG:initial}
\end{figure}

To fully connect the previous two sub-sections, consider a gluon from the minus-moving cascade generated by non-linear small-$x$ evolution of, say, a dipole, scattering on a single quark in the target. The scattering is considered forward, for simplicity, and is depicted in \fig{FIG:initial}. If the exchange is color-singlet, we have two diagrams at the eikonal order, as shown in \fig{FIG:initial}. 

Starting with the diagram A (on the left) in \fig{FIG:initial} we concentrate on the denominators of the propagators. Working in Feynman gauge, we write
\begin{align}\label{step1}
& \int \frac{d^4 l}{(2 \pi)^4} \, \frac{1}{(k+l)^2 + i \epsilon} \, \frac{1}{(l^2 + i \epsilon)^2} \, \frac{1}{(p-l)^2 + i \epsilon} = \int \frac{d^4 l}{(2 \pi)^4} \, \frac{1}{2 (k^- + l^-) \left( \frac{k_\perp^2}{2 k^-} + l^+ \right) - (\un k + \un l)^2 + i \epsilon} \, \frac{1}{(2 l^+ l^- - \un l^2 + i \epsilon)^2} \notag \\
& \times \, \frac{1}{- 2 (p^+ - l^+) \, l^- - \un l^2 + i \epsilon} \approx \frac{i}{4 p^+ k^-} \, \int \frac{d^2 l_\perp}{(2 \pi)^2} \, \frac{1}{(l_\perp^2)^2} \, \int\limits_{-k^-}^0 \frac{d l^-}{2 \pi} \, \frac{1}{l^- + \frac{l_\perp^2}{2 p^+} - i \epsilon}. 
\end{align}
We work in the frame where the (massless) quark carries momentum $p^\mu = (p^+, 0^-, \un 0)$ and assume that the incoming gluon is on mass shell, $k^2 =0$. Moreover, $p^+ >0$ and $k^- >0$. In arriving at the final result in \eq{step1}, we have also assumed that $k^- \gg l^-$, which is justified by the logarithmic integral over $l^-$ we have obtained. Our final conclusion below can be shown to apply if we relax this assumption as well.  

Diagram B in \fig{FIG:initial} is only different from the diagram A by the quark line propagator's denominator. Similar to the above, it gives
\begin{align}\label{step2}
& \int \frac{d^4 l}{(2 \pi)^4} \, \frac{1}{(k+l)^2 + i \epsilon} \, \frac{1}{(l^2 + i \epsilon)^2} \, \frac{1}{(p+l)^2 + i \epsilon} = \int \frac{d^4 l}{(2 \pi)^4} \, \frac{1}{2 (k^- + l^-) \left( \frac{k_\perp^2}{2 k^-} + l^+ \right) - (\un k + \un l)^2 + i \epsilon} \, \frac{1}{(2 l^+ l^- - \un l^2 + i \epsilon)^2} \notag \\
& \times \, \frac{1}{2 (p^+ + l^+) \, l^- - \un l^2 + i \epsilon} \approx - \frac{i}{4 p^+ k^-} \, \int \frac{d^2 l_\perp}{(2 \pi)^2} \, \frac{1}{(l_\perp^2)^2} \, \int\limits_{-k^-}^0 \frac{d l^-}{2 \pi} \, \frac{1}{l^- - \frac{l_\perp^2}{2 p^+} + i \epsilon}. 
\end{align}
In the case of an infinite $x^-$ light-cone Wilson line instead of the gluon projectile in \fig{FIG:initial}  the $(k+l)^2 + i \epsilon$ denominator will be effectively replaced by $l^+ + i \epsilon$: the rest of the calculation will proceed similarly, although the $l^-$ integral would have to be regulated in the ultraviolet (UV), perhaps by tilting the Wilson line away from the light cone.

We conclude that 
\begin{align}\label{step3}
A + B \sim \int\limits_{-k^-}^0 \, d l^- \left[ \frac{1}{l^- + \frac{l_\perp^2}{2 p^+} - i \epsilon} - \frac{1}{l^- - \frac{l_\perp^2}{2 p^+} + i \epsilon} \right] = \ln \left( \frac{l_\perp^2 - i \epsilon }{- 2 p^+ k^- + l_\perp^2 - i \epsilon } \right) - \ln \left( \frac{l_\perp^2 - i \epsilon }{2 p^+ k^- + l_\perp^2 - i \epsilon } \right) . 
\end{align}
Note that in the case when the projectile is a dipole, and the two $t$-channel gluons in \fig{FIG:initial} can couple to different lines in a dipole, there will be no difference between diagrams A and B, such that only one diagram will contribute the result in \eq{step3}.

Now, in the usual eikonal calculation, see, e.g., pp. 134-135 of \cite{Kovchegov:2012mbw}, we assume that $z \, s = 2 p^+ k^- \gg  l_\perp^2$ and simplify \eq{step3} as
\begin{align}\label{LO_simp}
A + B \sim - \ln \frac{zs}{l_\perp^2} + i \pi + \ln \frac{zs}{l_\perp^2} = i \pi. 
\end{align}
This is the origin of the energy-independent initial condition for the small-$x$ evolution equation \eqref{Neq}. 

However, if we remember that the gluon $k$ was generated by the small-$x$ evolution cascade, we would realize that we need to integrate A+B over $k^-$ with the logarithmic integration measure, 
\begin{align}\label{step4}
\int\limits_0^{p_2^-} \frac{d k^-}{k^-} \, (A+B), 
\end{align}
as in Eqs.~\eqref{Neq_int} and \eqref{Oeq_int}, since $z = k^-/p_2^-$. Furthermore, if we employ the simplification \eqref{LO_simp} in \eq{step4}, we will not generate any additional factors of $i \pi$. Doing so neglects the real part of the scattering amplitude $M^+_{q \bar q A}$ which results from summing such powers of $i \, \pi$. Therefore, to keep those $i \pi$'s, it appears necessary to employ the full exact expression \eqref{step3} in  \eq{step4}. 

To see what this gives us, we can consider $n$ steps of evolution, which yield 
\begin{align}\label{step5}
\int\limits_0^{p_2^-} \frac{d k_1^-}{k_1^-} \, \int\limits_0^{k_1^-} \frac{d k_2^-}{k_2^-} \ldots \int\limits_0^{k_{n-1}^-} \frac{d k_n^-}{k_n^-} \left[ \ln \left( \frac{l_\perp^2 - i \epsilon }{- 2 p^+ k_n^- + l_\perp^2 - i \epsilon } \right) - \ln \left( \frac{l_\perp^2 - i \epsilon }{2 p^+ k_n^- + l_\perp^2 - i \epsilon } \right) \right].
\end{align}
Note that we allow the evolution integrals to start at zero as the lower limit: the integrals are still convergent. 

The two terms in \eq{step5} are related by 
\begin{align}
\oone = \otwo  (p^+ \to - p^+),
\end{align}
or, equivalently, by
\begin{align}\label{sub}
\oone = \otwo  (s \to - s).
\end{align}
Calculating the second term first, we get
\begin{align}\label{step6}
& \otwo = \int\limits_0^{p_2^-} \frac{d k_1^-}{k_1^-} \, \int\limits_0^{k_1^-} \frac{d k_2^-}{k_2^-} \ldots \int\limits_0^{k_{n-1}^-} \frac{d k_n^-}{k_n^-}  \ln \left( \frac{l_\perp^2 - i \epsilon }{2 p^+ k_n^- + l_\perp^2 - i \epsilon } \right) \approx \int\limits_{l_\perp^2/(2 p^+)}^{p_2^-} \frac{d k_1^-}{k_1^-} \, \int\limits_{l_\perp^2/(2 p^+)}^{k_1^-} \frac{d k_2^-}{k_2^-} \ldots \int\limits_{l_\perp^2/(2 p^+)}^{k_{n-1}^-} \frac{d k_n^-}{k_n^-}  \ln \left( \frac{l_\perp^2  }{2 p^+ k_n^- - i \epsilon } \right) \notag \\
& = - \frac{1}{(n+1) !} \, \ln^{n+1}  \left( \frac{s - i \epsilon}{l_\perp^2} \right).
\end{align}
Using \eq{sub}, we arrive at
\begin{align}\label{log_diff}
\oone - \otwo = - \frac{1}{(n+1) !} \, \left[  \ln^{n+1}  \left( \frac{-s - i \epsilon}{l_\perp^2} \right)  -  \ln^{n+1}  \left( \frac{s - i \epsilon}{l_\perp^2} \right) \right]. 
\end{align}
For $n=0$ we readily reproduce \eq{LO_simp}. Note also that \eq{log_diff}, after being multiplied by Sign$(s)$ resulting from $s/|s|$ one gets from \eq{Adef}, becomes an even function of $s$. Hence, as a two-gluon $t$-channel exchange contribution to $D_{10}$, it survives in the positive-signature combination $M^+_{q {\bar q} A}$ and disappears in the negative-signature $M^-_{q {\bar q} A}$,
\begin{align}
i \, M^\pm_{q {\bar q} A} (s) = |s|  \, \left[  \left( D_{10} (s) - 1 \right) \pm \left( D_{10} (-s) - 1 \right) \right] = |s|  \, \left[  \left( D_{10} (s) - 1 \right) \pm \left( D_{01} (s) - 1 \right) \right],
\end{align}
as it should. 

Equation \eqref{log_diff} illustrates the terms we are trying to resum by including the signature factor. While at LLA one sums up leading logarithms of energy, that is, powers of $\as \ln s$, we see that some logarithms of energy come in as $\ln (-s - i \epsilon) = \ln s - i \pi$. The resummation parameter for such logarithms is $\as \ln (-s - i \epsilon) = \as  \ln s - \as  i \pi$. We see that in addition to resumming the usual LLA parameter $\as  \ln s$, keeping the signature factor effectively resums powers of $- \as i \pi$. While such powers of $- \as i \pi$, strictly-speaking, constitute non-logarithmic power-of-$\as$ higher-order corrections, their resummation is well-defined, as it allows one to reconstruct the real part of the LLA scattering amplitude $M (s)$, just like it would have been obtained using dispersion relations (see \cite{Kovchegov:2012mbw} and references therein).

Let us note that in the case of linear (BFKL) evolution, the terms in \eq{log_diff} can be summed over all integer $n\ge0$, after being multiplied by the BFKL intercept $\alpha_p - 1$ to the $n$th power (taken in the Mellin space of the $\gamma$ variable, where $\alpha_P - 1 = (\as \, N_c/\pi) \, \chi (\gamma)$ in leading-order (LO) BFKL evoltion, with $\chi (\gamma)$ the eigenvalue of the LO BFKL kernel \cite{Kuraev:1977fs,Balitsky:1978ic}). This gives
\begin{align}\label{exponentialtion}
    \sum_{n=0}^\infty \frac{(\alpha_P - 1)^n}{(n+1) !} \, \left[  \ln^{n+1}  \left( \frac{-s - i \epsilon}{l_\perp^2} \right)  -  \ln^{n+1}  \left( \frac{s - i \epsilon}{l_\perp^2} \right) \right] \sim \left( \frac{-s - i \epsilon}{l_\perp^2} \right)^{\alpha_P -1} -  \left( \frac{s - i \epsilon}{l_\perp^2} \right)^{\alpha_P -1},
\end{align}
a factor associated with the positive-signature exchange \cite{Forshaw:1997dc, Ioffe:2010zz}. The signature factor, in general, is defined by \cite{Forshaw:1997dc, Ioffe:2010zz} 
\begin{align}
    \xi_p (j) = - \frac{e^{- i \pi j} + p}{\sin \pi j},
\end{align}
where $j$ is the intercept (angular momentum in Regge theory) or the Mellin moment-space variable conjugate to $x$: $j = \alpha_P$ in \eq{exponentialtion}. Here, $p=+1$ corresponds to the positive signature and $p=-1$ corresponds to the negative signature. Our goal is to find a way to include the effect of this signature factor $\xi_p (j)$ into the nonlinear evolution of Eqs.~\eqref{Neq_int} and \eqref{Oeq_int}. 

What we have just shown is that by keeping naively energy-suppressed terms in the initial condition \eqref{step3} for the small-$x$ evolution (and not simplifying $A+B$ down to the standard \eq{LO_simp}) allows one to restore the signature factor in the BFKL evolution sourced by such an initial condition. The generalization to the non-linear case can now be done by analogy via modification of the initial conditions.  

Since we are not keeping track of corrections other than $i \pi$, we can replace $l_\perp^2 \to \Lambda^2$ with $\Lambda$ our IR cutoff. Based on the above, by comparing Eq.~\eqref{step3} to its standard approximation \eqref{LO_simp}, we see that to include the signature factor into the linear BFKL evolution in the dipole/shockwave formalism, the prescription is to multiply the initial condition/inhomogeneous term by 
\begin{align}\label{sig_factor}
\frac{\mbox{Sign} (zs)}{i \pi} \, \left[ \ln \left( \frac{\Lambda^2 - i \epsilon }{- z s + \Lambda^2 - i \epsilon } \right) - \ln \left( \frac{\Lambda^2 - i \epsilon }{z s + \Lambda^2 - i \epsilon } \right) \right] = \theta \left( |zs| - \Lambda^2 \right) + i \, \frac{\mbox{Sign} (zs)}{\pi}  \, \ln \left( \frac{|zs - \Lambda^2|}{|zs + \Lambda^2|} \right)
\end{align} 
and use the resulting $N_{10}^{(0)}$ in the linearized version of \eq{Neq_int}. The overall factor of Sign$(zs)$ on the left-hand side of \eq{sig_factor} comes from the fact that the eikonal two-gluon  exchange amplitude (or, to that matter, any eikonal amplitude) is proportional to $s$, while to make it energy-independent we divide it by $|s|$, per \eqref{Adef}. The factor in \eq{sig_factor} is equal to 1 for $|zs| \gg \Lambda^2$, thus restoring the original initial conditions in this high-energy limit. As we have seen above, not assuming that $|zs| \gg \Lambda^2$ allows one to keep the effects of the signature factor in small-$x$ evolution. Note also that the form of the expression on the right-hand side of \eq{sig_factor} clarifies the physical meaning of our factor on the left of the same equation: its real part is simply a $\theta$-function imposing the cutoff $z > \Lambda^2/s$ on the integral over $z$, as is done in the standard shockwave calculations. The imaginary part of the factor is strictly-speaking sub-eikonal: however, after (multiple) $z$-integral(s) in the small-$x$ evolution (written in an integral form), this part ceases to be sub-eikonal: rather, it becomes eikonal (power of energy suppression disappears) and generates the factors of $i \, \pi$, which, when assembled together, form the signature factor for the linear BFKL evolution. This justifies us keeping such sub-eikonal terms in our approximation for the initial conditions.

In the case of the non-linear (BK/JIMWLK) evolution, the GGM/MV initial condition \cite{Mueller:1989st, McLerran:1993ka, McLerran:1993ni, McLerran:1994vd} for the dipole amplitude 
\begin{align}
    N_{10}^{(0)} (z s) = 1 - \exp \left\{ - \frac{1}{4} \, x_{10}^2 \, Q_{s0}^2 \, \ln \left( \frac{1}{x_{10} \, \Lambda}  \right) \right\}
\end{align}
should be replaced by 
\begin{align}\label{N0}
    N_{10}^{(0)} (z s) = 1 - \exp \left\{ - \frac{1}{4} \, x_{10}^2 \, Q_{s0}^2 \, \ln \left( \frac{1}{x_{10} \, \Lambda}  \right) \, \frac{\mbox{Sign} (zs)}{i \pi} \, \left[ \ln \left( \frac{\Lambda^2 - i \epsilon }{- z s + \Lambda^2 - i \epsilon } \right) - \ln \left( \frac{\Lambda^2 - i \epsilon }{z s + \Lambda^2 - i \epsilon } \right)
 \right] \right\}
\end{align}
to be used as the inhomogeneous term in \eq{Neq_int}.

\begin{figure}[ht]
\centering
\includegraphics[width= 0.45 \textwidth]{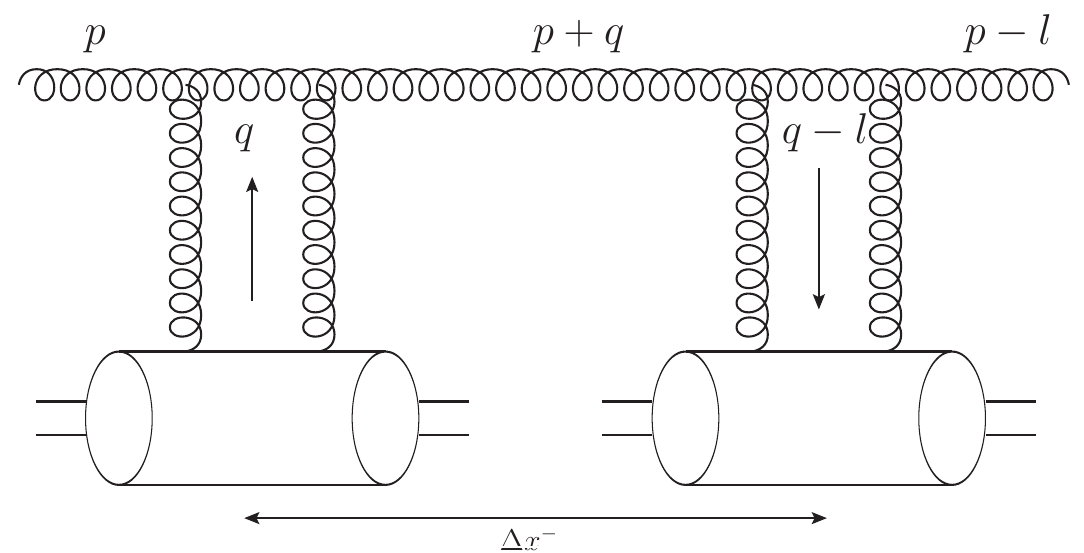}
\caption{Double scattering of a gluon in a target nucleus.}
\label{FIG:rescatterings}
\end{figure}

To justify the result \eqref{N0}, we note that here the two-gluon exchange cross section of a dipole scattering on a single nucleon in the nucleus (multiplied by appropriate factors) simply exponentiates \cite{Kovchegov:2012mbw}. Our above calculation of the two-gluon exchange showed that this cross section needs to be multiplied by the factor in \eq{sig_factor}. However, the factor of Sign$(zs)$ did not arise from the diagrammatic calculation; rather, it appeared due to the definition \eqref{Adef}. To show that this factor exponentiates as well, consider two consecutive scatterings of a gluon in a target nucleus, as depicted in \fig{FIG:rescatterings}. The incoming gluon has a large $p^-$ momentum, while the two scatterings are separated by the distance $\Delta x^-$ on the light cone. 

Repeating the standard calculation \cite{Kovchegov:2012mbw} yields for the longitudinal Fourier transform of the internal gluon propagator's denominator
\begin{align}
    \int\limits_{-\infty}^\infty \frac{d q^+}{2 \pi} \, \frac{e^{- i q^+ \Delta x^-}}{2 p^- q^+ + i \epsilon} = - \frac{i}{2 \, |p^-|} \, \left[ \theta (p^-) \, \theta (\Delta x^- ) +  \theta (-p^-) \, \theta (-\Delta x^- ) \right] .
\end{align}
We see that one obtains the absolute value of the gluon's minus momentum, $|p^-|$, in the denominator. This results in dividing the amplitude for the second scattering by $|s|$, giving us a factor of Sign$(zs)$ for each rescattering. We conclude that indeed this Sign$(zs)$ factor exponentiates as well, justifying \eq{N0}. Defining rapidity $Y = \ln (z s x_{10}^2)$ and employing the substitution in \eq{sig_factor} we reduce \eq{N0} to \eq{IC} in the physical regime of $z s >0$.

For the odderon amplitude in the scattering on a large nucleus, the initial condition including saturation effects reads \cite{Kovchegov:2012ga}
\begin{align}\label{odd_init_1}
    {\cal O}_{10}^{(0)} (z s) = o_{10}^{(0)} \,  \exp \left\{ - \frac{1}{4} \, x_{10}^2 \, Q_{s0}^2 \, \ln \left( \frac{1}{x_{10} \, \Lambda}  \right) \right\},
\end{align}
where $o_{10}^{(0)}$ is the lowest-order odderon amplitude due to an exchange of three gluons in the $d^{abc}$ color configuration. To include the signature factor effects, we have to perform an analysis similar to the one above, but only for the $d^{abc}$ three-gluon exchange. This analysis is detailed in Appendix~\ref{sec:look_odd} below. There, we obtain
\begin{align}\label{odd_init_3}
    & {\cal O}_{10}^{(0)} (z s) = o_{10}^{(0)}
    \, \frac{\mbox{Sign} (zs)}{(i \pi)^2} \, \left[ \ln \left( \frac{\Lambda^2 - i \epsilon }{- z s + \Lambda^2 - i \epsilon } \right) - \ln \left( \frac{\Lambda^2 - i \epsilon }{z s + \Lambda^2 - i \epsilon } \right)
  \right]^2 \\
 & \hspace{1.32cm} \times \exp \left\{ - \frac{1}{4} \, x_{10}^2 \, Q_{s0}^2 \, \ln \left( \frac{1}{x_{10} \, \Lambda}  \right) \, \left[ \theta (|zs| - \Lambda^2) + \frac{i}{\pi} \, \textrm{Sign} (z s)  \, \ln \left( \frac{|z s - \Lambda^2|}{|z s+\Lambda^2 | } \right) \right]  \right\}   \notag \\
 & \, = o_{10}^{(0)}  \, \mbox{Sign} (zs) \,  \left[ \theta (|zs| - \Lambda^2) + \frac{2 \, i}{\pi} \, \theta (|zs| - \Lambda^2) \, \textrm{Sign} (z s)  \, \ln \left( \frac{|z s - \Lambda^2|}{|z s+\Lambda^2 | } \right) - \frac{1}{\pi^2} \, \ln^2 \left( \frac{|z s - \Lambda^2|}{|z s+\Lambda^2 | } \right) \right] \notag  \\
   & \hspace{1.32cm}  \times \exp \left\{ - \frac{1}{4} \, x_{10}^2 \, Q_{s0}^2 \, \ln \left( \frac{1}{x_{10} \, \Lambda}  \right) \, \left[ \theta (|zs| - \Lambda^2) + \frac{i}{\pi} \, \textrm{Sign} (z s)  \, \ln \left( \frac{|z s - \Lambda^2|}{|z s+\Lambda^2 | } \right) \right]  \right\} .  \notag
\end{align}

Let us point out that the real part of the scattering amplitude $M$ (or $A$), via \eq{Adef}, corresponds to the (negative) imaginary part of the dipole amplitude $N$. The imaginary part of the odderon amplitude $\cal O$ corresponds to the imaginary part of the scattering amplitudes $M$ and $A$.


\subsection{Summary}

Let us summarize the results of this Section. In order to include the real part of the scattering amplitude $M$, or, equivalently, the imaginary part of the positive-signature dipole amplitude $N_{10} (s)$, one has to solve the non-linear evolution equation \cite{Balitsky:1995ub,Balitsky:1998ya,Kovchegov:1999yj,Kovchegov:1999ua} written in an integral form, 
\begin{align}\label{BK_int}
& N_{10} (z s) = N_{10}^{(0)} (z s) +  \frac{\as \, N_c}{2 \pi^2} \, \int\limits_{0}^z \frac{d z'}{z'} \,  \int d^2 x_2 \, \frac{x_{10}^2}{x_{21}^2 \, x_{20}^2} \, \left[ N_{12} (z' s) + N_{20} (z' s) - N_{10} (z' s) - N_{12} (z' s) \, N_{20} (z' s)  \right] ,
\end{align}
with the initial condition/inhomogeneous term given by 
\begin{align}\label{N0_2}
    N_{10}^{(0)} (z s) \, & =  1 - \exp \left\{ - \frac{1}{4} \, x_{10}^2 \, Q_{s0}^2 \, \ln \left( \frac{1}{x_{10} \, \Lambda}  \right) \, \frac{\mbox{Sign} (zs)}{i \pi} \, \left[ \ln \left( \frac{\Lambda^2 - i \epsilon }{- z s + \Lambda^2 - i \epsilon } \right) - \ln \left( \frac{\Lambda^2 - i \epsilon }{z s + \Lambda^2 - i \epsilon } \right)
 \right] \right\} \\
 & = 1 - \exp \left\{ - \frac{1}{4} \, x_{10}^2 \, Q_{s0}^2 \, \ln \left( \frac{1}{x_{10} \, \Lambda}  \right) \,  \left[ \theta \left( |zs| - \Lambda^2 \right) + i \, \frac{\mbox{Sign} (zs)}{\pi}  \, \ln \left( \frac{|zs - \Lambda^2|}{|zs + \Lambda^2|} \right)
 \right] \right\} . \notag
\end{align}
This prescription should replace the phenomenological R-factor used in the literature \cite{Kowalski:2006hc,Toll:2012mb,Mantysaari:2016jaz}. It is valid in the case when the odderon contribution is negligible: indeed the odderon amplitude is suppressed by one power of $\as$ in the small-$x$ power counting compared to the C-even (pomeron) amplitude \cite{Bartels:1999yt,Kovchegov:2003dm,Hatta:2005as}.

To also include the odderon effects and allow for the imaginary part to the negative-signature odderon amplitude ${\cal O}_{10} (s)$ one needs to solve the coupled integral equations \cite{Kovchegov:2003dm,Hatta:2005as} (see also \cite{Bartels:1999yt,Kovchegov:2012rz, Janik:1998xj, Caron-Huot:2013fea,Bartels:2013yga, Brower:2008cy,Avsar:2009hc,Brower:2014wha})
\begin{subequations}\label{NOeq_int}
\begin{align}
& N_{10} (z s) = N_{10}^{(0)} (z s) +  \frac{\as \, N_c}{2 \pi^2} \, \int\limits_{0}^z \frac{d z'}{z'} \,  \int d^2 x_2 \, \frac{x_{10}^2}{x_{21}^2 \, x_{20}^2} \, \left[ N_{12} (z' s) + N_{20} (z' s) - N_{10} (z' s) - N_{12} (z' s) \, N_{20} (z' s) \right.  \\
& \left. \hspace*{13cm} + {\cal O}_{12} (z' s) \, {\cal O}_{20} (z' s)  \right],   \notag \\
& {\cal O}_{10} (zs) = {\cal O}^{(0)}_{10} (zs) + \frac{\as \, N_c}{2 \pi^2} \, \int\limits_{0}^z \frac{d z'}{z'}  \, \int d^2 x_2 \, \frac{x_{10}^2}{x_{21}^2 \, x_{20}^2} \, \left[ S_{12} (z' s) \, {\cal O}_{20} (z' s) + {\cal O}_{12} (z' s) \, S_{20} (z' s)  - {\cal O}_{10} (z' s)  \right] ,
\end{align} 
\end{subequations}
with the initial condition/inhomogeneous term $N_{10}^{(0)}$ given by \eq{N0_2} and ${\cal O}^{(0)}_{10}$ given by \eq{odd_init_3} and $S_{10} (s) = 1 - N_{10} (s)$. Note that the inhomogeneous terms \eqref{N0_2} and \eqref{odd_init_3} make sure that the $z'$ integrals in Eqs.~\eqref{BK_int} and \eqref{NOeq_int} are convergent, and are effectively cut off by $z' > \Lambda^2/s$, in agreement with the conventional calculations where $\Lambda^2/s$ is chosen to be the lower limit of the $z'$ integral. 


It is instructive to study the high-energy asymptotics of the dipole amplitudes $N_{10}$ and ${\cal O}_{10}$ with the initial conditions allowing for non-zero imaginary parts for these dipole amplitudes, like those in Eqs.~\eqref{N0_2} and \eqref{odd_init_3}. 
The fixed points of Eqs.~\eqref{NOeq_int} are obtained by requiring that the integrands on their right-hand sides are zero for constant $N_{10}$ and ${\cal O}_{10}$. Just like in the case of purely real $N_{10}$ we get the fixed points at $N=0, {\cal O} = 0$ and $N=1, {\cal O} = 0$.\footnote{There are also fixed points at $N=1/2$ and ${\cal O} = \pm i/2$. However, these fixed points (along with any other constant and non-zero high-energy asymptotic expression for the odderon) violate \eq{O_zero} below. Therefore, ${\cal O} = \pm i/2$ cannot be the high-energy asymptotics of the odderon: this invalidates the fixed points at $N=1/2$ and ${\cal O} = \pm i/2$.}
The former is unstable, since, given a small fluctuation about $N=0$ the linear (BFKL) part of the evolution would drive the system away from the fixed point. The fixed point at $N=1, {\cal O} = 0$ is stable and represents the high-energy asymptotics of the dipole amplitudes $N$ and $\cal O$. This means that 
\begin{align}
    \lim_{s \to \infty} \textrm{Im} \left[ N_{10} (s) \right] = 0, \ \ \ \lim_{s \to \infty} \textrm{Im} \left[ {\cal O}_{10} (s) \right] = 0.
\end{align}
The imaginary parts of $N_{10}$ and ${\cal O}_{10}$ vanish in the black disk limit. Given that 
\begin{align}\label{O_zero}
    \int d^2 b_\perp \, {\cal O}_{10} (s) = 0,
\end{align}
we see that the total and elastic dipole--target cross sections are
\begin{align}
    \sigma_{tot}^{q {\bar q} A} = 2 \, \int d^2 b_\perp \, \textrm{Re} \left[ N_{10} (s) - i \, {\cal O}_{10} (s) \right] = 2 \, \int d^2 b_\perp \, \textrm{Re} \left[ N_{10} (s) \right], \ \ \ \sigma_{el}^{q {\bar q} A} = \int d^2 b_\perp \, \Big| N_{10} (s) - i \, {\cal O}_{10} (s) \Big|^2 
\end{align}
(for a dipole with the quark fixed at transverse position $\un x_1$ and the anti-quark fixed at $\un x_0$, as in \eq{Ddef}).
We conclude that the $N=1, {\cal O} = 0$ fixed point still implies that 
\begin{align}
    \lim_{s \to \infty} \, \frac{\sigma_{el}^{q {\bar q} A}}{\sigma_{tot}^{q {\bar q} A}} = \half, 
\end{align}
in agreement with conventional wisdom. 

We also would like to note that in the linear (BFKL) evolution regime, where \eq{exponentialtion} applies, our prescription above can be approximated by the relation
\begin{align}\label{ImRe}
 \textrm{Im} \left[ N_{10} (s) \right] \approx - \frac{\pi}{2} \, \frac{d}{d \ln s} \, \textrm{Re} \left[ N_{10} (s) \right],
\end{align}
valid at the first non-trivial order in $\as$ and often used in phenomenology \cite{Gotsman:2020mkd}. From our results presented here, it also follows that the relation \eqref{ImRe} is consistent with the $N_{10} (s) \to 1$ high-energy asymptotics of the dipole amplitude. However, we cannot verify the validity of \eq{ImRe} in the intermediate energy region between the BFKL evolution and the black-disk limit without fully solving the above evolution equation(s).   

\section{Reconciling the two parts of the paper}
\label{sec:rec}

A careful reader probably noticed that the definitions of dipole amplitudes differ in our Sections~\ref{sec:gpds} and \ref{sec:R_fac}. In the former, the dipole amplitudes are defined in Eqs.~\eqref{Ndef} and \eqref{Odef}, which contain no time ordering operator. Since the operator in the angle brackets of \eq{Ndef} is Hermitian, while the operator in the angle brackets of \eq{Odef} is anti-Hermitian and divided by $i$, both amplitudes $N_{10} (Y)$ and ${\cal O}_{10} (Y)$ are manifestly real. 

In Sec.~\ref{sec:R_fac}, we study an elastic scattering amplitude, and our definitions of the dipole amplitudes imply a time-ordering operator $T$; namely 
\begin{subequations}
\begin{align}
    & N_{10}^T (Y) \equiv 1 - \frac{1}{2 N_c} \, \left\langle \tord \tr \left[ V_{\un{x}_1} V_{\un{x}_0}^{\dagger} \right] + \tord \tr \left[ V_{\un{x}_0} V_{\un{x}_1}^{\dagger} \right]  \right\rangle (Y) ,\label{NdefT} \\
\label{OdefT}
    & {\cal O}_{10}^T (Y) \equiv \frac{1}{2 i N_c} \, \left\langle \tord \tr \left[ V_{\un{x}_1} V_{\un{x}_0}^{\dagger} \right] - \tord \tr \left[ V_{\un{x}_0} V_{\un{x}_1}^{\dagger} \right]  \right\rangle (Y) .
\end{align}
\end{subequations}
To accentuate the difference, we have added the superscript T to the amplitudes. Due to the time-ordering signs, the operators defining $N_{10}^T (Y)$ and ${\cal O}_{10}^T (Y)$ are no longer Hermitian and anti-Hermitian, respectively, and, therefore, these amplitudes are not necessarily real. This allowed us to calculate the imaginary parts of both $N_{10}^T (Y)$ and ${\cal O}_{10}^T (Y)$ in Sec.~\ref{sec:R_fac}. 

This difference between $N_{10} (Y)$ and ${\cal O}_{10} (Y)$ on one side and $N_{10}^T (Y)$ and ${\cal O}_{10}^T (Y)$ on the other appears to violate the equivalence between the two objects observed in \cite{Mueller:2012bn}. While the calculation in \cite{Mueller:2012bn} was eikonal, it included next-to-leading order (NLO) small-$x$ evolution, demonstrating that small-$x$ evolution at both leading order and at NLO leaves both types of eikonal amplitudes identical, 
\begin{align}\label{equal}
    N_{10} (Y) =  N_{10}^T (Y) \ \ \ \textrm{and} \ \ \ {\cal O}_{10} (Y) = {\cal O}_{10}^T (Y) .
\end{align}
However, the calculation in \cite{Mueller:2012bn} did not include the signature effects: we believe this is why we do not see the equalities \eqref{equal} in our present analysis. Rather, we conjecture that Eqs.~\eqref{equal} should be replaced by
\begin{align}\label{Re_equal}
    N_{10} (Y) =  \textrm{Re} \left[ N_{10}^T (Y) \right] \ \ \ \textrm{and} \ \ \ {\cal O}_{10} (Y) = \textrm{Re} \left[ {\cal O}_{10}^T (Y) \right] .
\end{align}

Another important consequence of the difference between $N_{10} (Y)$ and $N_{10}^T (Y)$ concerns \eq{gGPD}. Since $N_{10} (Y)$ is real, so is the gluon GPD $H^g$ given by \eq{gGPD}. (For the unpolarized target, the dipole amplitude $N_{10} (Y)$ is symmetric under $\un b \to - \un b$, such that the Fourier transform in \eq{gGPD} does not generate an imaginary part of $H^g$.) This is in accordance with the knwon fact that while GTMDs may have an imaginary part, the GPDs, as defined above, without the time-ordering sign, are real (see \cite{Diehl:2003ny} and references therein). At the same time, since GPDs are defined for elastic scattering amplitudes, the time-ordering sign is more natural for them (and follows from factorization theorems). Repeating the analysis of Sec.~\ref{sec:gpds} with time-ordering inserted in the definition of gluon GPD will result in \eq{gGPD} with the dipole amplitude $N_{10}^T (Y)$ instead of $N_{10} (Y)$. Since, as we have shown in Sec.~\ref{sec:R_fac}, $N_{10}^T (Y)$ has an imaginary part, it appears that so will the GPD $H^g$. (It is possible that a similar concern would apply to the quark GPD $H^q$: however, our derivation of $H^q$ at small-$x$ in Sec.~\ref{sec:gpds} assumed no time-ordering sign. Hence, to verify this concern the calculation would have to be redone assuming time-ordering in the definition of $H^q$, which is beyond the scope of this paper.) The possibility of a GPD with the time-ordering sign being different from the one without one seems to be in violation of the conclusion reached in \cite{Diehl:1998sm} (see also \cite{Landshoff:1970ff, Frankfurt:1997ha, Radyushkin:1997ki, Jaffe:1983hp}). Since the difference between $N_{10} (Y)$ and $N_{10}^T (Y)$ is beyond the LLA and resides in the factors of $i \pi$ which are suppressed by a logarithm of energy (they come in multiplied by $\as$ without a $\ln s$ enhancement), the difference in the $H^g$ with and without time ordering is also beyond the LLA approximation. However, the difference may still be non-zero, suppressed by only one power of a logarithm of energy (see, e.g., \eq{ImRe}). While it is possible that our derivation of \eq{gGPD} given above only applies in the LLA, it appears we have only neglected sub-eikonal (proportional to $x$) terms in this derivation, and, therefore, it should contain all the eikonal terms, at LLA and beyond. We note that in \cite{Muller:2013jur}, the  signature factor in the expression for the elastic cross section results from the conformal moment of the coefficient function: this appears to be different from our case, where the signature factor is a part of the GPD itself. We leave the resolution of this puzzle for future work.


\section{Conclusions} 

\label{sec:conc}

In this work, we have considered two physical phenomena which are often modeled by $R$-factors in small-$x$ phenomenology. We have studied the effects of longitudinal momentum transfer and the imaginary correction to the dipole amplitude at high energy. In the process, we derived expressions for the unpolarized quark and gluon GTMDs in the small-$x$ / shock wave formalism, obtaining results which agree with prior calculations in the literature when projected down to the GPD limit. The results of that calculation are given in Eqs.~\eqref{glueGTMD11}, \eqref{gGPD}, \eqref{q_GTMD_1} and \eqref{qGPD}. In the quark GPD case, we found a novel appearance of the spin-independent odderon exchange amplitude, confined to only a portion of the GPD phase space in $(x,\xi)$ --- the ERBL region. Both quark and gluon unpolarized GTMDs appear to depend on the odderon amplitude as well. 

The operator calculations at small $x$ yield the dipole amplitudes entering the expressions for quark and gluon GTMDs/GPDs without a specified rapidity with respect to the hadron. 
The first part of the above analysis concentrated on determining the value of rapidity $Y$ in the arguments of the dipole amplitudes $N_{10} (Y)$ and ${\cal O}_{10} (Y)$ entering the expressions for these distributions. We addressed the relation between the rapidity of the dipole amplitudes and the partonic longitudinal momentum fraction variables $x$ and $\xi$ by including skewness in the diagrams for LLA small-$x$ evolution. Restricting our analysis to the kinematics $|x| \ll |\xi| \ll 1$ or $|\xi| \ll |x| \ll 1$, we found that the result of including longitudinal momentum transfer in the shock wave formalism is to set the rapidity $Y$ in the argument of the dipole amplitudes resulting from the solution of the full non-linear evolution equations (BK or JIMWLK equations) at LLA equal to $Y = \ln \min \left\{ 1/|x| , 1/|\xi| \right\}$ (cf. \eq{Y}).

In the second half of the paper, we studied the contribution to the imaginary part of the dipole amplitude $N_{10} (Y)$, corresponding to the real part of the conventional elastic scattering amplitude $M$. By keeping the naively sub-eikonal terms in the logarithmic piece of the initial conditions, we modified the initial conditions for the small-$x$ evolution of the dipole to encode both the real and the leading imaginary parts. Explicitly, in \eq{N0_2} we give the modified initial condition for the GGM/MV model, allowing one to obtain the dominant imaginary contribution to the dipole amplitude by solving the non-linear evolution equation \eqref{BK_int} written in an integral form. The generalization including the odderon amplitude ${\cal O}_{10} (Y)$ into evolution is given by Eqs.~\eqref{NOeq_int} with the initial conditions given in Eqs.~\eqref{odd_init_3} and \eqref{N0_2}. 

We have thus constructed two specific prescriptions describing how both the $R$-factors can be systematically accounted for in small-$x$ calculations: our prescriptions are based on theoretical considerations and do not require modeling associated with the $R$-factors. Going forward, we hope that both prescriptions can be successfully implemented in phenomenology, including the analysis of the data to be generated at the Electron Ion Collider \cite{Accardi:2012qut,Boer:2011fh,Proceedings:2020eah,AbdulKhalek:2021gbh, Alexandrou:2026jpd}.


\section{Acknowledgments}

The authors are grateful to Markus Diehl, Zhongbo Kang, Genya Levin, Andreas Metz, Jani Penttala, Farid Salazar, and Alexey Vladimirov for informative discussions. 

This material is based upon work supported by
the U.S. Department of Energy, Office of Science, Office of Nuclear Physics under Award Number DE-SC0004286 and within the framework of the Saturated Glue (SURGE) Topical Theory Collaboration. The work is also supported by the Center for Nuclear Femtography, Southeastern Universities Research Association, Washington, D.C. and U.S. DOE Grant number DE-FG02-97ER41028, and also by the U.S. Department of Energy, Office of Science, Office of Nuclear Physics under Award Number ~DE-AC05-06OR23177 under which Jefferson Science Associates, LLC, manages and operates Jefferson Lab.


\appendix
\section{Simplified expression for the quark GPD $H^q$}
\label{sec:Hq}

To simplify \eq{qGPD} for the quark GPD $H^q$ at small $x$ and $\xi$, let us examine the $\un{k}$ and $\un{k}_1$ integrals. The last term in the curly brackets of \eq{q_GTMD_1} does not depend on $\un{k}$, so the $\un{k}$-integral, which can be extended to infinity at large $Q^2$, yields a factor of $\delta^{(2)}(\un{x}_{10})$ which sets the dipole amplitudes contribution equal to zero (both $N_{10}$ and ${\cal O}_{10}$ go to zero when $x_{10} \to 0$), making that term also zero. We are left with the contributions coming from the first and second terms in the curly brackets of \eq{q_GTMD_1}, where we need to consider integrals of the form
\begin{align}\label{int_type}
    &\int \dd[2]{k}_{\perp} \dd[2]{k}_{1 \perp} e^{-i (\un{k} +\un{k}_1) \vdot \un{x}_{10}} \left[ N_{10} (Y) - i \, {\cal O}_{10} (Y) \right] \left\{ \frac{ \un{k}_1 \vdot \left( \un{k} + \thalf \un \Delta \right) }{\left[ (x+\xi) \, z\, s + \un{k}_1^2 - i \epsilon \right] \left[ (x-\xi) \, z \, s + \left( \un{k} + \thalf \un \Delta \right)^2 - i \epsilon \right]} \right. \notag \\
   & \left.+ \frac{ \un{k}_1 \vdot \left( \un{k} - \thalf \un \Delta \right) }{\left[ (x+\xi) \, z\, s + \left( \un{k} - \thalf \un \Delta \right)^2 + i \epsilon \right] \left[ (x-\xi) \, z \, s + \un{k}_1^2 + i \epsilon \right]} \right\} . 
\end{align}
We can replace $\un{k} \rightarrow - \un{k}$, $\un{k}_1 \rightarrow - \un{k}_1$, and $\un{x}_1 \leftrightarrow \un{x}_0$ in the second term in curly brackets of \eq{int_type}, making use of $N_{10} = N_{01}$, and $\mathcal{O}_{10} = - \mathcal{O}_{01}$ to obtain
\begin{align}\label{GPD_proj}
    &\int \dd[2]{k}_{\perp} \dd[2]{k}_{1 \, \perp} e^{-i (\un{k} +\un{k}_1) \vdot \un{x}_{10}} \left\{ \frac{  \left[ N_{10} (Y) - i \, {\cal O}_{10} (Y) \right] \un{k}_1 \vdot \left( \un{k} + \thalf \un \Delta \right) }{\left[ (x+\xi) \, z\, s + \un{k}_1^2 - i \epsilon \right] \left[ (x-\xi) \, z \, s + \left( \un{k} + \thalf \un \Delta \right)^2 - i \epsilon \right]} \right.  \\
   & \left.+ \frac{  \left[ N_{10} (Y) + i \, {\cal O}_{10} (Y) \right] \un{k}_1 \vdot \left( \un{k} + \thalf \un \Delta \right) }{\left[ (x+\xi) \, z\, s + \left( \un{k} + \thalf \un \Delta \right)^2 + i \epsilon \right] \left[ (x-\xi) \, z \, s + \un{k}_1^2 + i \epsilon \right]} \right\} . \notag
\end{align}
Next, we perform the integrals over $\un{k}_1$ and $\un{k}$, obtaining
\begin{align}\label{qGPD_amp_ft}
   & - (2 \pi)^2 \, e^{i {\un \Delta} \cdot {\un x}_{10} /2} \, \bigg\{ N_{10} (Y) \Big[ \sqrt{(x + \xi) zs - i \epsilon} \sqrt{(x - \xi) zs - i \epsilon} \  \mbox{K}_1 ( x_{10} \, \sqrt{(x + \xi) zs - i \epsilon} ) \, \mbox{K}_1 ( x_{10} \, \sqrt{(x - \xi) zs - i \epsilon} ) \\
   &\hspace{2cm} + \sqrt{(x + \xi) zs + i \epsilon} \sqrt{(x - \xi) zs + i \epsilon} \ \mbox{K}_1 (x_{10} \, \sqrt{(x + \xi) zs + i \epsilon} ) \, \mbox{K}_1 ( x_{10} \, \sqrt{(x - \xi) zs + i \epsilon} ) \Big] \notag \\
   &- i \, \mathcal{O}_{10} (Y) \Big[ \sqrt{(x + \xi) zs - i \epsilon} \sqrt{(x - \xi) zs - i \epsilon} \ \mbox{K}_1 ( x_{10} \, \sqrt{(x + \xi) zs - i \epsilon} ) \, \mbox{K}_1 ( x_{10} \, \sqrt{(x - \xi) zs - i \epsilon} ) \notag \\
   &\hspace{2cm} - \sqrt{(x + \xi) zs + i \epsilon} \sqrt{(x - \xi) zs + i \epsilon} \ \mbox{K}_1 ( x_{10} \, \sqrt{(x + \xi) zs + i \epsilon} ) \, \mbox{K}_1 (x_{10} \, \sqrt{(x - \xi) zs + i \epsilon} ) \Big] \bigg\} . \notag
\end{align}
As $zs >0$, we can see that for $x>|\xi|$ the term proportional to $\mathcal{O}_{10} (Y)$ cancels as $\epsilon \rightarrow 0$. In the rest of the $(x,\xi)$ phase space we have no clear cancellation, and thus we have the curious result that there may be an Odderon contribution to the quark GPD in a limited part of its $(x,\xi)$ phase space.

We note that, as follows from the $z$-integral in \eq{q_GTMD_1}, the typical $z$-value is of the order-1, such that it does not significantly affect the value of rapidity $Y$ in the argument of the dipole amplitude. (As $\ln (zs) \approx \ln (s)$ for $z = {\cal O} (1)$, the rapidity $Y$, which usually depends on $\ln (zs)$, can be thought of as dependent only on $\ln (s)$ instead.) This allows us to neglect the dependence of $N_{10} (Y)$ and $\mathcal{O}_{10} (Y)$ on $z$, as a constant under the logarithm in the LLA approximation. In this simplified case where $N_{10} (Y)$ and $\mathcal{O}_{10} (Y)$ are independent of $z$, we can substitute \eq{qGPD_amp_ft} into Eqs.~\eqref{q_GTMD_1} and \eqref{qGPD} and allow the lower limit of $z$ integration to go to zero, obtaining
\begin{align}\label{qGPD_simp}
    H^q &(|x|\ll 1, |\xi| \ll 1, t, Q^2) = \frac{2 N_c \, s}{(2\pi)^4} \int \dd[2]{x}_{1 \, \perp} \dd[2]{x}_{0 \, \perp} e^{i \un{\Delta} \vdot  \un{x}_1} \\
   &\times \int\limits_{0}^1 \dd{z}  \, \Bigg{\{} N_{10} (Y) \Big[ \sqrt{(x + \xi) zs - i \epsilon} \sqrt{(x - \xi) zs - i \epsilon} \  \mbox{K}_1 ( x_{10} \, \sqrt{(x + \xi) zs - i \epsilon} ) \, \mbox{K}_1 ( x_{10} \, \sqrt{(x - \xi) zs - i \epsilon} ) \notag \\
   &\hspace{2cm} + \sqrt{(x + \xi) zs + i \epsilon} \sqrt{(x - \xi) zs + i \epsilon} \ \mbox{K}_1 (x_{10} \, \sqrt{(x + \xi) zs + i \epsilon} ) \, \mbox{K}_1 ( x_{10} \, \sqrt{(x - \xi) zs + i \epsilon} ) \Big] \notag \\
   &- i \, \mathcal{O}_{10} (Y) \Big[ \sqrt{(x + \xi) zs - i \epsilon} \sqrt{(x - \xi) zs - i \epsilon} \ \mbox{K}_1 ( x_{10} \, \sqrt{(x + \xi) zs - i \epsilon} ) \, \mbox{K}_1 ( x_{10} \, \sqrt{(x - \xi) zs - i \epsilon} ) \notag \\
   &\hspace{2cm} - \sqrt{(x + \xi) zs + i \epsilon} \sqrt{(x - \xi) zs + i \epsilon} \ \mbox{K}_1 ( x_{10} \, \sqrt{(x + \xi) zs + i \epsilon} ) \, \mbox{K}_1 (x_{10} \, \sqrt{(x - \xi) zs + i \epsilon} ) \Big] \Bigg{\}} . \notag
\end{align}

Following \cite{Bhattacharya:2025fnz}, we change the integration variable from $z$ to $z s$, and, assuming that for very large $s$ the upper limit of the resulting integral can be replaced by infinity, we write
\begin{align}\label{qGPD_simp2}
    H^q &(|x|\ll 1, |\xi| \ll 1, t, Q^2) =  \frac{2 N_c \, }{(2\pi)^4} \int \dd[2]{x}_{1 \, \perp} \dd[2]{x}_{0 \, \perp} e^{i \un{\Delta} \vdot  \un{x}_1} \\
   &\times \int\limits_{0}^\infty \dd{(zs)} zs  \, \Bigg{\{} N_{10} (Y) \Big[ \sqrt{x + \xi - i \epsilon} \sqrt{x - \xi - i \epsilon} \,  \mbox{K}_1 ( x_{10} \, \sqrt{zs} \sqrt{x + \xi - i \epsilon} ) \, \mbox{K}_1 (x_{10} \, \sqrt{zs} \sqrt{x - \xi - i \epsilon} ) \notag \\
   &\hspace{2cm} + \sqrt{x + \xi + i \epsilon} \sqrt{x - \xi + i \epsilon} \, \mbox{K}_1 ( x_{10} \, \sqrt{zs} \sqrt{x + \xi + i \epsilon} ) \, \mbox{K}_1 (x_{10} \, \sqrt{zs} \sqrt{x - \xi + i \epsilon} ) \Big] \notag \\
   &- i \mathcal{O}_{10} (Y) \Big[ \sqrt{x + \xi - i \epsilon} \sqrt{x - \xi - i \epsilon} \, \mbox{K}_1 (x_{10} \, \sqrt{zs} \sqrt{x + \xi - i \epsilon} ) \, \mbox{K}_1 (x_{10} \, \sqrt{zs} \sqrt{x - \xi - i \epsilon} ) \notag \\
   &\hspace{2cm} - \sqrt{x + \xi + i \epsilon} \sqrt{x - \xi + i \epsilon} \, \mbox{K}_1 ( x_{10} \, \sqrt{zs} \sqrt{x + \xi + i \epsilon} ) \, \mbox{K}_1 ( x_{10} \, \sqrt{zs} \sqrt{x - \xi + i \epsilon} ) \Big] \Bigg{\}}  \notag \\
   &=  \frac{N_c \, }{(2\pi)^4} \int \dd[2]{x}_{1 \, \perp} \dd[2]{x}_{0 \, \perp} \frac{e^{i \un{\Delta} \vdot  \un{x}_1}}{x_{10}^4} \, \frac{1}{\xi^3} \notag \\
   &\times \, \Bigg{\{} N_{10} (Y) \Bigg[ (x + \xi - i \epsilon)^2 - (x - \xi - i \epsilon)^2 + 2 ( x + \xi - i \epsilon) ( x - \xi - i \epsilon) \ln \left( \frac{x - \xi - i \epsilon}{x + \xi - i \epsilon} \right) \notag \\
   &\hspace{2cm} + (x + \xi + i \epsilon)^2 - (x - \xi + i \epsilon)^2 + 2 ( x + \xi + i \epsilon) ( x - \xi + i \epsilon) \ln \left( \frac{x - \xi + i \epsilon}{x + \xi + i \epsilon} \right)  \Bigg] \notag \\
   &- i \mathcal{O}_{10} (Y) \Bigg[ (x + \xi - i \epsilon)^2 - (x - \xi - i \epsilon)^2 + 2 ( x + \xi - i \epsilon) ( x - \xi - i \epsilon) \ln \left( \frac{x - \xi - i \epsilon}{x + \xi - i \epsilon} \right) \notag \\
   &\hspace{2cm} -  \left[ (x + \xi + i \epsilon)^2 - (x - \xi + i \epsilon)^2 + 2 ( x + \xi + i \epsilon) ( x - \xi + i \epsilon) \ln \left( \frac{x - \xi + i \epsilon}{x + \xi + i \epsilon} \right) \right] \Bigg] \Bigg{\}} . \notag
\end{align}
Outside of the logarithms we can safely take $\epsilon \rightarrow 0$ regardless of the values of $x$ and $\xi$. This leads to \eq{qGPD_simp3} in the main text.

\section{Imaginary corrections to the Born-level odderon amplitude}
\label{sec:look_odd}

The initial condition for the dipole amplitude and odderon amplitude (before including multiple rescatterings) is given by the lowest-order two-gluon and three-gluon exchange diagrams. Considering the scattering amplitude of a quark-quark scattering with $t$-channel gluon exchanges, we see that a convenient way to obtain these amplitudes is by using dispersion relations. For instance, instead of calculating directly the two-gluon exchange diagrams, we would only need to calculate the imaginary part of these diagrams given by the cut diagram. The cut diagram would only consist of two one-gluon exchange diagrams (see \fig{2 gluon cut} below). In this way we can build up the amplitudes iteratively.

\begin{figure}[h]
    \centering
    \includegraphics[width=0.24\linewidth]{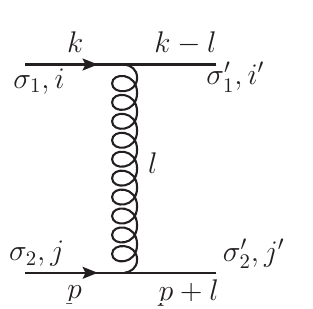}
    \caption{One-gluon exchange diagrams for a quark scattering on a quark at high energy.}
    \label{1 gluon}
\end{figure}

The high-energy amplitude for quark-quark scattering via one t-channel gluon exchange (shown in \fig{1 gluon}) is, in the eikonal approximation,
\begin{equation}\label{1gluon}
    M_{1g}(s,t) = -8\pi\as\frac{s}{l_{\perp}^{2}}(t^{a})_{i'i}(t^{a})_{j'j},
\end{equation}
where $s=2p^{+}k^{-}$ is the center-of-mass energy. The imaginary part of the quark-quark scattering amplitude via two-gluon exchange is given by the cut diagram shown in \fig{2 gluon cut}, which can be obtained using Cutkosky rules and the one-gluon amplitude in \eq{1gluon}:
\begin{figure}[h]
    \centering
    \includegraphics[width=0.3\linewidth]{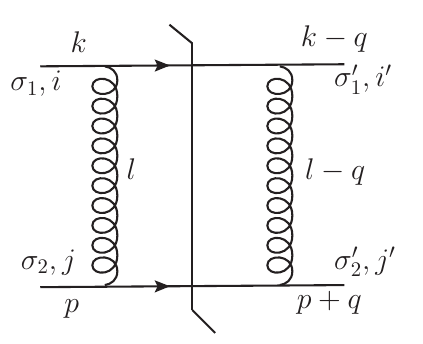}
    \caption{Imaginary part of a two-gluon exchange scattering amplitude for a quark scattering on another quark.}
    \label{2 gluon cut}
\end{figure}
\begin{equation}
    2\Im M_{2g}(s, t=-\un{q}^{2})=\int\frac{d^{4}l}{(2\pi)^{4}} \, M_{1g}(s,-\un{l}^{2}) M_{1g}^{*}(s,-(\un{l}-\un{q})^{2}) \, 2\pi\delta((k-l)^{2}) \, 2 \pi \delta((p+l)^{2}).
\end{equation}
This gives
\begin{equation}\label{2 gluon imaginary}
    \Im M_{2g} (s, t=-\un{q}^{2}) =4\as^{2}\,s \, (t^{a}t^{b})_{i'i}(t^{a}t^{b})_{j'j}\int  \frac{d^{2} l_\perp}{\un{l}^{2}(\un{l}-\un{q})^{2}}.
\end{equation}
Since $\Im{M_{2g}}\propto s$, we shall use the double-subtracted dispersion relation to obtain (the leading part of) $M_{2g}$ from its imaginary part \cite{Kovchegov:2012mbw}:
\begin{align}\label{2 gluon}
    M_{2g}(s,t=-\un{q}^{2})&= M_{2g}(0, t= - \un q^2) + s \, \pd_s M_{2g} (0, t= - \un q^2)\nonumber\\
    &\hspace{1cm} +\frac{1}{\pi}\left\lbrace s^{2}\int_{\Lambda^{2}}^{\infty}ds^{\prime}\frac{\Im M_{2g}(s',t)}{(s^{\prime}-i\epsilon)^{2}(s^{\prime}-s-i\epsilon)}+u^{2}\int_{\Lambda^{2}}^{\infty}du'\frac{\Im M_{2g}(u',t)}{(u'-i\epsilon)^{2}(u'-u-i\epsilon)}\right\rbrace\nonumber\\
    &=-\frac{4\as^{2}}{\pi} \, s \, (t^{a}t^{b})_{i'i}(t^{a}t^{b})_{j'j}\int \frac{d^{2} l_\perp}{\un{l}^{2}(\un{l}-\un{q})^{2}}\left[ \ln \left( \frac{-s+\Lambda^{2}-i\epsilon}{\Lambda^{2}-i\epsilon} \right) - \ln \left( \frac{s+\Lambda^{2}-i\epsilon}{\Lambda^{2}-i\epsilon} \right) \right].
\end{align}
In the last step, we have neglected $M_{2g}(0, t= - \un q^2) + s \, \pd_s M_{2g} (0, t= - \un q^2)$ and assumed that the color factors in the two terms are the same, in anticipation of calculating the odderon contribution.
The structure of \eq{2 gluon} agrees with that in \eq{step3}. We see that the two-gluon exchange scattering amplitude aside from the color factor is (see Eq.~(3.49) in \cite{Kovchegov:2012mbw})
\begin{align}
    M_{2g} (s, t= - \un l^2) = - \frac{4 \, \as^2 \, s}{\pi} \, \int \frac{d^2 k_\perp}{\un k^2 (\un l - \un k)^2} \, \left[ \ln \left( \frac{- s + \Lambda^2 - i \epsilon}{\Lambda^2 - i \epsilon} \right) - \ln \left( \frac{s + \Lambda^2 -i \epsilon}{\Lambda^2 - i \epsilon} \right) \right].
\end{align}

\begin{figure}[h]
    \centering
    \includegraphics[width=0.5\linewidth]{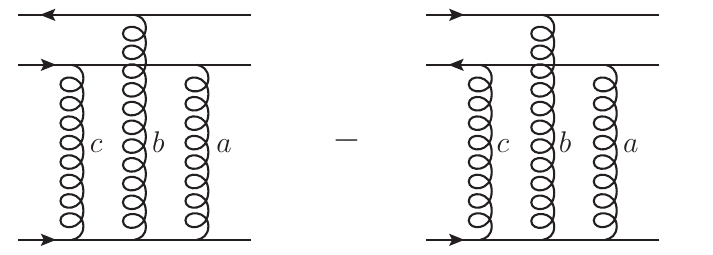}
    \caption{Diagrammatic representation of the odderon exchange at the lowest 3-gluon order.} 
    \label{3 gluon color factor}
\end{figure}

The odderon exchange at the lowest 3-gluon order is illustrated in \fig{3 gluon color factor}. Since odderon is defined to be the antisymmtric part of the dipole S-matrix, the color factor of every possible configuration of gluon exchanges takes the form $\tr[t^{a}t^{b}t^{c}]\tr[t^{a}\{t^{b},t^{c}\}]=\frac{1}{8}(d^{abc})^{2}$. Therefore, in calculating the odderon amplitude we can drop the color factors for simplicity as they are all the same.

\begin{figure}[h]
    \centering
    \includegraphics[width=0.5\linewidth]{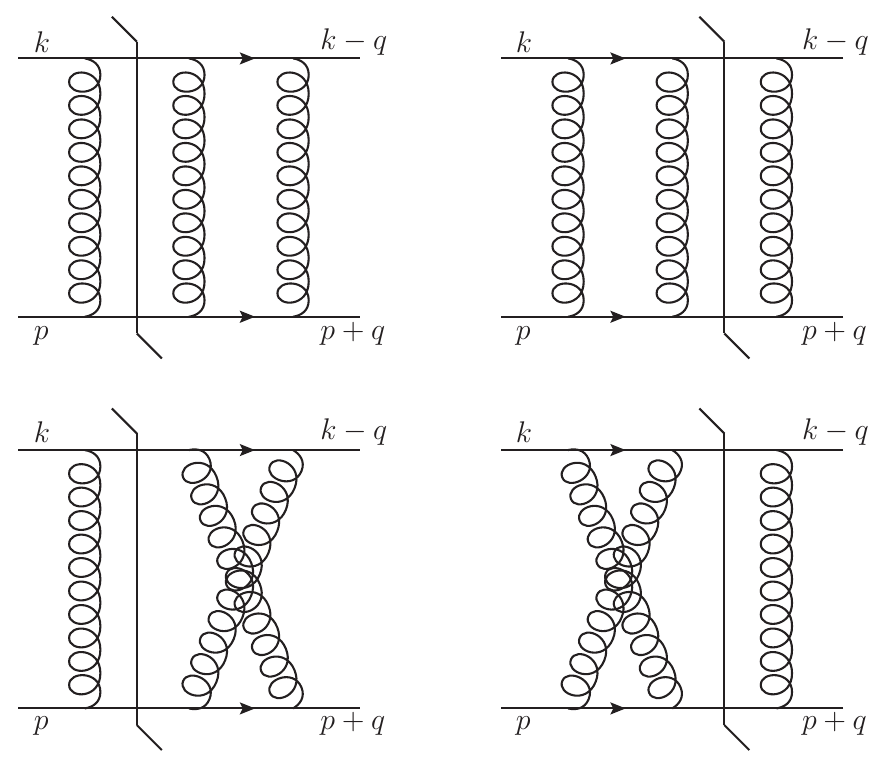}
    \caption{Cut diagrams of the odderon exchange at the lowest 3-gluon order.}
    \label{3 gluon cuts}
\end{figure}

Similar to how we obtain the 2-gluon exchange amplitude, our method is to reconstruct the odderon amplitude from its imaginary part. The imaginary part is given by all the possible cuts between the $t$-channel gluon exchanges, as shown in \fig{3 gluon cuts} (see more on this below). Gluing the 1-gluon and 2-gluon exchange amplitudes together we get the Im part of the 3-gluon exchange contribution (with the color factor removed)

\begin{align}\label{IM3}
    \textrm{Im} \, M_{3g} (s, t= - \un q^2) & \, = \frac{2 \, \as^3 \, s}{\pi^2} \, \int \frac{d^2 k_\perp \, d^2 l_\perp}{\un k^2 (\un l - \un k)^2 (\un l - \un q)^2 } \, \textrm{Sign} (s) \, \left[ \ln \left( \frac{- s + \Lambda^2 - i \epsilon}{\Lambda^2 - i \epsilon} \right) - \ln \left( \frac{s + \Lambda^2 - i \epsilon}{\Lambda^2 - i \epsilon} \right) + \cc \right] \, \theta (|s| - \Lambda^2) \notag \\
    & = \frac{4 \, \as^3 \, s}{\pi^2} \, \int \frac{d^2 k_\perp \, d^2 l_\perp}{\un k^2 (\un l - \un k)^2 (\un l - \un q)^2 } \, \textrm{Sign} (s) \,  \ln \left( \frac{|s - \Lambda^2|}{|s+\Lambda^2 | } \right) \, \theta (|s| - \Lambda^2) . 
\end{align}
The $u$-channel contribution is obtained by reversing the order of the vertices attaching the gluons to the projectile and then by applying crossing symmetry to change the quark into an anti-quark in the projectile in \fig{3 gluon cuts}. This procedure would then require us to change the sign of all three quark-gluon vertices while keeping the $d^{abc}$ color factor the same. Hence, Sign$(s)$ is added to account for the additional sign change when we do the $s \to u \approx -s$ replacement to obtain the $u$-channel contribution to (the imaginary part of the) odderon amplitude. The theta-function describes the threshold above which the scattering is possible (in both the $s$- and $u$-channels), which is taken to be the same $\Lambda^2$ for simplicity.

Squaring \eq{sig_factor} and putting $z=1$ for simplicity we get
\begin{align}\label{sig_factor2}
- \frac{1}{\pi^2} \, \left[ \ln \left( \frac{\Lambda^2 - i \epsilon }{- s + \Lambda^2 - i \epsilon } \right) - \ln \left( \frac{\Lambda^2 - i \epsilon }{s + \Lambda^2 - i \epsilon } \right) \right]^2 = & \ \theta \left( |s| - \Lambda^2 \right) + 2 \, i \, \theta \left( |s| - \Lambda^2 \right) \, \frac{\mbox{Sign} (s)}{\pi}  \, \ln \left( \frac{|s - \Lambda^2|}{|s + \Lambda^2|} \right) \notag \\
& - \frac{1}{\pi^2} \, \ln^2 \left( \frac{|s - \Lambda^2|}{|s + \Lambda^2|} \right) .
\end{align}
From \eq{sig_factor2} we see that the imaginary part in \eq{IM3} may arise from the following full expression for $M_{3g}$, 
\begin{align}\label{odd_sig}
    M_{3g} (s, t= - \un q^2) = \frac{4 \, \as^3 \, s}{\pi^2} \, \int \frac{d^2 k_\perp \, d^2 l_\perp}{\un k^2 (\un l - \un k)^2 (\un l - \un q)^2 } \, \left( - \frac{1}{2 \pi} \right) \, \left[ \ln \left( \frac{\Lambda^2 - i \epsilon }{- s + \Lambda^2 - i \epsilon } \right) - \ln \left( \frac{\Lambda^2 - i \epsilon }{s + \Lambda^2 - i \epsilon } \right) \right]^2 .
\end{align}
It appears impossible to reconstruct the full Re~$M_{3g}$ using dispersion relations, as they generate only the sub-eikonal contribution to Re~$M_{3g}$. Employing the function \eqref{sig_factor2} in the derivation of the single-subtracted dispersion relation, one can see that the semi-circle at infinity would contain the eikonal contribution to Re~$M_{3g}$ and cannot be discarded, making it impossible to construct Re~$M_{3g}$ from Im~$M_{3g}$ only. For the double-subtracted dispersion relation, like that used in \eqref{2 gluon}, the integral again gives the sub-eikonal contribution to Re~$M_{3g}$ only, with the needed eikonal contribution likely residing in the $s \, \pd_s M_{3g} (0,t)$   term. Hence, while we could not reconstruct the full Re~$M_{3g}$ using dispersion relations, we have found the analytic function \eqref{sig_factor2} whose imaginary part is the same (up to a multiplicative factor) as that in \eq{IM3} and which contains Re~$M_{3g}$ as well. The latter happens to be eikonal, as expected for the odderon. Therefore, equation~\eqref{odd_sig} is our well-motivated conjecture for the full odderon amplitude at the 3-gluon exchange level: it contains the correction we have included in \eq{odd_init_3} in the main text. 

\begin{figure}[h]
    \centering
    \includegraphics[width=0.5\linewidth]{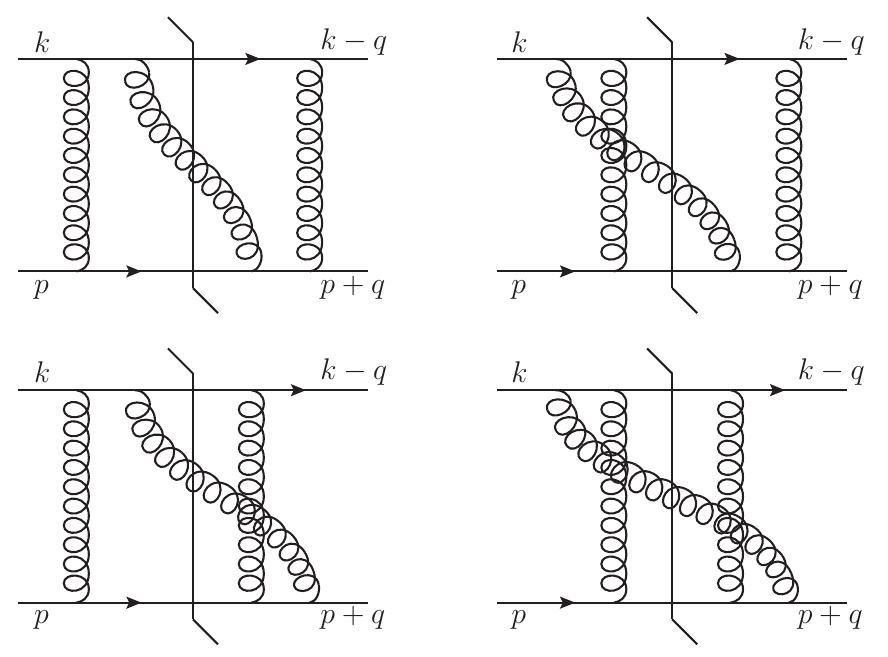}
    \caption{Diagrams with the cuts that go through one of the gluons. These diagrams give an overall sub-eikonal contribution due to the gluons being in the $d^{abc}$ color state.}
    \label{cutting through gluon}
\end{figure}

We also notice that some of the 3-gluon exchange diagrams may contain a cut going through a gluon line. For instance, for the diagrams in \fig{3 gluon color factor}, when all three gluons connect to either the quark or the anti-quark line in the dipole, it is possible to place a cut going through the gluon line with the color index $b$ in the figure. This is depicted in \fig{cutting through gluon}. A straightforward calculation of these diagrams, which we will not present here, leads to the following contribution to the imaginary part of the odderon amplitude coming from such a cut,
\begin{align}\label{IM30}
    \textrm{Im} \, M_{3g} (s, t= - \un q^2) \supset \theta \left( |s| - \Lambda^2 \right) \, \mbox{Sign} (s) \, \ln \left( \frac{|s| - \Lambda^2}{\Lambda^2} \right) .
\end{align}
This contribution is missing an overall factor of $s$ and is, therefore, sub-eikonal. (While our calculation gave us terms with either $\ln ((|s| - \Lambda^2)/\Lambda^2)$ or $\ln ((|s| + \Lambda^2)/\Lambda^2)$, we have replaced the latter logarithm with the former in \eq{IM30} within the present sub-eikonal accuracy.) The amplitude which leads to the imaginary part in \eq{IM30} can be easily identified as proportional to 
\begin{align}
    M_{3g} (s, t= - \un q^2) \supset \ln^2 \left( \frac{\Lambda^2 - i \epsilon }{- s + \Lambda^2 - i \epsilon } \right) - \ln^2 \left( \frac{\Lambda^2 - i \epsilon }{s + \Lambda^2 - i \epsilon } \right) .
\end{align}
This contribution is also sub-eikonal; it can be neglected at high energies compared to that in \eq{odd_sig} above, as a sub-eikonal correction to the (real part of the) odderon scattering amplitude.


\providecommand{\href}[2]{#2}\begingroup\raggedright\endgroup



\begin{thebibliography}{100}

\bibitem{Muller:1994ses}
D.~M{\"u}ller, D.~Robaschik, B.~Geyer, F.~M. Dittes and J.~Ho{\v{r}}ej{\v{s}}i, \emph{{Wave functions, evolution equations and evolution kernels from light ray operators of QCD}}, \href{https://doi.org/10.1002/prop.2190420202}{\emph{Fortsch. Phys.} {\bfseries 42} (1994) 101--141}, [\href{https://arxiv.org/abs/hep-ph/9812448}{{\ttfamily hep-ph/9812448}}].

\bibitem{Ji:1996ek}
X.-D. Ji, \emph{{Gauge-Invariant Decomposition of Nucleon Spin}}, \href{https://doi.org/10.1103/PhysRevLett.78.610}{\emph{Phys. Rev. Lett.} {\bfseries 78} (1997) 610--613}, [\href{https://arxiv.org/abs/hep-ph/9603249}{{\ttfamily hep-ph/9603249}}].

\bibitem{Radyushkin:1996nd}
A.~V. Radyushkin, \emph{{Scaling limit of deeply virtual Compton scattering}}, \href{https://doi.org/10.1016/0370-2693(96)00528-X}{\emph{Phys. Lett. B} {\bfseries 380} (1996) 417--425}, [\href{https://arxiv.org/abs/hep-ph/9604317}{{\ttfamily hep-ph/9604317}}].

\bibitem{Ji:1996nm}
X.-D. Ji, \emph{{Deeply virtual Compton scattering}}, \href{https://doi.org/10.1103/PhysRevD.55.7114}{\emph{Phys. Rev. D} {\bfseries 55} (1997) 7114--7125}, [\href{https://arxiv.org/abs/hep-ph/9609381}{{\ttfamily hep-ph/9609381}}].

\bibitem{Ji:1998pc}
X.-D. Ji, \emph{{Off forward parton distributions}}, \href{https://doi.org/10.1088/0954-3899/24/7/002}{\emph{J. Phys. G} {\bfseries 24} (1998) 1181--1205}, [\href{https://arxiv.org/abs/hep-ph/9807358}{{\ttfamily hep-ph/9807358}}].

\bibitem{Berger:2001xd}
E.~R. Berger, M.~Diehl and B.~Pire, \emph{{Time - like Compton scattering: Exclusive photoproduction of lepton pairs}}, \href{https://doi.org/10.1007/s100520200917}{\emph{Eur. Phys. J. C} {\bfseries 23} (2002) 675--689}, [\href{https://arxiv.org/abs/hep-ph/0110062}{{\ttfamily hep-ph/0110062}}].

\bibitem{Polyakov:2002wz}
M.~V. Polyakov and A.~G. Shuvaev, \emph{{On'dual' parametrizations of generalized parton distributions}},  \href{https://arxiv.org/abs/hep-ph/0207153}{{\ttfamily hep-ph/0207153}}.

\bibitem{Ivanov:2002jj}
D.~Y. Ivanov, B.~Pire, L.~Szymanowski and O.~V. Teryaev, \emph{{Probing chiral odd GPD's in diffractive electroproduction of two vector mesons}}, \href{https://doi.org/10.1016/S0370-2693(02)02856-3}{\emph{Phys. Lett. B} {\bfseries 550} (2002) 65--76}, [\href{https://arxiv.org/abs/hep-ph/0209300}{{\ttfamily hep-ph/0209300}}].

\bibitem{Diehl:2003ny}
M.~Diehl, \emph{{Generalized parton distributions}}, \href{https://doi.org/10.1016/j.physrep.2003.08.002}{\emph{Phys. Rept.} {\bfseries 388} (2003) 41--277}, [\href{https://arxiv.org/abs/hep-ph/0307382}{{\ttfamily hep-ph/0307382}}].

\bibitem{Guidal:2004nd}
M.~Guidal, M.~V. Polyakov, A.~V. Radyushkin and M.~Vanderhaeghen, \emph{{Nucleon form-factors from generalized parton distributions}}, \href{https://doi.org/10.1103/PhysRevD.72.054013}{\emph{Phys. Rev. D} {\bfseries 72} (2005) 054013}, [\href{https://arxiv.org/abs/hep-ph/0410251}{{\ttfamily hep-ph/0410251}}].

\bibitem{Belitsky:2005qn}
A.~V. Belitsky and A.~V. Radyushkin, \emph{{Unraveling hadron structure with generalized parton distributions}}, \href{https://doi.org/10.1016/j.physrep.2005.06.002}{\emph{Phys. Rept.} {\bfseries 418} (2005) 1--387}, [\href{https://arxiv.org/abs/hep-ph/0504030}{{\ttfamily hep-ph/0504030}}].

\bibitem{Goloskokov:2005sd}
S.~V. Goloskokov and P.~Kroll, \emph{{Vector meson electroproduction at small Bjorken-x and generalized parton distributions}}, \href{https://doi.org/10.1140/epjc/s2005-02298-5}{\emph{Eur. Phys. J. C} {\bfseries 42} (2005) 281--301}, [\href{https://arxiv.org/abs/hep-ph/0501242}{{\ttfamily hep-ph/0501242}}].

\bibitem{Mueller:2005ed}
D.~Mueller and A.~Schafer, \emph{{Complex conformal spin partial wave expansion of generalized parton distributions and distribution amplitudes}}, \href{https://doi.org/10.1016/j.nuclphysb.2006.01.019}{\emph{Nucl. Phys. B} {\bfseries 739} (2006) 1--59}, [\href{https://arxiv.org/abs/hep-ph/0509204}{{\ttfamily hep-ph/0509204}}].

\bibitem{Enberg:2006he}
R.~Enberg, B.~Pire and L.~Szymanowski, \emph{{Transversity GPD in photo- and electroproduction of two vector mesons}}, \href{https://doi.org/10.1140/epjc/s2006-02545-3}{\emph{Eur. Phys. J. C} {\bfseries 47} (2006) 87--94}, [\href{https://arxiv.org/abs/hep-ph/0601138}{{\ttfamily hep-ph/0601138}}].

\bibitem{Kumericki:2007sa}
K.~Kumericki, D.~Mueller and K.~Passek-Kumericki, \emph{{Towards a fitting procedure for deeply virtual Compton scattering at next-to-leading order and beyond}}, \href{https://doi.org/10.1016/j.nuclphysb.2007.10.029}{\emph{Nucl. Phys. B} {\bfseries 794} (2008) 244--323}, [\href{https://arxiv.org/abs/hep-ph/0703179}{{\ttfamily hep-ph/0703179}}].

\bibitem{Kumericki:2009uq}
K.~Kumeri{\v{c}}ki and D.~Mueller, \emph{{Deeply virtual Compton scattering at small $x_B$ and the access to the GPD H}}, \href{https://doi.org/10.1016/j.nuclphysb.2010.07.015}{\emph{Nucl. Phys. B} {\bfseries 841} (2010) 1--58}, [\href{https://arxiv.org/abs/0904.0458}{{\ttfamily 0904.0458}}].

\bibitem{Goldstein:2010gu}
G.~R. Goldstein, J.~O. Hernandez and S.~Liuti, \emph{{Flexible Parametrization of Generalized Parton Distributions from Deeply Virtual Compton Scattering Observables}}, \href{https://doi.org/10.1103/PhysRevD.84.034007}{\emph{Phys. Rev. D} {\bfseries 84} (2011) 034007}, [\href{https://arxiv.org/abs/1012.3776}{{\ttfamily 1012.3776}}].

\bibitem{ElBeiyad:2010pji}
M.~El~Beiyad, B.~Pire, M.~Segond, L.~Szymanowski and S.~Wallon, \emph{{Photoproduction of a pi rhoT pair with a large invariant mass and transversity generalized parton distribution}}, \href{https://doi.org/10.1016/j.physletb.2010.02.086}{\emph{Phys. Lett. B} {\bfseries 688} (2010) 154--167}, [\href{https://arxiv.org/abs/1001.4491}{{\ttfamily 1001.4491}}].

\bibitem{Gonzalez-Hernandez:2012xap}
J.~O. Gonzalez-Hernandez, S.~Liuti, G.~R. Goldstein and K.~Kathuria, \emph{{Interpretation of the Flavor Dependence of Nucleon Form Factors in a Generalized Parton Distribution Model}}, \href{https://doi.org/10.1103/PhysRevC.88.065206}{\emph{Phys. Rev. C} {\bfseries 88} (2013) 065206}, [\href{https://arxiv.org/abs/1206.1876}{{\ttfamily 1206.1876}}].

\bibitem{Kumericki:2016ehc}
K.~Kumericki, S.~Liuti and H.~Moutarde, \emph{{GPD phenomenology and DVCS fitting}: {Entering the high-precision era}}, \href{https://doi.org/10.1140/epja/i2016-16157-3}{\emph{Eur. Phys. J. A} {\bfseries 52} (2016) 157}, [\href{https://arxiv.org/abs/1602.02763}{{\ttfamily 1602.02763}}].

\bibitem{Dupre:2016mai}
R.~Dupre, M.~Guidal and M.~Vanderhaeghen, \emph{{Tomographic image of the proton}}, \href{https://doi.org/10.1103/PhysRevD.95.011501}{\emph{Phys. Rev. D} {\bfseries 95} (2017) 011501}, [\href{https://arxiv.org/abs/1606.07821}{{\ttfamily 1606.07821}}].

\bibitem{Berthou:2015oaw}
B.~Berthou et~al., \emph{{PARTONS: PARtonic Tomography Of Nucleon Software}: {A computing framework for the phenomenology of Generalized Parton Distributions}}, \href{https://doi.org/10.1140/epjc/s10052-018-5948-0}{\emph{Eur. Phys. J. C} {\bfseries 78} (2018) 478}, [\href{https://arxiv.org/abs/1512.06174}{{\ttfamily 1512.06174}}].

\bibitem{Boussarie:2016qop}
R.~Boussarie, B.~Pire, L.~Szymanowski and S.~Wallon, \emph{{Exclusive photoproduction of a $\gamma\,\rho$ pair with a large invariant mass}}, \href{https://doi.org/10.1007/JHEP02(2017)054}{\emph{JHEP} {\bfseries 02} (2017) 054}, [\href{https://arxiv.org/abs/1609.03830}{{\ttfamily 1609.03830}}].

\bibitem{Pedrak:2017cpp}
A.~Pedrak, B.~Pire, L.~Szymanowski and J.~Wagner, \emph{{Hard photoproduction of a diphoton with a large invariant mass}}, \href{https://doi.org/10.1103/PhysRevD.96.074008}{\emph{Phys. Rev. D} {\bfseries 96} (2017) 074008}, [\href{https://arxiv.org/abs/1708.01043}{{\ttfamily 1708.01043}}].

\bibitem{Duplancic:2018bum}
G.~Duplan{\v{c}}i{\'c}, K.~Passek-Kumeri{\v{c}}ki, B.~Pire, L.~Szymanowski and S.~Wallon, \emph{{Probing axial quark generalized parton distributions through exclusive photoproduction of a $\gamma\,\pi^\pm$ pair with a large invariant mass}}, \href{https://doi.org/10.1007/JHEP11(2018)179}{\emph{JHEP} {\bfseries 11} (2018) 179}, [\href{https://arxiv.org/abs/1809.08104}{{\ttfamily 1809.08104}}].

\bibitem{Moutarde:2019tqa}
H.~Moutarde, P.~Sznajder and J.~Wagner, \emph{{Unbiased determination of DVCS Compton Form Factors}}, \href{https://doi.org/10.1140/epjc/s10052-019-7117-5}{\emph{Eur. Phys. J. C} {\bfseries 79} (2019) 614}, [\href{https://arxiv.org/abs/1905.02089}{{\ttfamily 1905.02089}}].

\bibitem{Lin:2020rxa}
H.-W. Lin, \emph{{Nucleon Tomography and Generalized Parton Distribution at Physical Pion Mass from Lattice QCD}}, \href{https://doi.org/10.1103/PhysRevLett.127.182001}{\emph{Phys. Rev. Lett.} {\bfseries 127} (2021) 182001}, [\href{https://arxiv.org/abs/2008.12474}{{\ttfamily 2008.12474}}].

\bibitem{Pedrak:2020mfm}
A.~Pedrak, B.~Pire, L.~Szymanowski and J.~Wagner, \emph{{Electroproduction of a large invariant mass photon pair}}, \href{https://doi.org/10.1103/PhysRevD.101.114027}{\emph{Phys. Rev. D} {\bfseries 101} (2020) 114027}, [\href{https://arxiv.org/abs/2003.03263}{{\ttfamily 2003.03263}}].

\bibitem{Hashamipour:2021kes}
H.~Hashamipour, M.~Goharipour, K.~Azizi and S.~V. Goloskokov, \emph{{Determination of the generalized parton distributions through the analysis of the world electron scattering data considering two-photon exchange corrections}}, \href{https://doi.org/10.1103/PhysRevD.105.054002}{\emph{Phys. Rev. D} {\bfseries 105} (2022) 054002}, [\href{https://arxiv.org/abs/2111.02030}{{\ttfamily 2111.02030}}].

\bibitem{Dutrieux:2021wll}
H.~Dutrieux, H.~Dutrieux, O.~Grocholski, O.~Grocholski, H.~Moutarde, H.~Moutarde et~al., \emph{{Artificial neural network modelling of generalised parton distributions}}, \href{https://doi.org/10.1140/epjc/s10052-022-10211-5}{\emph{Eur. Phys. J. C} {\bfseries 82} (2022) 252}, [\href{https://arxiv.org/abs/2112.10528}{{\ttfamily 2112.10528}}].

\bibitem{Grocholski:2021man}
O.~Grocholski, B.~Pire, P.~Sznajder, L.~Szymanowski and J.~Wagner, \emph{{Collinear factorization of diphoton photoproduction at next to leading order}}, \href{https://doi.org/10.1103/PhysRevD.104.114006}{\emph{Phys. Rev. D} {\bfseries 104} (2021) 114006}, [\href{https://arxiv.org/abs/2110.00048}{{\ttfamily 2110.00048}}].

\bibitem{Grocholski:2022rqj}
O.~Grocholski, B.~Pire, P.~Sznajder, L.~Szymanowski and J.~Wagner, \emph{{Phenomenology of diphoton photoproduction at next-to-leading order}}, \href{https://doi.org/10.1103/PhysRevD.105.094025}{\emph{Phys. Rev. D} {\bfseries 105} (2022) 094025}, [\href{https://arxiv.org/abs/2204.00396}{{\ttfamily 2204.00396}}].

\bibitem{Qiu:2022bpq}
J.-W. Qiu and Z.~Yu, \emph{{Exclusive production of a pair of high transverse momentum photons in pion-nucleon collisions for extracting generalized parton distributions}}, \href{https://doi.org/10.1007/JHEP08(2022)103}{\emph{JHEP} {\bfseries 08} (2022) 103}, [\href{https://arxiv.org/abs/2205.07846}{{\ttfamily 2205.07846}}].

\bibitem{Guo:2022upw}
Y.~Guo, X.~Ji and K.~Shiells, \emph{{Generalized parton distributions through universal moment parameterization: zero skewness case}}, \href{https://doi.org/10.1007/JHEP09(2022)215}{\emph{JHEP} {\bfseries 09} (2022) 215}, [\href{https://arxiv.org/abs/2207.05768}{{\ttfamily 2207.05768}}].

\bibitem{Duplancic:2022ffo}
G.~Duplan{\v{c}}i{\'c}, S.~Nabeebaccus, K.~Passek-Kumeri{\v{c}}ki, B.~Pire, L.~Szymanowski and S.~Wallon, \emph{{Accessing chiral-even quark generalised parton distributions in the exclusive photoproduction of a $ \gamma \pi ^{\pm} $ pair with large invariant mass in both fixed-target and collider experiments}}, \href{https://doi.org/10.1007/JHEP03(2023)241}{\emph{JHEP} {\bfseries 03} (2023) 241}, [\href{https://arxiv.org/abs/2212.00655}{{\ttfamily 2212.00655}}].

\bibitem{Qiu:2023mrm}
J.-W. Qiu and Z.~Yu, \emph{{Extraction of the Parton Momentum-Fraction Dependence of Generalized Parton Distributions from Exclusive Photoproduction}}, \href{https://doi.org/10.1103/PhysRevLett.131.161902}{\emph{Phys. Rev. Lett.} {\bfseries 131} (2023) 161902}, [\href{https://arxiv.org/abs/2305.15397}{{\ttfamily 2305.15397}}].

\bibitem{Deja:2023ahc}
K.~Deja, V.~Martinez-Fernandez, B.~Pire, P.~Sznajder and J.~Wagner, \emph{{Phenomenology of double deeply virtual Compton scattering in the era of new experiments}}, \href{https://doi.org/10.1103/PhysRevD.107.094035}{\emph{Phys. Rev. D} {\bfseries 107} (2023) 094035}, [\href{https://arxiv.org/abs/2303.13668}{{\ttfamily 2303.13668}}].

\bibitem{Duplancic:2023kwe}
G.~Duplan{\v{c}}i{\'c}, S.~Nabeebaccus, K.~Passek-Kumeri{\v{c}}ki, B.~Pire, L.~Szymanowski and S.~Wallon, \emph{{Probing chiral-even and chiral-odd leading twist quark generalized parton distributions through the exclusive photoproduction of a {\ensuremath{\gamma}}{\ensuremath{\rho}} pair}}, \href{https://doi.org/10.1103/PhysRevD.107.094023}{\emph{Phys. Rev. D} {\bfseries 107} (2023) 094023}, [\href{https://arxiv.org/abs/2302.12026}{{\ttfamily 2302.12026}}].

\bibitem{Guo:2023ahv}
Y.~Guo, X.~Ji, M.~G. Santiago, K.~Shiells and J.~Yang, \emph{{Generalized parton distributions through universal moment parameterization: non-zero skewness case}}, \href{https://doi.org/10.1007/JHEP05(2023)150}{\emph{JHEP} {\bfseries 05} (2023) 150}, [\href{https://arxiv.org/abs/2302.07279}{{\ttfamily 2302.07279}}].

\bibitem{Qiu:2024mny}
J.-W. Qiu and Z.~Yu, \emph{{Extracting transition generalized parton distributions from hard exclusive pion-nucleon scattering}}, \href{https://doi.org/10.1103/PhysRevD.109.074023}{\emph{Phys. Rev. D} {\bfseries 109} (2024) 074023}, [\href{https://arxiv.org/abs/2401.13207}{{\ttfamily 2401.13207}}].

\bibitem{Goharipour:2024atx}
{\scshape MMGPDs} collaboration, M.~Goharipour, H.~Hashamipour, F.~Irani and K.~Azizi, \emph{{Impact of JLab data on the determination of GPDs at zero skewness and new insights from transition form factors $N\rightarrow \Delta$}}, \href{https://doi.org/10.1103/PhysRevD.109.074042}{\emph{Phys. Rev. D} {\bfseries 109} (2024) 074042}, [\href{https://arxiv.org/abs/2403.19384}{{\ttfamily 2403.19384}}].

\bibitem{Siddikov:2024blb}
M.~Siddikov, \emph{{Exclusive photoproduction of {\ensuremath{\eta}}c{\ensuremath{\gamma}} pairs with large invariant mass}}, \href{https://doi.org/10.1103/PhysRevD.110.056043}{\emph{Phys. Rev. D} {\bfseries 110} (2024) 056043}, [\href{https://arxiv.org/abs/2408.01822}{{\ttfamily 2408.01822}}].

\bibitem{Almaeen:2024guo}
M.~Almaeen, T.~Alghamdi, B.~Kriesten, D.~Adams, Y.~Li, H.-W. Lin et~al., \emph{{VAIM-CFF: a variational autoencoder inverse mapper solution to Compton form factor extraction from deeply virtual exclusive reactions}}, \href{https://doi.org/10.1140/epjc/s10052-025-14091-3}{\emph{Eur. Phys. J. C} {\bfseries 85} (2025) 499}, [\href{https://arxiv.org/abs/2405.05826}{{\ttfamily 2405.05826}}].

\bibitem{Guo:2024wxy}
Y.~Guo, X.~Ji, M.~G. Santiago, J.~Yang and H.-C. Zhang, \emph{{Small-x gluon GPD constrained from deeply virtual J/{\ensuremath{\psi}} production and gluon PDF through universal-moment parametrization}}, \href{https://doi.org/10.1103/np1g-6c2z}{\emph{Phys. Rev. D} {\bfseries 112} (2025) 054036}, [\href{https://arxiv.org/abs/2409.17231}{{\ttfamily 2409.17231}}].

\bibitem{Guo:2025muf}
Y.~Guo, F.~P. Aslan, X.~Ji and M.~G. Santiago, \emph{{First Global Extraction of Generalized Parton Distributions from Experiment and Lattice Data with Next-to-Leading-Order Accuracy}}, \href{https://doi.org/10.1103/qct5-y7rp}{\emph{Phys. Rev. Lett.} {\bfseries 135} (2025) 261903}, [\href{https://arxiv.org/abs/2509.08037}{{\ttfamily 2509.08037}}].

\bibitem{Hatta:2022bxn}
Y.~Hatta and J.~Zhou, \emph{{Small-$x$ evolution of the gluon GPD $E_g$}}, \href{https://doi.org/10.1103/PhysRevLett.129.252002}{\emph{Phys. Rev. Lett.} {\bfseries 129} (2022) 252002}, [\href{https://arxiv.org/abs/2207.03378}{{\ttfamily 2207.03378}}].

\bibitem{Bhattacharya:2025fnz}
S.~Bhattacharya, C.-Q. He, Z.-B. Kang, D.~Padilla and J.~Penttala, \emph{{Parton distributions in the shockwave formalism}},  \href{https://arxiv.org/abs/2510.02254}{{\ttfamily 2510.02254}}.

\bibitem{Kovchegov:2025yyl}
Y.~V. Kovchegov, M.~G. Santiago and H.~Sun, \emph{{Unpolarized GPDs at small $x$ and non-zero skewness}},  \href{https://arxiv.org/abs/2512.10086}{{\ttfamily 2512.10086}}.

\bibitem{Gribov:1984tu}
L.~V. Gribov, E.~M. Levin and M.~G. Ryskin, \emph{{Semihard Processes in QCD}}, {\emph{Phys. Rept.} {\bfseries 100} (1983) 1--150}.

\bibitem{Iancu:2003xm}
E.~Iancu and R.~Venugopalan, \emph{{The Color glass condensate and high-energy scattering in QCD}}, \href{https://doi.org/10.1142/97898127955330005}{\emph{Quark-gluon plasma 4, edited by R.C. Hwa and X.-N. Wang} (2003) 249--363}, [\href{https://arxiv.org/abs/hep-ph/0303204}{{\ttfamily hep-ph/0303204}}].

\bibitem{Weigert:2005us}
H.~Weigert, \emph{Evolution at small {$x_{bj}$: The Color Glass Condensate}}, {\emph{Prog. Part. Nucl. Phys.} {\bfseries 55} (2005) 461--565}, [\href{https://arxiv.org/abs/hep-ph/0501087}{{\ttfamily hep-ph/0501087}}].

\bibitem{JalilianMarian:2005jf}
J.~Jalilian-Marian and Y.~V. Kovchegov, \emph{{Saturation physics and deuteron-Gold collisions at RHIC}}, \href{https://doi.org/10.1016/j.ppnp.2005.07.002}{\emph{Prog. Part. Nucl. Phys.} {\bfseries 56} (2006) 104--231}, [\href{https://arxiv.org/abs/hep-ph/0505052}{{\ttfamily hep-ph/0505052}}].

\bibitem{Gelis:2010nm}
F.~Gelis, E.~Iancu, J.~Jalilian-Marian and R.~Venugopalan, \emph{{The Color Glass Condensate}}, \href{https://doi.org/10.1146/annurev.nucl.010909.083629}{\emph{Ann.Rev.Nucl.Part.Sci.} {\bfseries 60} (2010) 463--489}, [\href{https://arxiv.org/abs/1002.0333}{{\ttfamily 1002.0333}}].

\bibitem{Albacete:2014fwa}
J.~L. Albacete and C.~Marquet, \emph{{Gluon saturation and initial conditions for relativistic heavy ion collisions}}, \href{https://doi.org/10.1016/j.ppnp.2014.01.004}{\emph{Prog.Part.Nucl.Phys.} {\bfseries 76} (2014) 1--42}, [\href{https://arxiv.org/abs/1401.4866}{{\ttfamily 1401.4866}}].

\bibitem{Kovchegov:2012mbw}
Y.~V. Kovchegov and E.~Levin, \emph{{Quantum chromodynamics at high energy}}, vol.~33.
\newblock Cambridge University Press, 2012.

\bibitem{Morreale:2021pnn}
A.~Morreale and F.~Salazar, \emph{{Mining for Gluon Saturation at Colliders}}, \href{https://doi.org/10.3390/universe7080312}{\emph{Universe} {\bfseries 7} (2021) 312}, [\href{https://arxiv.org/abs/2108.08254}{{\ttfamily 2108.08254}}].

\bibitem{Wallon:2023asa}
S.~Wallon, \emph{{The QCD Shockwave Approach at NLO: Towards Precision Physics in Gluonic Saturation}}, \href{https://doi.org/10.5506/APhysPolBSupp.16.5-A26}{\emph{Acta Phys. Polon. Supp.} {\bfseries 16} (2023) 26}, [\href{https://arxiv.org/abs/2302.04526}{{\ttfamily 2302.04526}}].

\bibitem{JalilianMarian:2004da}
J.~Jalilian-Marian and Y.~V. Kovchegov, \emph{{Inclusive two-gluon and valence quark-gluon production in DIS and p A}}, {\emph{Phys. Rev.} {\bfseries D70} (2004) 114017}, [\href{https://arxiv.org/abs/hep-ph/0405266}{{\ttfamily hep-ph/0405266}}].

\bibitem{Burkardt:2000za}
M.~Burkardt, \emph{{Impact parameter dependent parton distributions and off forward parton distributions for zeta ---{\ensuremath{>}} 0}}, \href{https://doi.org/10.1103/PhysRevD.62.071503}{\emph{Phys. Rev. D} {\bfseries 62} (2000) 071503}, [\href{https://arxiv.org/abs/hep-ph/0005108}{{\ttfamily hep-ph/0005108}}].

\bibitem{Burkardt:2002hr}
M.~Burkardt, \emph{{Impact parameter space interpretation for generalized parton distributions}}, \href{https://doi.org/10.1142/S0217751X03012370}{\emph{Int. J. Mod. Phys. A} {\bfseries 18} (2003) 173--208}, [\href{https://arxiv.org/abs/hep-ph/0207047}{{\ttfamily hep-ph/0207047}}].

\bibitem{Ji:2003ak}
X.-d. Ji, \emph{{Viewing the proton through 'color' filters}}, \href{https://doi.org/10.1103/PhysRevLett.91.062001}{\emph{Phys. Rev. Lett.} {\bfseries 91} (2003) 062001}, [\href{https://arxiv.org/abs/hep-ph/0304037}{{\ttfamily hep-ph/0304037}}].

\bibitem{Belitsky:2003nz}
A.~V. Belitsky, X.-d. Ji and F.~Yuan, \emph{{Quark imaging in the proton via quantum phase space distributions}}, \href{https://doi.org/10.1103/PhysRevD.69.074014}{\emph{Phys. Rev.} {\bfseries D69} (2004) 074014}, [\href{https://arxiv.org/abs/hep-ph/0307383}{{\ttfamily hep-ph/0307383}}].

\bibitem{Ji:1994av}
X.-D. Ji, \emph{{A QCD analysis of the mass structure of the nucleon}}, \href{https://doi.org/10.1103/PhysRevLett.74.1071}{\emph{Phys. Rev. Lett.} {\bfseries 74} (1995) 1071--1074}, [\href{https://arxiv.org/abs/hep-ph/9410274}{{\ttfamily hep-ph/9410274}}].

\bibitem{Polyakov:2002yz}
M.~V. Polyakov, \emph{{Generalized parton distributions and strong forces inside nucleons and nuclei}}, \href{https://doi.org/10.1016/S0370-2693(03)00036-4}{\emph{Phys. Lett. B} {\bfseries 555} (2003) 57--62}, [\href{https://arxiv.org/abs/hep-ph/0210165}{{\ttfamily hep-ph/0210165}}].

\bibitem{Burkert:2018bqq}
V.~D. Burkert, L.~Elouadrhiri and F.~X. Girod, \emph{{The pressure distribution inside the proton}}, \href{https://doi.org/10.1038/s41586-018-0060-z}{\emph{Nature} {\bfseries 557} (2018) 396--399}.

\bibitem{Kumericki:2019ddg}
K.~Kumeri{\v{c}}ki, \emph{{Measurability of pressure inside the proton}}, \href{https://doi.org/10.1038/s41586-019-1211-6}{\emph{Nature} {\bfseries 570} (2019) E1--E2}.

\bibitem{Ji:2025qax}
X.~Ji and C.~Yang, \emph{{A Journey of Seeking Pressures and Forces in the Nucleon}},  \href{https://arxiv.org/abs/2508.16727}{{\ttfamily 2508.16727}}.

\bibitem{Mueller:1999wm}
A.~H. Mueller, \emph{Parton saturation at small x and in large nuclei}, {\emph{Nucl. Phys.} {\bfseries B558} (1999) 285--303}, [\href{https://arxiv.org/abs/hep-ph/9904404}{{\ttfamily hep-ph/9904404}}].

\bibitem{Marquet:2009ca}
C.~Marquet, B.-W. Xiao and F.~Yuan, \emph{{Semi-inclusive Deep Inelastic Scattering at small x}}, \href{https://doi.org/10.1016/j.physletb.2009.10.099}{\emph{Phys. Lett.} {\bfseries B682} (2009) 207--211}, [\href{https://arxiv.org/abs/0906.1454}{{\ttfamily 0906.1454}}].

\bibitem{Dominguez:2011wm}
F.~Dominguez, C.~Marquet, B.-W. Xiao and F.~Yuan, \emph{{Universality of Unintegrated Gluon Distributions at small x}}, \href{https://doi.org/10.1103/PhysRevD.83.105005}{\emph{Phys.Rev.} {\bfseries D83} (2011) 105005}, [\href{https://arxiv.org/abs/1101.0715}{{\ttfamily 1101.0715}}].

\bibitem{Kovchegov:2015zha}
Y.~V. Kovchegov and M.~D. Sievert, \emph{{Calculating TMDs of a Large Nucleus: Quasi-Classical Approximation and Quantum Evolution}}, \href{https://doi.org/10.1016/j.nuclphysb.2015.12.008}{\emph{Nucl. Phys.} {\bfseries B903} (2016) 164--203}, [\href{https://arxiv.org/abs/1505.01176}{{\ttfamily 1505.01176}}].

\bibitem{Kotko:2015ura}
P.~Kotko, K.~Kutak, C.~Marquet, E.~Petreska, S.~Sapeta and A.~van Hameren, \emph{{Improved TMD factorization for forward dijet production in dilute-dense hadronic collisions}}, \href{https://doi.org/10.1007/JHEP09(2015)106}{\emph{JHEP} {\bfseries 09} (2015) 106}, [\href{https://arxiv.org/abs/1503.03421}{{\ttfamily 1503.03421}}].

\bibitem{Marquet:2016cgx}
C.~Marquet, E.~Petreska and C.~Roiesnel, \emph{{Transverse-momentum-dependent gluon distributions from JIMWLK evolution}}, \href{https://doi.org/10.1007/JHEP10(2016)065}{\emph{JHEP} {\bfseries 10} (2016) 065}, [\href{https://arxiv.org/abs/1608.02577}{{\ttfamily 1608.02577}}].

\bibitem{Hatta:2016aoc}
Y.~Hatta, Y.~Nakagawa, F.~Yuan, Y.~Zhao and B.~Xiao, \emph{{Gluon orbital angular momentum at small-$x$}}, \href{https://doi.org/10.1103/PhysRevD.95.114032}{\emph{Phys. Rev.} {\bfseries D95} (2017) 114032}, [\href{https://arxiv.org/abs/1612.02445}{{\ttfamily 1612.02445}}].

\bibitem{vanHameren:2016ftb}
A.~van Hameren, P.~Kotko, K.~Kutak, C.~Marquet, E.~Petreska and S.~Sapeta, \emph{{Forward di-jet production in p+Pb collisions in the small-x improved TMD factorization framework}}, \href{https://doi.org/10.1007/JHEP12(2016)034}{\emph{JHEP} {\bfseries 12} (2016) 034}, [\href{https://arxiv.org/abs/1607.03121}{{\ttfamily 1607.03121}}].

\bibitem{Kovchegov:2015pbl}
Y.~V. Kovchegov, D.~Pitonyak and M.~D. Sievert, \emph{{Helicity Evolution at Small-x}}, \href{https://doi.org/10.1007/JHEP01(2016)072}{\emph{JHEP} {\bfseries 01} (2016) 072}, [\href{https://arxiv.org/abs/1511.06737}{{\ttfamily 1511.06737}}].

\bibitem{Kovchegov:2018znm}
Y.~V. Kovchegov and M.~D. Sievert, \emph{{Small-$x$ Helicity Evolution: an Operator Treatment}}, \href{https://doi.org/10.1103/PhysRevD.99.054032}{\emph{Phys. Rev.} {\bfseries D99} (2019) 054032}, [\href{https://arxiv.org/abs/1808.09010}{{\ttfamily 1808.09010}}].

\bibitem{Kovchegov:2017lsr}
Y.~V. Kovchegov, D.~Pitonyak and M.~D. Sievert, \emph{{Small-$x$ Asymptotics of the Gluon Helicity Distribution}}, \href{https://doi.org/10.1007/JHEP10(2017)198}{\emph{JHEP} {\bfseries 10} (2017) 198}, [\href{https://arxiv.org/abs/1706.04236}{{\ttfamily 1706.04236}}].

\bibitem{Kovchegov:2021iyc}
Y.~V. Kovchegov and M.~G. Santiago, \emph{{Quark sivers function at small x: spin-dependent odderon and the sub-eikonal evolution}}, \href{https://doi.org/10.1007/JHEP11(2021)200}{\emph{JHEP} {\bfseries 11} (2021) 200}, [\href{https://arxiv.org/abs/2108.03667}{{\ttfamily 2108.03667}}].

\bibitem{Chirilli:2021lif}
G.~A. Chirilli, \emph{{High-energy operator product expansion at sub-eikonal level}}, \href{https://doi.org/10.1007/JHEP06(2021)096}{\emph{JHEP} {\bfseries 06} (2021) 096}, [\href{https://arxiv.org/abs/2101.12744}{{\ttfamily 2101.12744}}].

\bibitem{Cougoulic:2022gbk}
F.~Cougoulic, Y.~V. Kovchegov, A.~Tarasov and Y.~Tawabutr, \emph{{Quark and gluon helicity evolution at small x: revised and updated}}, \href{https://doi.org/10.1007/JHEP07(2022)095}{\emph{JHEP} {\bfseries 07} (2022) 095}, [\href{https://arxiv.org/abs/2204.11898}{{\ttfamily 2204.11898}}].

\bibitem{Kovchegov:2022kyy}
Y.~V. Kovchegov and M.~G. Santiago, \emph{{T-odd leading-twist quark TMDs at small x}}, \href{https://doi.org/10.1007/JHEP11(2022)098}{\emph{JHEP} {\bfseries 11} (2022) 098}, [\href{https://arxiv.org/abs/2209.03538}{{\ttfamily 2209.03538}}].

\bibitem{Kovchegov:2024aus}
Y.~V. Kovchegov and M.~Li, \emph{{Gluon double-spin asymmetry in the longitudinally polarized p + p collisions}}, \href{https://doi.org/10.1007/JHEP05(2024)177}{\emph{JHEP} {\bfseries 05} (2024) 177}, [\href{https://arxiv.org/abs/2403.06959}{{\ttfamily 2403.06959}}].

\bibitem{Hauksson:2024bvv}
S.~Hauksson, E.~Iancu, A.~H. Mueller, D.~N. Triantafyllopoulos and S.~Y. Wei, \emph{{TMD factorisation for diffractive jets in photon-nucleus interactions}}, \href{https://doi.org/10.1007/JHEP06(2024)180}{\emph{JHEP} {\bfseries 06} (2024) 180}, [\href{https://arxiv.org/abs/2402.14748}{{\ttfamily 2402.14748}}].

\bibitem{Balitsky:2024ozy}
I.~Balitsky, \emph{{$1/Q^2$ power corrections to TMD factorization for Drell-Yan hadronic tensor}}, \href{https://doi.org/10.1016/j.nuclphysb.2024.116658}{\emph{Nucl. Phys. B} {\bfseries 1006} (2024) 116658}, [\href{https://arxiv.org/abs/2404.15116}{{\ttfamily 2404.15116}}].

\bibitem{Borden:2024bxa}
J.~Borden, Y.~V. Kovchegov and M.~Li, \emph{{Helicity evolution at small x: quark to gluon and gluon to quark transition operators}}, \href{https://doi.org/10.1007/JHEP09(2024)037}{\emph{JHEP} {\bfseries 09} (2024) 037}, [\href{https://arxiv.org/abs/2406.11647}{{\ttfamily 2406.11647}}].

\bibitem{Kovchegov:2025gcg}
Y.~V. Kovchegov and M.~Li, \emph{{Weizs{\"a}cker-Williams gluon helicity distribution and inclusive dijet production in longitudinally polarized electron-proton collisions}}, \href{https://doi.org/10.1007/JHEP08(2025)206}{\emph{JHEP} {\bfseries 08} (2025) 206}, [\href{https://arxiv.org/abs/2504.12979}{{\ttfamily 2504.12979}}].

\bibitem{Balitsky:2026nux}
I.~Balitsky and A.~Prokudin, \emph{{Next-to-next-to-leading power corrections to unpolarized Semi-Inclusive Deep Inelastic Scattering}},  \href{https://arxiv.org/abs/2601.18882}{{\ttfamily 2601.18882}}.

\bibitem{Boussarie:2023xun}
R.~Boussarie and Y.~Mehtar-Tani, \emph{{Low and moderate x gluon contribution to exclusive Compton scattering processes}}, \href{https://doi.org/10.1007/JHEP10(2024)056}{\emph{JHEP} {\bfseries 10} (2024) 056}, [\href{https://arxiv.org/abs/2309.16576}{{\ttfamily 2309.16576}}].

\bibitem{Kovchegov:1999ji}
Y.~V. Kovchegov and E.~Levin, \emph{Diffractive dissociation including multiple pomeron exchanges in high parton density {QCD}}, {\emph{Nucl. Phys.} {\bfseries B577} (2000) 221--239}, [\href{https://arxiv.org/abs/hep-ph/9911523}{{\ttfamily hep-ph/9911523}}].

\bibitem{Kovner:2001vi}
A.~Kovner and U.~A. Wiedemann, \emph{{Eikonal evolution and gluon radiation}}, {\emph{Phys. Rev.} {\bfseries D64} (2001) 114002}, [\href{https://arxiv.org/abs/hep-ph/0106240}{{\ttfamily hep-ph/0106240}}].

\bibitem{Hentschinski:2005er}
M.~Hentschinski, H.~Weigert and A.~Schafer, \emph{{Extension of the color glass condensate approach to diffractive reactions}}, \href{https://doi.org/10.1103/PhysRevD.73.051501}{\emph{Phys. Rev.} {\bfseries D73} (2006) 051501}, [\href{https://arxiv.org/abs/hep-ph/0509272}{{\ttfamily hep-ph/0509272}}].

\bibitem{Kovner:2006ge}
A.~Kovner, M.~Lublinsky and H.~Weigert, \emph{{Treading on the cut: Semi inclusive observables at high energy}}, \href{https://doi.org/10.1103/PhysRevD.74.114023}{\emph{Phys. Rev.} {\bfseries D74} (2006) 114023}, [\href{https://arxiv.org/abs/hep-ph/0608258}{{\ttfamily hep-ph/0608258}}].

\bibitem{Hatta:2006hs}
Y.~Hatta, E.~Iancu, C.~Marquet, G.~Soyez and D.~N. Triantafyllopoulos, \emph{{Diffusive scaling and the high-energy limit of deep inelastic scattering in QCD at large N(c)}}, \href{https://doi.org/10.1016/j.nuclphysa.2006.04.003}{\emph{Nucl. Phys.} {\bfseries A773} (2006) 95--155}, [\href{https://arxiv.org/abs/hep-ph/0601150}{{\ttfamily hep-ph/0601150}}].

\bibitem{Kowalski:2006hc}
H.~Kowalski, L.~Motyka and G.~Watt, \emph{{Exclusive diffractive processes at HERA within the dipole picture}}, \href{https://doi.org/10.1103/PhysRevD.74.074016}{\emph{Phys. Rev. D} {\bfseries 74} (2006) 074016}, [\href{https://arxiv.org/abs/hep-ph/0606272}{{\ttfamily hep-ph/0606272}}].

\bibitem{Rezaeian:2012ji}
A.~H. Rezaeian, M.~Siddikov, M.~Van~de Klundert and R.~Venugopalan, \emph{{Analysis of combined HERA data in the Impact-Parameter dependent Saturation model}}, \href{https://doi.org/10.1103/PhysRevD.87.034002}{\emph{Phys. Rev. D} {\bfseries 87} (2013) 034002}, [\href{https://arxiv.org/abs/1212.2974}{{\ttfamily 1212.2974}}].

\bibitem{Hatta:2017cte}
Y.~Hatta, B.-W. Xiao and F.~Yuan, \emph{{Gluon Tomography from Deeply Virtual Compton Scattering at Small-x}}, \href{https://doi.org/10.1103/PhysRevD.95.114026}{\emph{Phys. Rev. D} {\bfseries 95} (2017) 114026}, [\href{https://arxiv.org/abs/1703.02085}{{\ttfamily 1703.02085}}].

\bibitem{Mantysaari:2021ryb}
H.~M{\"a}ntysaari and J.~Penttala, \emph{{Exclusive heavy vector meson production at next-to-leading order in the dipole picture}}, \href{https://doi.org/10.1016/j.physletb.2021.136723}{\emph{Phys. Lett. B} {\bfseries 823} (2021) 136723}, [\href{https://arxiv.org/abs/2104.02349}{{\ttfamily 2104.02349}}].

\bibitem{Mantysaari:2022kdm}
H.~M{\"a}ntysaari and J.~Penttala, \emph{{Complete calculation of exclusive heavy vector meson production at next-to-leading order in the dipole picture}}, \href{https://doi.org/10.1007/JHEP08(2022)247}{\emph{JHEP} {\bfseries 08} (2022) 247}, [\href{https://arxiv.org/abs/2204.14031}{{\ttfamily 2204.14031}}].

\bibitem{Toll:2012mb}
T.~Toll and T.~Ullrich, \emph{{Exclusive diffractive processes in electron-ion collisions}}, \href{https://doi.org/10.1103/PhysRevC.87.024913}{\emph{Phys. Rev. C} {\bfseries 87} (2013) 024913}, [\href{https://arxiv.org/abs/1211.3048}{{\ttfamily 1211.3048}}].

\bibitem{Mantysaari:2016jaz}
H.~M{\"a}ntysaari and B.~Schenke, \emph{{Revealing proton shape fluctuations with incoherent diffraction at high energy}}, \href{https://doi.org/10.1103/PhysRevD.94.034042}{\emph{Phys. Rev. D} {\bfseries 94} (2016) 034042}, [\href{https://arxiv.org/abs/1607.01711}{{\ttfamily 1607.01711}}].

\bibitem{Mueller:1989st}
A.~H. Mueller, \emph{{Small x Behavior and Parton Saturation: A QCD Model}}, {\emph{Nucl. Phys.} {\bfseries B335} (1990) 115}.

\bibitem{McLerran:1993ka}
L.~D. McLerran and R.~Venugopalan, \emph{Gluon distribution functions for very large nuclei at small transverse momentum}, {\emph{Phys. Rev.} {\bfseries D49} (1994) 3352--3355}, [\href{https://arxiv.org/abs/hep-ph/9311205}{{\ttfamily hep-ph/9311205}}].

\bibitem{McLerran:1993ni}
L.~D. McLerran and R.~Venugopalan, \emph{Computing quark and gluon distribution functions for very large nuclei}, {\emph{Phys. Rev.} {\bfseries D49} (1994) 2233--2241}, [\href{https://arxiv.org/abs/hep-ph/9309289}{{\ttfamily hep-ph/9309289}}].

\bibitem{McLerran:1994vd}
L.~D. McLerran and R.~Venugopalan, \emph{Green's functions in the color field of a large nucleus}, {\emph{Phys. Rev.} {\bfseries D50} (1994) 2225--2233}, [\href{https://arxiv.org/abs/hep-ph/9402335}{{\ttfamily hep-ph/9402335}}].

\bibitem{Balitsky:1995ub}
I.~Balitsky, \emph{{Operator expansion for high-energy scattering}}, \href{https://doi.org/10.1016/0550-3213(95)00638-9}{\emph{Nucl. Phys.} {\bfseries B463} (1996) 99--160}, [\href{https://arxiv.org/abs/hep-ph/9509348}{{\ttfamily hep-ph/9509348}}].

\bibitem{Balitsky:1998ya}
I.~Balitsky, \emph{Factorization and high-energy effective action}, {\emph{Phys. Rev.} {\bfseries D60} (1999) 014020}, [\href{https://arxiv.org/abs/hep-ph/9812311}{{\ttfamily hep-ph/9812311}}].

\bibitem{Kovchegov:1999yj}
Y.~V. Kovchegov, \emph{Small-x {$F_2$} structure function of a nucleus including multiple pomeron exchanges}, {\emph{Phys. Rev.} {\bfseries D60} (1999) 034008}, [\href{https://arxiv.org/abs/hep-ph/9901281}{{\ttfamily hep-ph/9901281}}].

\bibitem{Kovchegov:1999ua}
Y.~V. Kovchegov, \emph{Unitarization of the {BFKL} pomeron on a nucleus}, {\emph{Phys. Rev.} {\bfseries D61} (2000) 074018}, [\href{https://arxiv.org/abs/hep-ph/9905214}{{\ttfamily hep-ph/9905214}}].

\bibitem{Bhattacharya:2018lgm}
S.~Bhattacharya, A.~Metz, V.~K. Ojha, J.-Y. Tsai and J.~Zhou, \emph{{Exclusive double quarkonium production and generalized TMDs of gluons}}, \href{https://doi.org/10.1016/j.physletb.2022.137383}{\emph{Phys. Lett. B} {\bfseries 833} (2022) 137383}, [\href{https://arxiv.org/abs/1802.10550}{{\ttfamily 1802.10550}}].

\bibitem{Kovchegov:2018zeq}
Y.~V. Kovchegov and M.~D. Sievert, \emph{{Valence Quark Transversity at Small $x$}}, \href{https://doi.org/10.1103/PhysRevD.99.054033}{\emph{Phys. Rev.} {\bfseries D99} (2019) 054033}, [\href{https://arxiv.org/abs/1808.10354}{{\ttfamily 1808.10354}}].

\bibitem{Kovchegov:2019rrz}
Y.~V. Kovchegov, \emph{{Orbital Angular Momentum at Small $x$}}, \href{https://doi.org/10.1007/JHEP03(2019)174}{\emph{JHEP} {\bfseries 03} (2019) 174}, [\href{https://arxiv.org/abs/1901.07453}{{\ttfamily 1901.07453}}].

\bibitem{Kovchegov:1996ty}
Y.~V. Kovchegov, \emph{Non-abelian {Weizs\"{a}cker-Williams} field and a two- dimensional effective color charge density for a very large nucleus}, {\emph{Phys. Rev.} {\bfseries D54} (1996) 5463--5469}, [\href{https://arxiv.org/abs/hep-ph/9605446}{{\ttfamily hep-ph/9605446}}].

\bibitem{Jalilian-Marian:1997dw}
J.~Jalilian-Marian, A.~Kovner and H.~Weigert, \emph{The {Wilson} renormalization group for low x physics: Gluon evolution at finite parton density}, {\emph{Phys. Rev.} {\bfseries D59} (1998) 014015}, [\href{https://arxiv.org/abs/hep-ph/9709432}{{\ttfamily hep-ph/9709432}}].

\bibitem{Wu:2017rry}
B.~Wu and Y.~V. Kovchegov, \emph{{Time-dependent observables in heavy ion collisions. Part I. Setting up the formalism}}, \href{https://doi.org/10.1007/JHEP03(2018)158}{\emph{JHEP} {\bfseries 03} (2018) 158}, [\href{https://arxiv.org/abs/1709.02866}{{\ttfamily 1709.02866}}].

\bibitem{Cougoulic:2020tbc}
F.~Cougoulic and Y.~V. Kovchegov, \emph{{Helicity-dependent extension of the McLerran-Venugopalan model}}, \href{https://doi.org/10.1016/j.nuclphysa.2020.122051}{\emph{Nucl. Phys. A} {\bfseries 1004} (2020) 122051}, [\href{https://arxiv.org/abs/2005.14688}{{\ttfamily 2005.14688}}].

\bibitem{Kovchegov:2013cva}
Y.~V. Kovchegov and M.~D. Sievert, \emph{{Sivers function in the quasiclassical approximation}}, \href{https://doi.org/10.1103/PhysRevD.89.054035}{\emph{Phys. Rev.} {\bfseries D89} (2014) 054035}, [\href{https://arxiv.org/abs/1310.5028}{{\ttfamily 1310.5028}}].

\bibitem{Collins:1992kk}
J.~C. Collins, \emph{{Fragmentation of transversely polarized quarks probed in transverse momentum distributions}}, \href{https://doi.org/10.1016/0550-3213(93)90262-N}{\emph{Nucl.Phys.} {\bfseries B396} (1993) 161--182}, [\href{https://arxiv.org/abs/hep-ph/9208213}{{\ttfamily hep-ph/9208213}}].

\bibitem{Collins:2002kn}
J.~C. Collins, \emph{{Leading twist single transverse-spin asymmetries: Drell-Yan and deep inelastic scattering}}, \href{https://doi.org/10.1016/S0370-2693(02)01819-1}{\emph{Phys.Lett.} {\bfseries B536} (2002) 43--48}, [\href{https://arxiv.org/abs/hep-ph/0204004}{{\ttfamily hep-ph/0204004}}].

\bibitem{Brodsky:2002rv}
S.~J. Brodsky, D.~S. Hwang and I.~Schmidt, \emph{{Initial state interactions and single spin asymmetries in Drell-Yan processes}}, \href{https://doi.org/10.1016/S0550-3213(02)00617-X}{\emph{Nucl.Phys.} {\bfseries B642} (2002) 344--356}, [\href{https://arxiv.org/abs/hep-ph/0206259}{{\ttfamily hep-ph/0206259}}].

\bibitem{Jalilian-Marian:1997xn}
J.~Jalilian-Marian, A.~Kovner, L.~D. McLerran and H.~Weigert, \emph{The intrinsic glue distribution at very small x}, {\emph{Phys. Rev.} {\bfseries D55} (1997) 5414--5428}, [\href{https://arxiv.org/abs/hep-ph/9606337}{{\ttfamily hep-ph/9606337}}].

\bibitem{Kovchegov:1997pc}
Y.~V. Kovchegov, \emph{{Quantum structure of the non-Abelian Weizs\"{a}cker-Williams field for a very large nucleus}}, {\emph{Phys. Rev.} {\bfseries D55} (1997) 5445--5455}, [\href{https://arxiv.org/abs/hep-ph/9701229}{{\ttfamily hep-ph/9701229}}].

\bibitem{Kovchegov:1998bi}
Y.~V. Kovchegov and A.~H. Mueller, \emph{Gluon production in current nucleus and nucleon nucleus collisions in a quasi-classical approximation}, {\emph{Nucl. Phys.} {\bfseries B529} (1998) 451--479}, [\href{https://arxiv.org/abs/hep-ph/9802440}{{\ttfamily hep-ph/9802440}}].

\bibitem{Brodsky:2002ue}
S.~J. Brodsky, P.~Hoyer, N.~Marchal, S.~Peigne and F.~Sannino, \emph{{Structure functions are not parton probabilities}}, \href{https://doi.org/10.1103/PhysRevD.65.114025}{\emph{Phys. Rev. D} {\bfseries 65} (2002) 114025}, [\href{https://arxiv.org/abs/hep-ph/0104291}{{\ttfamily hep-ph/0104291}}].

\bibitem{Belitsky:2002sm}
A.~V. Belitsky, X.~Ji and F.~Yuan, \emph{{Final state interactions and gauge invariant parton distributions}}, \href{https://doi.org/10.1016/S0550-3213(03)00121-4}{\emph{Nucl.Phys.} {\bfseries B656} (2003) 165--198}, [\href{https://arxiv.org/abs/hep-ph/0208038}{{\ttfamily hep-ph/0208038}}].

\bibitem{Chirilli:2015fza}
G.~A. Chirilli, Y.~V. Kovchegov and D.~E. Wertepny, \emph{{Regularization of the Light-Cone Gauge Gluon Propagator Singularities Using Sub-Gauge Conditions}}, \href{https://doi.org/10.1007/JHEP12(2015)138}{\emph{JHEP} {\bfseries 12} (2015) 138}, [\href{https://arxiv.org/abs/1508.07962}{{\ttfamily 1508.07962}}].

\bibitem{Kovchegov:2003dm}
Y.~V. Kovchegov, L.~Szymanowski and S.~Wallon, \emph{{Perturbative odderon in the dipole model}}, \href{https://doi.org/10.1016/j.physletb.2004.02.036}{\emph{Phys.Lett.} {\bfseries B586} (2004) 267--281}, [\href{https://arxiv.org/abs/hep-ph/0309281}{{\ttfamily hep-ph/0309281}}].

\bibitem{Hatta:2005as}
Y.~Hatta, E.~Iancu, K.~Itakura and L.~McLerran, \emph{{Odderon in the color glass condensate}}, \href{https://doi.org/10.1016/j.nuclphysa.2005.05.163}{\emph{Nucl.Phys.} {\bfseries A760} (2005) 172--207}, [\href{https://arxiv.org/abs/hep-ph/0501171}{{\ttfamily hep-ph/0501171}}].

\bibitem{Kovchegov:2012ga}
Y.~V. Kovchegov and M.~D. Sievert, \emph{{A New Mechanism for Generating a Single Transverse Spin Asymmetry}}, \href{https://doi.org/10.1103/PhysRevD.86.034028}{\emph{Phys.Rev.} {\bfseries D86} (2012) 034028}, [\href{https://arxiv.org/abs/1201.5890}{{\ttfamily 1201.5890}}].

\bibitem{Boussarie:2023izj}
R.~Boussarie et~al., \emph{{TMD Handbook}},  \href{https://arxiv.org/abs/2304.03302}{{\ttfamily 2304.03302}}.

\bibitem{delRio:2024vvq}
O.~del Rio, A.~Prokudin, I.~Scimemi and A.~Vladimirov, \emph{{Transverse momentum moments}}, \href{https://doi.org/10.1103/PhysRevD.110.016003}{\emph{Phys. Rev. D} {\bfseries 110} (2024) 016003}, [\href{https://arxiv.org/abs/2402.01836}{{\ttfamily 2402.01836}}].

\bibitem{Meissner:2009ww}
S.~Meissner, A.~Metz and M.~Schlegel, \emph{{Generalized parton correlation functions for a spin-1/2 hadron}}, \href{https://doi.org/10.1088/1126-6708/2009/08/056}{\emph{JHEP} {\bfseries 08} (2009) 056}, [\href{https://arxiv.org/abs/0906.5323}{{\ttfamily 0906.5323}}].

\bibitem{Collins:2011zzd}
J.~Collins, \emph{{Foundations of perturbative QCD}}, vol.~32.
\newblock Cambridge University Press, 11, 2013.

\bibitem{Dokshitzer:1977sg}
Y.~L. Dokshitzer, \emph{{Calculation of the Structure Functions for Deep Inelastic Scattering and $e^+ e^-$ Annihilation by Perturbation Theory in Quantum Chromodynamics}}, {\emph{Sov. Phys. JETP} {\bfseries 46} (1977) 641--653}.

\bibitem{Gribov:1972ri}
V.~N. Gribov and L.~N. Lipatov, \emph{{Deep inelastic e p scattering in perturbation theory}}, {\emph{Sov. J. Nucl. Phys.} {\bfseries 15} (1972) 438--450}.

\bibitem{Altarelli:1977zs}
G.~Altarelli and G.~Parisi, \emph{{Asymptotic Freedom in Parton Language}}, \href{https://doi.org/10.1016/0550-3213(77)90384-4}{\emph{Nucl. Phys.} {\bfseries B126} (1977) 298}.

\bibitem{Efremov:1978rn}
A.~V. Efremov and A.~V. Radyushkin, \emph{{Asymptotical Behavior of Pion Electromagnetic Form-Factor in QCD}}, \href{https://doi.org/10.1007/BF01032111}{\emph{Theor. Math. Phys.} {\bfseries 42} (1980) 97--110}.

\bibitem{Lepage:1979zb}
G.~P. Lepage and S.~J. Brodsky, \emph{{Exclusive Processes in Quantum Chromodynamics: Evolution Equations for Hadronic Wave Functions and the Form-Factors of Mesons}}, \href{https://doi.org/10.1016/0370-2693(79)90554-9}{\emph{Phys. Lett. B} {\bfseries 87} (1979) 359--365}.

\bibitem{Jalilian-Marian:1997jx}
J.~Jalilian-Marian, A.~Kovner, A.~Leonidov and H.~Weigert, \emph{The {BFKL} equation from the {Wilson} renormalization group}, {\emph{Nucl. Phys.} {\bfseries B504} (1997) 415--431}, [\href{https://arxiv.org/abs/hep-ph/9701284}{{\ttfamily hep-ph/9701284}}].

\bibitem{Jalilian-Marian:1997gr}
J.~Jalilian-Marian, A.~Kovner, A.~Leonidov and H.~Weigert, \emph{The {Wilson} renormalization group for low x physics: Towards the high density regime}, {\emph{Phys. Rev.} {\bfseries D59} (1998) 014014}, [\href{https://arxiv.org/abs/hep-ph/9706377}{{\ttfamily hep-ph/9706377}}].

\bibitem{Iancu:2001ad}
E.~Iancu, A.~Leonidov and L.~D. McLerran, \emph{{The renormalization group equation for the color glass condensate}}, \href{https://doi.org/10.1016/S0370-2693(01)00524-X}{\emph{Phys. Lett.} {\bfseries B510} (2001) 133--144}.

\bibitem{Iancu:2000hn}
E.~Iancu, A.~Leonidov and L.~D. McLerran, \emph{Nonlinear gluon evolution in the color glass condensate. {I}}, {\emph{Nucl. Phys.} {\bfseries A692} (2001) 583--645}, [\href{https://arxiv.org/abs/hep-ph/0011241}{{\ttfamily hep-ph/0011241}}].

\bibitem{Bartels:1999yt}
J.~Bartels, L.~Lipatov and G.~Vacca, \emph{{A New odderon solution in perturbative QCD}}, \href{https://doi.org/10.1016/S0370-2693(00)00221-5}{\emph{Phys.Lett.} {\bfseries B477} (2000) 178--186}, [\href{https://arxiv.org/abs/hep-ph/9912423}{{\ttfamily hep-ph/9912423}}].

\bibitem{Bartels:1980pe}
J.~Bartels, \emph{{High-Energy Behavior in a Nonabelian Gauge Theory. 2. First Corrections to $T(n \to m)$ Beyond the Leading LNS Approximation}}, \href{https://doi.org/10.1016/0550-3213(80)90019-X}{\emph{Nucl.Phys.} {\bfseries B175} (1980) 365}.

\bibitem{Kwiecinski:1980wb}
J.~Kwiecinski and M.~Praszalowicz, \emph{{Three Gluon Integral Equation and Odd c Singlet Regge Singularities in QCD}}, \href{https://doi.org/10.1016/0370-2693(80)90909-0}{\emph{Phys.Lett.} {\bfseries B94} (1980) 413}.

\bibitem{Jaroszewicz:1980mq}
T.~Jaroszewicz, \emph{{Infrared Divergences and Regge Behavior in QCD}}, {\emph{Acta Phys. Polon. B} {\bfseries 11} (1980) 965}.

\bibitem{Kovchegov:2012rz}
Y.~V. Kovchegov, \emph{{Running Coupling Evolution for Diffractive Dissociation and the NLO Odderon Intercept}}, \href{https://doi.org/10.1063/1.4802180}{\emph{AIP Conf. Proc.} {\bfseries 1523} (2013) 335--338}, [\href{https://arxiv.org/abs/1212.2113}{{\ttfamily 1212.2113}}].

\bibitem{Janik:1998xj}
R.~A. Janik and J.~Wosiek, \emph{{Solution of the odderon problem}}, \href{https://doi.org/10.1103/PhysRevLett.82.1092}{\emph{Phys. Rev. Lett.} {\bfseries 82} (1999) 1092--1095}, [\href{https://arxiv.org/abs/hep-th/9802100}{{\ttfamily hep-th/9802100}}].

\bibitem{Caron-Huot:2013fea}
S.~Caron-Huot, \emph{{When does the gluon reggeize?}}, \href{https://doi.org/10.1007/JHEP05(2015)093}{\emph{JHEP} {\bfseries 05} (2015) 093}, [\href{https://arxiv.org/abs/1309.6521}{{\ttfamily 1309.6521}}].

\bibitem{Bartels:2013yga}
J.~Bartels and G.~P. Vacca, \emph{{Generalized Bootstrap Equations and possible implications for the NLO Odderon}}, \href{https://doi.org/10.1140/epjc/s10052-013-2602-8}{\emph{Eur. Phys. J. C} {\bfseries 73} (2013) 2602}, [\href{https://arxiv.org/abs/1307.3985}{{\ttfamily 1307.3985}}].

\bibitem{Brower:2008cy}
R.~C. Brower, M.~Djuric and C.-I. Tan, \emph{{Odderon in gauge/string duality}}, \href{https://doi.org/10.1088/1126-6708/2009/07/063}{\emph{JHEP} {\bfseries 0907} (2009) 063}, [\href{https://arxiv.org/abs/0812.0354}{{\ttfamily 0812.0354}}].

\bibitem{Avsar:2009hc}
E.~Avsar, Y.~Hatta and T.~Matsuo, \emph{{Odderon in baryon-baryon scattering from the AdS/CFT correspondence}}, \href{https://doi.org/10.1007/JHEP03(2010)037}{\emph{JHEP} {\bfseries 03} (2010) 037}, [\href{https://arxiv.org/abs/0912.3806}{{\ttfamily 0912.3806}}].

\bibitem{Brower:2014wha}
R.~C. Brower, M.~S. Costa, M.~Djuri\'c, T.~Raben and C.-I. Tan, \emph{{Strong Coupling Expansion for the Conformal Pomeron/Odderon Trajectories}}, \href{https://doi.org/10.1007/JHEP02(2015)104}{\emph{JHEP} {\bfseries 02} (2015) 104}, [\href{https://arxiv.org/abs/1409.2730}{{\ttfamily 1409.2730}}].

\bibitem{Kuraev:1977fs}
E.~A. Kuraev, L.~N. Lipatov and V.~S. Fadin, \emph{{The Pomeranchuk singlularity in non-Abelian gauge theories}}, {\emph{Sov. Phys. JETP} {\bfseries 45} (1977) 199--204}.

\bibitem{Balitsky:1978ic}
I.~Balitsky and L.~Lipatov, \emph{{The Pomeranchuk Singularity in Quantum Chromodynamics}}, {\emph{Sov.J.Nucl.Phys.} {\bfseries 28} (1978) 822--829}.

\bibitem{Bartels:1981jh}
J.~Bartels and M.~Loewe, \emph{{The Nonforward {QCD} Ladder Diagrams}}, \href{https://doi.org/10.1007/BF01558265}{\emph{Z. Phys. C} {\bfseries 12} (1982) 263}.

\bibitem{Lepage:1980fj}
G.~P. Lepage and S.~J. Brodsky, \emph{Exclusive processes in perturbative quantum chromodynamics}, {\emph{Phys. Rev.} {\bfseries D22} (1980) 2157}.

\bibitem{Brodsky:1997de}
S.~J. Brodsky, H.-C. Pauli and S.~S. Pinsky, \emph{{Quantum chromodynamics and other field theories on the light cone}}, \href{https://doi.org/10.1016/S0370-1573(97)00089-6}{\emph{Phys.Rept.} {\bfseries 301} (1998) 299--486}, [\href{https://arxiv.org/abs/hep-ph/9705477}{{\ttfamily hep-ph/9705477}}].

\bibitem{Kovchegov:2016zex}
Y.~V. Kovchegov, D.~Pitonyak and M.~D. Sievert, \emph{{Helicity Evolution at Small $x$: Flavor Singlet and Non-Singlet Observables}}, \href{https://doi.org/10.1103/PhysRevD.95.014033}{\emph{Phys. Rev.} {\bfseries D95} (2017) 014033}, [\href{https://arxiv.org/abs/1610.06197}{{\ttfamily 1610.06197}}].

\bibitem{Cougoulic:2019aja}
F.~Cougoulic and Y.~V. Kovchegov, \emph{{Helicity-dependent generalization of the JIMWLK evolution}}, \href{https://doi.org/10.1103/PhysRevD.100.114020}{\emph{Phys. Rev.} {\bfseries D100} (2019) 114020}, [\href{https://arxiv.org/abs/1910.04268}{{\ttfamily 1910.04268}}].

\bibitem{Mueller:1994rr}
A.~H. Mueller, \emph{Soft gluons in the infinite momentum wave function and the {BFKL} pomeron}, {\emph{Nucl. Phys.} {\bfseries B415} (1994) 373--385}.

\bibitem{Mueller:1994jq}
A.~H. Mueller and B.~Patel, \emph{Single and double {BFKL} pomeron exchange and a dipole picture of high-energy hard processes}, {\emph{Nucl. Phys.} {\bfseries B425} (1994) 471--488}, [\href{https://arxiv.org/abs/hep-ph/9403256}{{\ttfamily hep-ph/9403256}}].

\bibitem{Mueller:1995gb}
A.~H. Mueller, \emph{Unitarity and the {BFKL} pomeron}, {\emph{Nucl. Phys.} {\bfseries B437} (1995) 107--126}, [\href{https://arxiv.org/abs/hep-ph/9408245}{{\ttfamily hep-ph/9408245}}].

\bibitem{Weigert:2000gi}
H.~Weigert, \emph{Unitarity at small {B}jorken x}, {\emph{Nucl. Phys.} {\bfseries A703} (2002) 823--860}, [\href{https://arxiv.org/abs/hep-ph/0004044}{{\ttfamily hep-ph/0004044}}].

\bibitem{Ferreiro:2001qy}
E.~Ferreiro, E.~Iancu, A.~Leonidov and L.~McLerran, \emph{Nonlinear gluon evolution in the color glass condensate. {II}}, {\emph{Nucl. Phys.} {\bfseries A703} (2002) 489--538}, [\href{https://arxiv.org/abs/hep-ph/0109115}{{\ttfamily hep-ph/0109115}}].

\bibitem{Kovner:2005qj}
A.~Kovner and M.~Lublinsky, \emph{{Odderon and seven Pomerons: QCD Reggeon field theory from JIMWLK evolution}}, \href{https://doi.org/10.1088/1126-6708/2007/02/058}{\emph{JHEP} {\bfseries 0702} (2007) 058}, [\href{https://arxiv.org/abs/hep-ph/0512316}{{\ttfamily hep-ph/0512316}}].

\bibitem{Kirschner:1983di}
R.~Kirschner and L.~Lipatov, \emph{{Double Logarithmic Asymptotics and Regge Singularities of Quark Amplitudes with Flavor Exchange}}, \href{https://doi.org/10.1016/0550-3213(83)90178-5}{\emph{Nucl.Phys.} {\bfseries B213} (1983) 122--148}.

\bibitem{Borden:2023ugd}
J.~Borden and Y.~V. Kovchegov, \emph{{Analytic solution for the revised helicity evolution at small x and large Nc: New resummed gluon-gluon polarized anomalous dimension and intercept}}, \href{https://doi.org/10.1103/PhysRevD.108.014001}{\emph{Phys. Rev. D} {\bfseries 108} (2023) 014001}, [\href{https://arxiv.org/abs/2304.06161}{{\ttfamily 2304.06161}}].

\bibitem{Adamiak:2023okq}
D.~Adamiak, Y.~V. Kovchegov and Y.~Tawabutr, \emph{{Helicity evolution at small x: Revised asymptotic results at large Nc and Nf}}, \href{https://doi.org/10.1103/PhysRevD.108.054005}{\emph{Phys. Rev. D} {\bfseries 108} (2023) 054005}, [\href{https://arxiv.org/abs/2306.01651}{{\ttfamily 2306.01651}}].

\bibitem{Borden:2025ehe}
J.~Borden and Y.~V. Kovchegov, \emph{{Analytic Solution for the Helicity Evolution Equations at Small $x$ and Large $N_c\&N_f$}},  \href{https://arxiv.org/abs/2508.00195}{{\ttfamily 2508.00195}}.

\bibitem{Itakura:2003jp}
K.~Itakura, Y.~V. Kovchegov, L.~McLerran and D.~Teaney, \emph{{Baryon stopping and valence quark distribution at small x}}, \href{https://doi.org/10.1016/j.nuclphysa.2003.10.016}{\emph{Nucl. Phys.} {\bfseries A730} (2004) 160--190}, [\href{https://arxiv.org/abs/hep-ph/0305332}{{\ttfamily hep-ph/0305332}}].

\bibitem{Forshaw:1997dc}
J.~R. Forshaw and D.~A. Ross, \emph{{Quantum chromodynamics and the pomeron}}.
\newblock Cambridge University Press, Cambridge, UK, 2004.

\bibitem{Ioffe:2010zz}
B.~L. Ioffe, V.~S. Fadin and L.~N. Lipatov, \emph{{Quantum chromodynamics: Perturbative and nonperturbative aspects}}.
\newblock Cambridge Univ. Press, 2010, \href{https://doi.org/10.1017/CBO9780511711817}{10.1017/CBO9780511711817}.

\bibitem{Gotsman:2020mkd}
E.~Gotsman, E.~Levin and I.~Potashnikova, \emph{{New parton model for the soft interactions at high energies: The odderon}}, \href{https://doi.org/10.1103/PhysRevD.101.094021}{\emph{Phys. Rev. D} {\bfseries 101} (2020) 094021}, [\href{https://arxiv.org/abs/2003.09155}{{\ttfamily 2003.09155}}].

\bibitem{Mueller:2012bn}
A.~Mueller and S.~Munier, \emph{{$p_{\perp}$-broadening and production processes versus dipole/quadrupole amplitudes at next-to-leading order}}, \href{https://doi.org/10.1016/j.nuclphysa.2012.08.005}{\emph{Nucl.Phys.} {\bfseries A893} (2012) 43--86}, [\href{https://arxiv.org/abs/1206.1333}{{\ttfamily 1206.1333}}].

\bibitem{Diehl:1998sm}
M.~Diehl and T.~Gousset, \emph{{Time ordering in off diagonal parton distributions}}, \href{https://doi.org/10.1016/S0370-2693(98)00439-0}{\emph{Phys. Lett. B} {\bfseries 428} (1998) 359--370}, [\href{https://arxiv.org/abs/hep-ph/9801233}{{\ttfamily hep-ph/9801233}}].

\bibitem{Landshoff:1970ff}
P.~V. Landshoff, J.~C. Polkinghorne and R.~D. Short, \emph{{a Nonperturbative parton model of current interactions}}, \href{https://doi.org/10.1016/0550-3213(71)90375-0}{\emph{Nucl. Phys. B} {\bfseries 28} (1971) 225--239}.

\bibitem{Frankfurt:1997ha}
L.~Frankfurt, A.~Freund, V.~Guzey and M.~Strikman, \emph{{Nondiagonal parton distribution in the leading logarithmic approximation}}, \href{https://doi.org/10.1016/S0370-2693(97)01152-0}{\emph{Phys. Lett. B} {\bfseries 418} (1998) 345--354}, [\href{https://arxiv.org/abs/hep-ph/9703449}{{\ttfamily hep-ph/9703449}}].

\bibitem{Radyushkin:1997ki}
A.~V. Radyushkin, \emph{{Nonforward parton distributions}}, \href{https://doi.org/10.1103/PhysRevD.56.5524}{\emph{Phys. Rev. D} {\bfseries 56} (1997) 5524--5557}, [\href{https://arxiv.org/abs/hep-ph/9704207}{{\ttfamily hep-ph/9704207}}].

\bibitem{Jaffe:1983hp}
R.~L. Jaffe, \emph{{Parton Distribution Functions for Twist Four}}, \href{https://doi.org/10.1016/0550-3213(83)90361-9}{\emph{Nucl. Phys. B} {\bfseries 229} (1983) 205--230}.

\bibitem{Muller:2013jur}
D.~M{\"u}ller, T.~Lautenschlager, K.~Passek-Kumericki and A.~Schaefer, \emph{{Towards a fitting procedure to deeply virtual meson production - the next-to-leading order case}}, \href{https://doi.org/10.1016/j.nuclphysb.2014.04.012}{\emph{Nucl. Phys. B} {\bfseries 884} (2014) 438--546}, [\href{https://arxiv.org/abs/1310.5394}{{\ttfamily 1310.5394}}].

\bibitem{Accardi:2012qut}
A.~Accardi et~al., \emph{{Electron Ion Collider: The Next QCD Frontier}}, \href{https://doi.org/10.1140/epja/i2016-16268-9}{\emph{Eur. Phys. J.} {\bfseries A52} (2016) 268}, [\href{https://arxiv.org/abs/1212.1701}{{\ttfamily 1212.1701}}].

\bibitem{Boer:2011fh}
D.~Boer et~al., \emph{{Gluons and the quark sea at high energies: Distributions, polarization, tomography}},  \href{https://arxiv.org/abs/1108.1713}{{\ttfamily 1108.1713}}.

\bibitem{Proceedings:2020eah}
A.~Prokudin, Y.~Hatta, Y.~Kovchegov and C.~Marquet, eds., \emph{{Proceedings, Probing Nucleons and Nuclei in High Energy Collisions: Dedicated to the Physics of the Electron Ion Collider}: {Seattle (WA), United States, October 1 - November 16, 2018}}, WSP, 2020.
\newblock 10.1142/11684.

\bibitem{AbdulKhalek:2021gbh}
R.~Abdul~Khalek et~al., \emph{{Science Requirements and Detector Concepts for the Electron-Ion Collider}: {EIC Yellow Report}}, \href{https://doi.org/10.1016/j.nuclphysa.2022.122447}{\emph{Nucl. Phys. A} {\bfseries 1026} (2022) 122447}, [\href{https://arxiv.org/abs/2103.05419}{{\ttfamily 2103.05419}}].

\bibitem{Alexandrou:2026jpd}
C.~Alexandrou et~al., \emph{{Precision QCD with the Electron-Ion Collider}},  \href{https://arxiv.org/abs/2604.04765}{{\ttfamily 2604.04765}}.

\end{thebibliography}
\end{document}